\documentclass[11pt,prd,letterpaper,amsmath,amssymb,final,nofootinbib
,unsortedaddress,superscriptaddress
]{revtex4-1}

\pdfoutput=1

\usepackage[top=0.9in, bottom=0.9in, left = 0.9in, right=0.9in]{geometry}

\usepackage{graphicx}
\usepackage{dcolumn}
\usepackage{bbm}
\usepackage{textcomp}
\usepackage{color}
\usepackage{gensymb}
\usepackage{wasysym}
\usepackage{amssymb}
\usepackage{enumerate}
\usepackage{geometry}
\usepackage{setspace}
\begin{document}

\linespread{1.4}


\title{ Advanced Scintillator Detector Concept (ASDC): \\ {\small A Concept Paper on the Physics Potential of Water-Based Liquid Scintillator}}

\author{J.~R.~Alonso}
\affiliation{Massachusetts Institute of Technology, Cambridge, MA 02139, USA}

\author{N.~Barros}
\affiliation{Department of Physics and Astronomy, University of Pennsylvania, Philadelphia, PA 19104, USA}

\author{M.~Bergevin}
\affiliation{Physics Department, University of California, Davis CA 95616, USA}

\author{A.~Bernstein}
\affiliation{Lawrence Livermore National Laboratory, Livermore, CA 94550, USA}

\author{L.~Bignell}
\affiliation{Brookhaven National Laboratory, Upton, NY 11973, USA}

\author{E.~Blucher}
\affiliation{Enrico Fermi Institute, University of Chicago, Chicago, IL 60637, USA}

\author{F.~Calaprice}
\affiliation{Department of Physics, Princeton University, NJ 08544, USA}

\author{J.~M.~Conrad}
\affiliation{Massachusetts Institute of Technology, Cambridge, MA 02139, USA}

\author{F.~B.~Descamps}
\affiliation{Lawrence Berkeley National Laboratory, Berkeley, CA 94720, USA}

\author{M.~V.~Diwan}
\affiliation{Brookhaven National Laboratory, Upton, NY 11973, USA}

\author{D.~A.~Dwyer}
\affiliation{Lawrence Berkeley National Laboratory, Berkeley, CA 94720, USA}

\author{S.~T.~Dye}
\affiliation{Department of Natural Sciences, Hawaii Pacific University, Kaneohe, Hawaii 96744, USA}

\author{A.~Elagin}
\affiliation{Enrico Fermi Institute, University of Chicago, Chicago, IL 60637, USA}

\author{P.~Feng}
\affiliation{Sandia National Laboratories, Livermore, CA 94550, USA}

\author{C.~Grant}
\affiliation{Physics Department, University of California, Davis CA 95616, USA}

\author{S.~Grullon}
\affiliation{Department of Physics and Astronomy, University of Pennsylvania, Philadelphia, PA 19104, USA}

\author{S.~Hans}
\affiliation{Brookhaven National Laboratory, Upton, NY 11973, USA}

\author{D.~E.~Jaffe}
\affiliation{Brookhaven National Laboratory, Upton, NY 11973, USA}

\author{S.~H.~Kettell}
\affiliation{Brookhaven National Laboratory, Upton, NY 11973, USA}

\author{J.~R.~Klein}
\affiliation{Department of Physics and Astronomy, University of Pennsylvania, Philadelphia, PA 19104, USA}

\author{K.~Lande}
\affiliation{Department of Physics and Astronomy, University of Pennsylvania, Philadelphia, PA 19104, USA}

\author{J.~G.~Learned}
\affiliation{Department of Physics and Astronomy, University of Hawaii at Manoa, Honolulu, HI 96922 USA}

\author{K.~B.~Luk}
\affiliation{Lawrence Berkeley National Laboratory, Berkeley, CA 94720, USA}
\affiliation{Department of Physics, University of California, Berkeley, CA 94720, USA}

\author{J.~Maricic}
\affiliation{Department of Physics and Astronomy, University of Hawaii at Manoa, Honolulu, HI 96922 USA}

\author{P.~Marleau}
\affiliation{Sandia National Laboratories, Livermore, CA 94550, USA}

\author{A.~Mastbaum}
\affiliation{Department of Physics and Astronomy, University of Pennsylvania, Philadelphia, PA 19104, USA}

\author{W.~F.~McDonough}
\affiliation{Department of Geology, University of Maryland, College Park, MD 20742, USA}

\author{L.~Oberauer}
\affiliation{TUM, Physik-Department, James-Franck-Str. 1, 85748 Garching, Germany}

\author{G.~D.~Orebi Gann\footnote{Corresponding Author, email: gorebigann@lbl.gov}}
\affiliation{Lawrence Berkeley National Laboratory, Berkeley, CA 94720, USA}
\affiliation{Department of Physics, University of California, Berkeley, CA 94720, USA}

\author{R.~Rosero}
\affiliation{Brookhaven National Laboratory, Upton, NY 11973, USA}

\author{S.~D.~Rountree}
\affiliation{Department of Physics, Virginia Polytechnic Institute and State University, Blacksburg, VA 24061, USA}

\author{M.~C.~Sanchez}
\affiliation{Department of Physics and Astronomy, Iowa State University, Ames, IA 50011, USA}

\author{M.~H.~Shaevitz}
\affiliation{Department of Physics, Columbia University, New York, NY 10027, USA}

\author{T.~M.~Shokair}
\affiliation{Department of Nuclear Engineering, University of California, Berkeley, CA 94720, USA}

\author{M.~B.~Smy}
\affiliation{Department of Physics and Astronomy,
University of California, Irvine, CA 92697, USA}

\author{A.~Stahl}
\affiliation{RWTH Aachen University, Physikzentrum, 52074 Aachen, Germany}

\author{M.~Strait}
\affiliation{Enrico Fermi Institute, University of Chicago, Chicago, IL 60637, USA}

\author{R.~Svoboda}
\affiliation{Physics Department, University of California, Davis CA 95616, USA}

\author{N.~Tolich}
\affiliation{Center for Experimental Nuclear Physics and Astrophysics, and Department of Physics, University of Washington, Seattle, WA 98195, USA}

\author{M.~R.~Vagins}
\affiliation{Department of Physics and Astronomy,
University of California, Irvine, CA 92697, USA}

\author{K.~A.~van~Bibber}
\affiliation{Department of Nuclear Engineering, University of California, Berkeley, CA 94720, USA}

\author{B.~Viren}
\affiliation{Brookhaven National Laboratory, Upton, NY 11973, USA}

\author{R.~B.~Vogelaar}
\affiliation{Department of Physics, Virginia Polytechnic Institute and State University, Blacksburg, VA 24061, USA}

\author{M.~J.~Wetstein}
\affiliation{Enrico Fermi Institute, University of Chicago, Chicago, IL 60637, USA}

\author{L.~Winslow}
\affiliation{Massachusetts Institute of Technology, Cambridge, MA 02139, USA}

\author{B.~Wonsak}
\affiliation{Institute for Experimental Physics, University of Hamburg, Germany}

\author{E.~T.~Worcester}
\affiliation{Brookhaven National Laboratory, Upton, NY 11973, USA}

\author{M.~Wurm}
\affiliation{Institute of Physics \& EC PRISMA, Johannes Gutenberg-University Mainz, 55128 Mainz, Germany}

\author{M.~Yeh}
\affiliation{Brookhaven National Laboratory, Upton, NY 11973, USA}

\author{C.~Zhang}
\affiliation{Brookhaven National Laboratory, Upton, NY 11973, USA}



\maketitle

%
%
%
%
%
%
%



\section*{Executive Summary}

The recent development of Water-based Liquid Scintillator (WbLS), and the
concurrent development of high-efficiency and high-precision-timing light
sensors, has opened up the possibility for a new kind of large-scale detector
capable of a very broad program of physics.   The program would include
determination of the neutrino mass hierarchy and observation of CP violation
with long-baseline neutrinos, searches for proton decay, ultra-precise solar
neutrino measurements, geo- and supernova neutrinos including diffuse supernova
antineutrinos, and neutrinoless double beta decay.  
We outline here the basic requirements of the Advanced Scintillation Detector
Concept (ASDC), which combines the use of WbLS, doping with a number of potential isotopes for a range of physics goals, high efficiency and ultra-fast
timing photosensors, and a deep underground location.  We are considering such a detector at the Long Baseline
Neutrino Facility (LBNF) far site, where the ASDC could operate in conjunction with the liquid argon tracking detector proposed by the LBNE collaboration.  The goal is the deployment of a
30--100 kiloton-scale detector, the basic elements of which are being developed
now in experiments such as WATCHMAN, ANNIE, SNO+, and EGADS.

The ASDC combines the benefits of both water Cherenkov detection and pure
liquid scintillator in a single detector.  WbLS uniquely offers the high light
yield and low threshold of scintillator with the 
directionality of a Cherenkov detector~\cite{wbls}.  Initial light absorbance measurements 
promise a long attenuation length, perhaps even close to the attenuation length of pure water at wavelengths above 400~nm. 
The use of high-sensitivity photomultipliers or high-precision timing
measurement devices such as the newly-developed Large Area Picosecond
Photo-Detectors (LAPPDs)~\cite{lappd1, lappd2, lappd3, lappd4, lappd5} would allow excellent separation of the prompt
Cherenkov light from the delayed scintillation, enabling a strong long-baseline
program, especially when combined with planned large liquid argon detectors.
Thus an affordable, large-scale detector capable of physics from below 1 MeV to
many GeV is possible.

WbLS chemistry also allows loading of metallic ions as an additional target for
particle detection, including $^6$Li for short-baseline reactor anomaly
studies, $^7$Li for charged-current solar neutrino detection, $^{\rm nat}$Gd
for neutron tagging enhancement, 
or $^{\rm nat}$Pb for
total-absorption calorimetry or solar neutrino studies.  Loading of isotopes that undergo double beta decay would allow the ASDC to pursue a
program of  neutrinoless double-beta decay ($0\nu\beta\beta$). 
Double beta decay isotopes that cannot be loaded in pure scintillator, due to
their hydrophilic nature, would be accessible with WbLS technology. With the
large size and high fractional loading possible, 
$0\nu\beta\beta$ target masses of tens of tons or more could be achieved
with this technique, offering a practical way to push sensitivities to
lepton number violation toward the normal hierarchy region. 

The formula and principle of a mass-produced WbLS have been developed and
demonstrated at the BNL Liquid Scintillator Development Facility.  Both bench-top
measurements and low-intensity proton beams have been utilized at incident
energies above and below Cherenkov threshold. 
Different metal-doped WbLS samples have been produced with high
stability (including Li, Gd, Te, Pb, Zr), with loadings of a few tenths to
several percent for different experimental requirements,  and are continuously
monitored at different temperature ranges over time. A further study for
prototyping large-scale liquid production and deployment at the ton-scale is in
high demand, which would allow a direct measurement of the separation of Cherenkov and
scintillation light, and attenuation measurements at longer scales. The instrumentation for large-scale liquid production
is still under design. The assembly of  a 1-ton WbLS prototype is ongoing and
expected to be ready for  deployment in 2015.  In addition, a large scale ($\sim$3~kT) deployment of WbLS is intended for  the second phase of the planned
WATCHMAN detector.

Development of fast timing solutions, such as LAPPDs, with single-photon timing resolution better than 50~ps would
revolutionize the reconstruction of neutrino interactions in liquid detectors.  This would provide a significant enhancement to  the ASDC, where separation of scintillation and
Cherenkov components would provide multiple handles for particle identification and
reconstruction. LAPPD precision timing would allow improved discrimination
between quasi-elastic (QE) and non-QE events in the GeV energy range, and
between $\nu_{e}$ charged current events and neutral current induced
$\pi^{0}$'s. The large size of an ASDC would provide sensitivity to the second
oscillation maximum in long-baseline neutrino oscillation, greatly
complementing the capabilities of a large liquid argon TPC. Adaptation of
LAPPDs for underwater use and deployment of a significant number for
$in$--$situ$ measurement of atmospheric neutrinos is planned for the first
phase of the WATCHMAN detector and also for the ANNIE neutrino test beam at
Fermilab.

The ASDC is the ultimate goal of a new research effort to design a massive detector with  unique
capabilities for exploring physics below the Cherenkov threshold, 
 and the potential to load metallic isotopes for different physics applications. 
This novel approach to neutrino detection could transform the field, allowing the construction of a single large-scale detector with the 
unique ability to do both conventional neutrino physics and a  complementary set of rare event searches.  The combination of these physics opportunities makes the ASDC a powerful scientific instrument for the next-generation of experiments, with the capability to simultaneously resolve many
of the major open questions in neutrino physics. 
A vigorous and forward-looking program of R\&D
towards the ASDC is called for that would lead to a proposal for such a detector
and program at the LBNF.

\newpage

\tableofcontents
\setcounter{tocdepth}{5}
\newpage

\section{Introduction}
The recent development of Water-based Liquid Scintillator (WbLS), and the
concurrent development of high-efficiency and high-precision-timing light
sensors, has opened up the possibility for a new kind of large-scale detector
capable of a very broad program of physics. 
 In order to improve sensitivity to unprecedented levels, detectors will need low energy thresholds, good energy resolution, and high-precision separation of signal from background events.  
An approach that has achieved great success in the past, in particular in neutrino physics, is to house a light-producing target in a large-scale detector.  Both water and liquid scintillator (LS) have been used to great effect by experiments such as Super-Kamiokande~\cite{sk}, SNO~\cite{sno}, 
KamLAND~\cite{kl}, and Borexino~\cite{Bor}. 
Incident neutrinos create electrons either by elastic scattering or through
charged-current interactions, which in turn generate light in the target.  The
Cherenkov light produced in water has a unique cone-like topology, which
provides these detectors with excellent determination of particle direction,
something not possible with isotropic scintillation light.  However, the
factor of 50 higher light yield in LS significantly lowers the achievable
energy threshold and improves the energy resolution.  Light propagation is also
critical, since it impacts overall photon collection.   Current LS detectors
are limited to kilotonne scales by attenuation of light in the target, and also
by target cost.  

A WbLS target allows for the high light yield
and correspondingly high energy resolution and low threshold of a scintillator
experiment in a directionally sensitive detector.  The admixture of water also
increases the attenuation length, therefore allowing the possibility of a very
large detector. The development of new, high-precision timing devices such as
LAPPDs~\cite{lappd1, lappd2, lappd3, lappd4, lappd5} could enable a high degree of separation between the prompt Cherenkov
light and the delayed scintillation light, offering many advantages including a
powerful long-baseline program.  The ability to load hydrophilic ions further
increases the physics program, including the potential for a $0\nu\beta\beta$
search with an isotopic mass on the scale of tens of tons.

This document discusses the potential of the Advanced Scintillation Detector Concept (ASDC) for a broad range of physics goals.  A further program of R\&D is required in order to realize the full potential of the new WbLS target medium.  The results from this program will be a necessary input to the optimization of the final detector configuration in order to maximize the physics potential.  (This is discussed in more detail in Section~\ref{s:detector}.)  The following sections refer to different options for potential target sizes, isotopic loadings, and coverage, depending on the requirements of each physics topic.

We present some of the characteristics of  the WbLS technology in
Section~\ref{s:wbls}, and then discuss the potential for a broad physics
program in Section~\ref{s:physics}, including the detector requirements and
sensitivities in each case.  A long-baseline program is presented in
Section~\ref{s:lbl}; a $0\nu\beta\beta$ search in Section~\ref{s:0nbb};   solar
neutrino detection in Section~\ref{s:solar};   geo-neutrinos in
Section~\ref{s:geo};   supernova-burst neutrino detection in Section~\ref{s:sn};
detection of the diffuse neutrino flux from distant supernova in
Section~\ref{s:dsnb};   proton decay in Section~\ref{s:decay}; and   sterile
neutrinos in Section~\ref{s:sterile}.  Section~\ref{s:detector} concludes with
a summary of an R\&D program aimed at developing the requirements for  a single detector with maximal  physics potential.

\section{Water-Based Liquid Scintillator Target}\label{s:wbls}

Water-based liquid scintillator (WbLS) is a novel scintillation medium for large liquid detectors, in which scintillating organic molecules and water are co-mixed using surfactants.  While pure water detectors have been employed in large neutrino detectors in the past, they are primarily sensitive to high energy interactions creating particles above the Cherenkov threshold. On the other hand, pure scintillator detectors (isotropy without directionality) are sensitive to low-energy events and have been successfully deployed for reactor electron anti-neutrino ($\bar{\nu}_e$) detection through the inverse beta decay (IBD) process ($\bar{\nu}_e+p\rightarrow n+e^-$).  The usage of large volumes of scintillation liquid requires special material handling and safety regulations and could be limited by the increasing cost of oil. 

\begin{figure}[htbp]
\begin{center}
\includegraphics[width=0.5\textwidth]{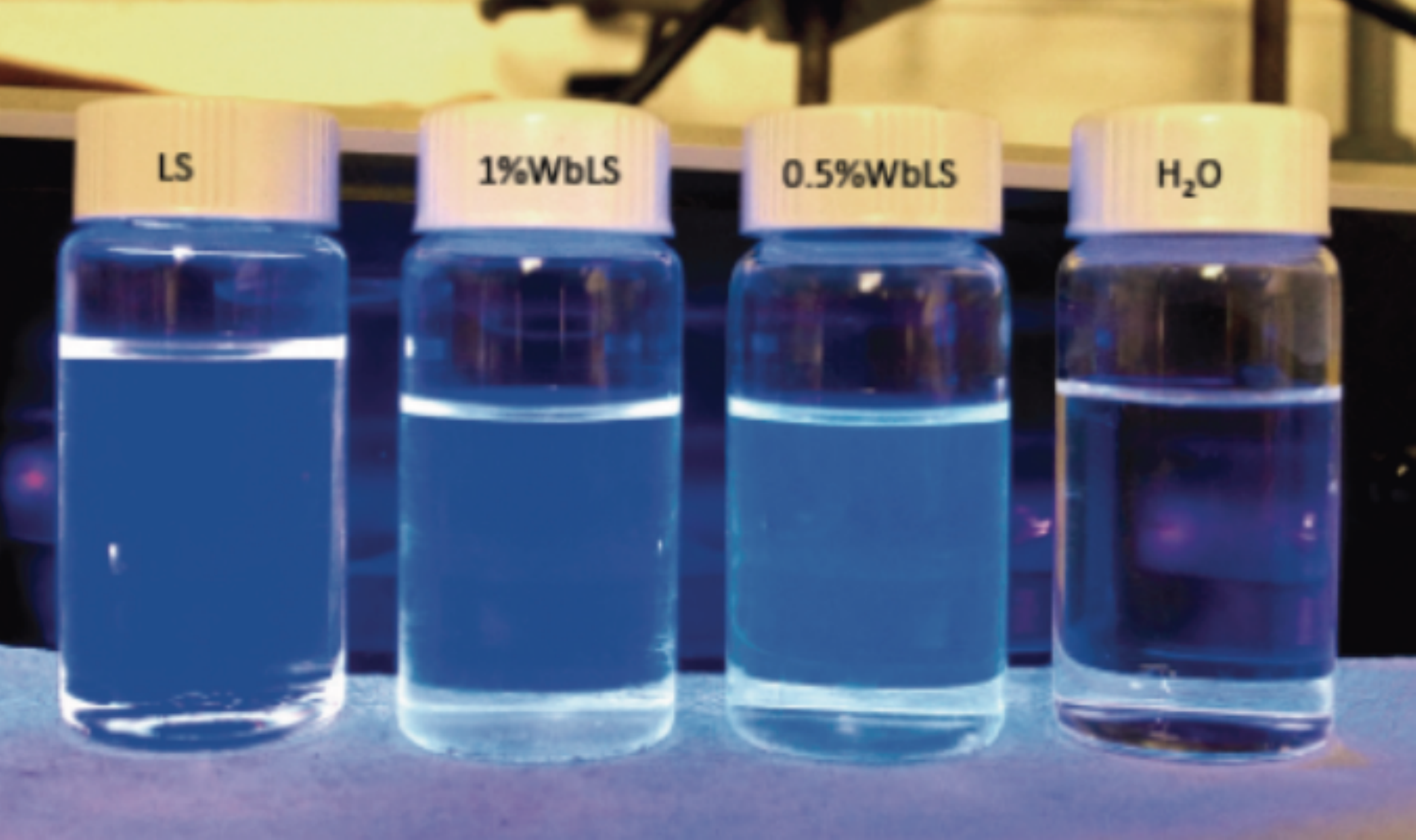}
\caption{Samples of WbLS in comparison with pure LS and water, illuminated by UV light ($\lambda = 250$nm).  From left to right: pure LS, water with 1\% LS, water with 0.5\% LS, and pure water.  The 1\% solution has a light yield of $\sim$100 optical photons per MeV.}
\label{f:wbls}
\end{center}
\end{figure}

One of the great strengths of the WbLS material is the degree to which the target properties can be tuned to optimize the physics program.  
These systems are flexible, with the ability to form scintillation solutions over a wide range of compositions that can be tuned to meet detector needs ranging from almost pure water to almost pure organic scintillator for different nuclear and particle physics experiments.  Scintillator  can be blended into water to make a scintillating water Cherenkov detector; alternatively, admixtures of water can be added to pure organic scintillator to add directionality and metal-loading capability.  
The scintillator can be made up of a single or a mixed solvent with possibilities including linear alkyl benzene (LAB), pseudocumene (PC), di-isopropylnaphthalene (DIN), phenylxylylethane (PXE), and phenylcyclohexane (PCH).  
Different wavelength shifters and other additives can be used to adjust the timing of different scintillation components.  Figure~\ref{f:wbls} shows some samples of scintillating water illuminated by UV light, in comparison to pure LS and pure water.

While a small admixture of water in a pure scintillator detector, coupled with precision timing to separate Cherenkov and scintillation light, could similarly allow a broad physics program, such a detector would be a substantial investment, requiring large volumes of scintillator and high coverage using state-of-the-art light detectors.  A water-like detector with a small fraction of scintillator would provide many of the same advantages at a significantly reduced cost.  Such a detector would be a critical enhancement over previous massive Cherenkov detectors, with the capability of scintillation detection making it sensitive to low-energy physics below the Cherenkov threshold. 
The addition of scintillator has a few direct consequences: ionizing particles below the Cherenkov threshold in water ($\beta$ $<$ 0.75) become detectable by their scintillation light; the overall light collection for all particles is increased, increasing resolution and allowing lower thresholds; and particle identification is improved through the comparison of scintillation to Cherenkov components, and through use of the time profile of scintillation light, which has a different form for excitation by $e.g.$ $\alpha$ and $\beta$ particles.

Deployment of such a scintillating water detector at a deep underground laboratory with careful controls of low-energy radioactive background could provide a cost-effective detector with extensive physics potentials in neutrino and rare-event physics for the next-generation frontier experiments. For example, the search for the $p \rightarrow K^+ \bar{\nu}$ mode in large water Cherenkov detectors is hindered by the kinematics of the initial $K^+$  (105~MeV energy, below the Cherenkov threshold) with the best sensitivity of $\tau (p \rightarrow K^+ \bar{\nu}) > 2.8 \times 10^{33}$~yr at 90\% C.L. from Super Kamiokande. An enhancement to a SK-like water Cherenkov detector by the dissolution of scintillation liquid in the ultra-pure water would allow detection of the $K^+$ and further suppression of the atmospheric $\nu_{\mu}$-induced $\mu^+$ background, and could extend the limit of the proton lifetime to $\tau > 2 \times 10^{34}$~yr at 90\% C.L over a 10-yr run~\cite{wbls}. 

Furthermore, metallic ions of interest can be straightforwardly loaded into WbLS (e.g. $^6$Li for short-baseline reactor anomalies, $^7$Li for solar neutrino, $^{\rm nat}$Gd for neutron tagging enhancement or medical imaging, or $^{\rm nat}$Pb for total-absorption calorimeter or solar neutrino). The options for isotopic loading are discussed in more detail in the relevant physics sections (Secs.~\ref{s:lbl}~--~\ref{s:sterile}) and in Section~\ref{s:detector}.  A similar water-based detector could also serve as a near detector for T2K (or Hyper-K) or be used for detection of diffuse neutrino flux from distant past supernovae. Another extended application of the WbLS is neutrinoless double beta decay (DBD) for the determination of neutrino mass and Majorana or Dirac nature. Some DBD isotopes that were not previously favorable for scintillator-based detectors  due to their hydrophilic nature (difficult to load in organic solvents), would now be accessible using the WbLS technology. An increase of double beta decay target mass to the Òtens of tonsÓ scale might be possible, providing a practical and cost-effective way of searching for DBD should the mass hierarchy turn out to be normal.  The following sections discuss each of these potential physics applications in more detail.  Critically, the target material can be adapted to pursue the most promising physics program as the field evolves with new experimental data. This is a major advantage of the ASDC over other detector technologies.

The development work at Brookhaven National Laboratory has already demonstrated stable WbLS cocktails with a wide range of properties, including:
\begin{itemize}
\item Light yields from 100 -- 6000 photons per MeV (up to 2500 in a water-like detector with 20\% LS; higher yields require a higher fraction of scintillator); 
\item Scattering lengths from a few meters to tens of meters;
\item A range of timing profiles, including a fast component ($\sim$3~ns) for proton decay and slower (10s of ns) components for improved Cherenkov / scintillation separation.
\end{itemize}
Measurements of intrinsic light yields at low energies have shown that they
scale roughly with the fraction of scintillator: a 4\% scintillator mixture has
a light-yield four times that of water alone.

Several improvements have been made over years of developments. The formula and principles of a mass-produced WbLS were developed and demonstrated at the BNL Liquid Scintillator Development Facility using  bench-top measurements and low-intensity proton beams at incident energies both above and below the Cherenkov threshold. Different metal-doped WbLS samples (i.e. Li, Gd, Te, Pb, Zr), from a few tenths of a percent to several percent for different experimental requirements, have been produced with high stability and are continuously monitored in different temperature ranges over time. 
The prototyping of liquid production and deployment at the ton-scale for direct measurement of Cherenkov and scintillation separation is planned. 
The assembly of a 1-ton WbLS prototype is ongoing at BNL and expected to be ready for liquid deployment in 2015.  Finally, R\&D for WbLS deployment at the kT scale has been proposed for the WATCHMAN detector~\cite{wm}.

\section{Physics Sensitivities and Detector Requirements}\label{s:physics}

\subsection{Long baseline}\label{s:lbl}				

Neutrino oscillation results from mixing between neutrino flavor 
and mass eigenstates, which can be described by a complex unitary matrix 
depending only on three mixing angles and a CP-violating phase, $\delta_{CP}$. These 
parameters and the two mass differences, $\Delta m^{2}_{21}$ and 
$\Delta m^{2}_{32}$, fully describe neutrino oscillations in the three-flavor 
scenario. The values of the three mixing angles and the value of 
$\Delta m^{2}_{21}$ have been measured experimentally. The absolute 
value of $\Delta m^{2}_{32}$ is known, but its sign has not yet been 
determined. The value of $\delta_{CP}$ is unknown. 

Determination of the neutrino mass hierarchy, or equivalently, the sign
of $\Delta m^{2}_{32}$, is important to fully understand the three-flavor
model, to resolve degeneracy between CP and matter-induced asymmetries
in long-baseline neutrino oscillation experiments,
and to guide the required sensitivity of neutrinoless double beta decay
experiments. Discovery of CP violation in neutrinos could
provide a mechanism for explaining the baryon asymmetry of the universe. 
Qualitative differences between the quark 
and neutrino 
mixing 
matrices suggest that different physics and potentially different mass 
scales in the two sectors may be present, which motivates precise
measurement of the parameters governing neutrino oscillation.
Precise measurement of mixing parameters also allows for 
unitarity tests and tests of models in which sum rules predict relations 
among the parameters.

An experiment studying neutrino
appearance and disappearance in an intense $\nu_\mu$ beam with a massive
neutrino detector located at a distance of order 1000 km allows 
determination of the neutrino mass hierarchy,
precision measurement of $\delta_{CP}$, and precision measurement of
multiple neutrino mixing angles in a single experiment. Next-generation
long-baseline experiments plan to use either ring-imaging water-Cherenkov
detectors (WCDs) or liquid argon time projection chambers (LAr TPCs) to
detect neutrino interactions. WCDs are less expensive per kiloton, 
making it feasible to build a very large far detector, with a target mass
of hundreds of kilotons.
LAr TPCs are expected to provide excellent efficiency and background
rejection, allowing similar sensitivity to a WCD with a less massive far
detector due to improved signal-to-noise performance. Here we explore the possibility of using the WbLS target in the ASDC in place of
water in a ring-imaging Cherenkov detector.

\begin{figure}[!h]
\begin{center}
\includegraphics[width=0.47\textwidth]{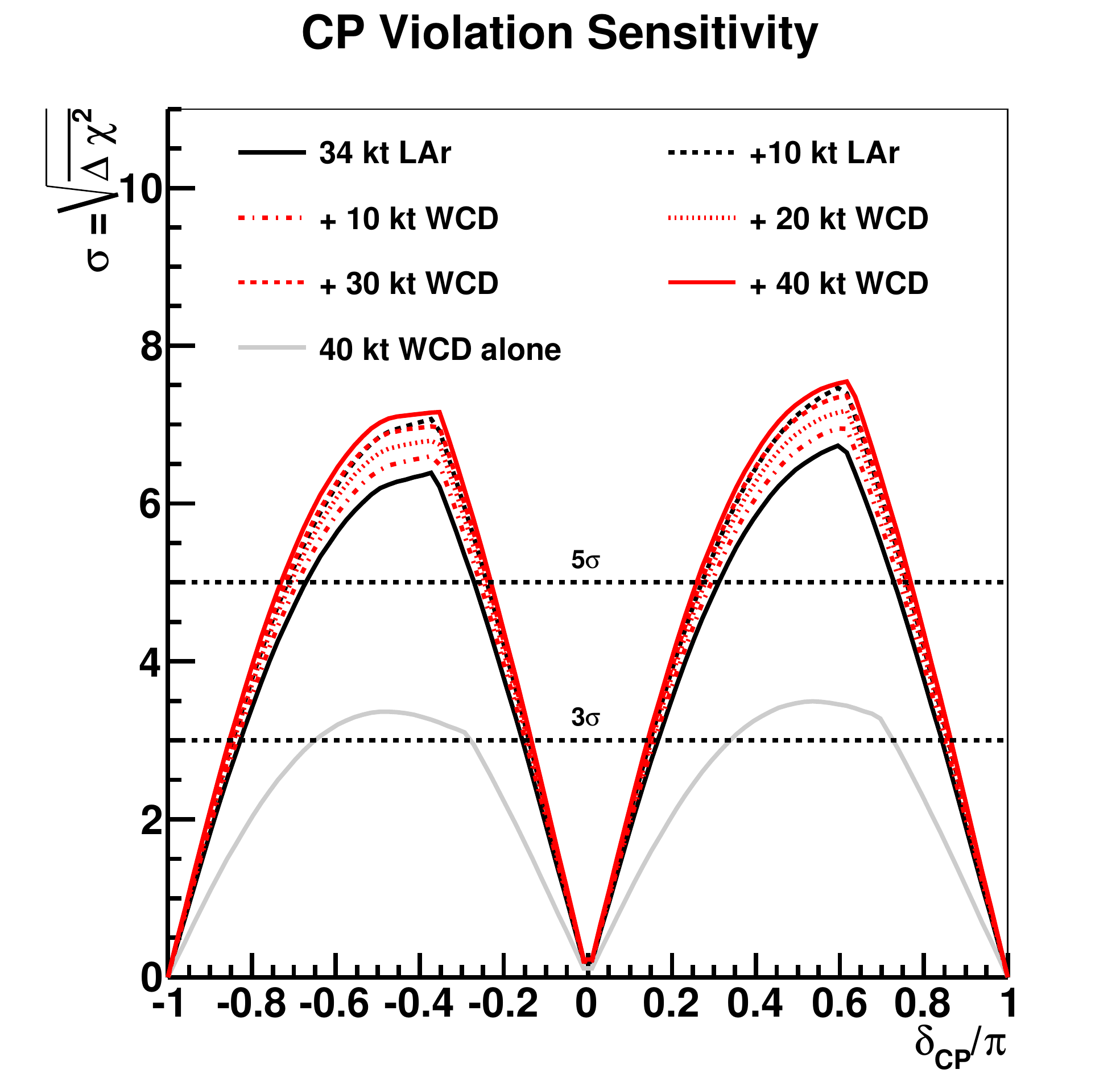}
\includegraphics[width=0.47\textwidth]{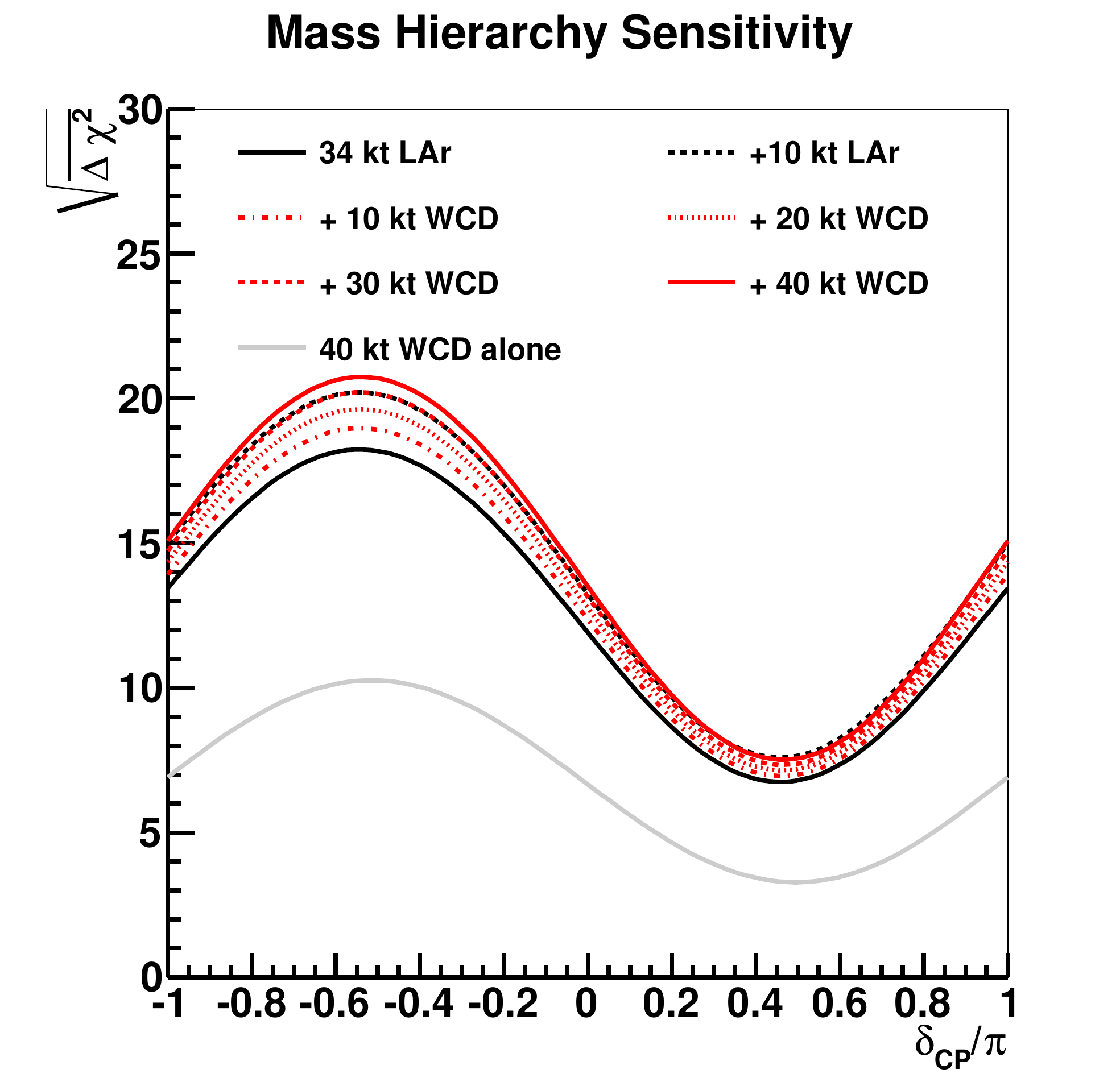}
\caption{Significance of sensitivity to CP violation (left) and neutrino
mass hierarchy (right), as a function of the true value of $\delta_{CP}$, 
for a variety of detector combinations. A 34-kT 
liquid argon TPC (solid black) is
compared to a 34-kT LAr TPC in combination with an additional 10-kT LAr TPC
(dashed black) and an additional 10--40 kT WCD (red dashed-solid). The
sensitivity of a 40-kT WCD alone (gray) is also shown. All detector masses are
fiducial.  The normal hierarchy is assumed, and oscillation parameters and uncertainties are taken from a recent global fit~\cite{global}.   The absolute sensitivity is dependent on these parameters, in particular the choice of $\theta_{23}$, but the relative comparison is unaffected.}
\label{fig:lbne_beamphys}
\vspace{-0.5cm}
\end{center}
\end{figure}

The sensitivity of a large-scale ASDC detector to long-baseline neutrino
oscillation physics as an additional far detector, at a 1300-km baseline
 in the LBNF\cite{lbne-sci-opp} beam, may be estimated using 
GLoBES\cite{globes1,globes2} configurations developed for the LBNE
WCD\cite{lbne-wcd-cdr}. 
Figure~\ref{fig:lbne_beamphys} shows the sensitivity to CP violation, i.e:
$\delta_{CP} \neq$ 0 or $\pi$, and the neutrino mass hierarchy,
for various masses of WCD, in combination
with a 34-kT LAr TPC, assuming
ring-imaging reconstruction performance similar to 
Super-Kamiokande\cite{superk_atmospheric}. A 30-40 kT fiducial-mass WCD
in the LBNF beam provides sensitivity equivalent to or better than an 
additional ten kilotons of fiducial mass in the LAr TPC. 

A WbLS-filled detector differs
from a WCD in that the absorption length will be somewhat shorter,
Cherenkov light will be absorbed and re-emitted isotropically by the
scintillator, and there will be direct production of scintillation light. 
These differences have both benefits and possible downsides. 
There is the potential for the ring-imaging performance of the
WbLS-filled detector to be degraded if the isotropic scintillation light, either from direct
scintillation or absorption/re-emission, obscures or smears out the Cherenkov
rings. 
Experience from Super-Kamiokande I and II shows no degradation of
long-baseline beam-physics sensitivity for a factor of two reduction in 
Cherenkov light yield. Potential reduction in $\nu_{e}$ appearance efficiency
due to degradation of electron-muon particle identification would significantly
impact long-baseline beam-physics sensitivity. 

The ASDC may produce enhancements relative to WCD sensitivity,
particularly if it is possible to separately detect scintillation and 
Cherenkov light. This includes improved particle ID using comparison 
of Cherenkov and scintillation light yield and detection of neutrons and low-energy charged hadrons produced 
by neutrino interactions. The ``wrong-sign''
component of the antineutrino signal in LBNE is $\sim$25\% assuming a true
normal hierarchy;
neutron detection could potentially improve sensitivity to 
the neutrino-antineutrino asymmetry from CP violation by 
allowing for discrimination
between neutrino and antineutrino interactions.

\begin{figure}[!h]
\begin{center}
\includegraphics[width=0.47\textwidth]{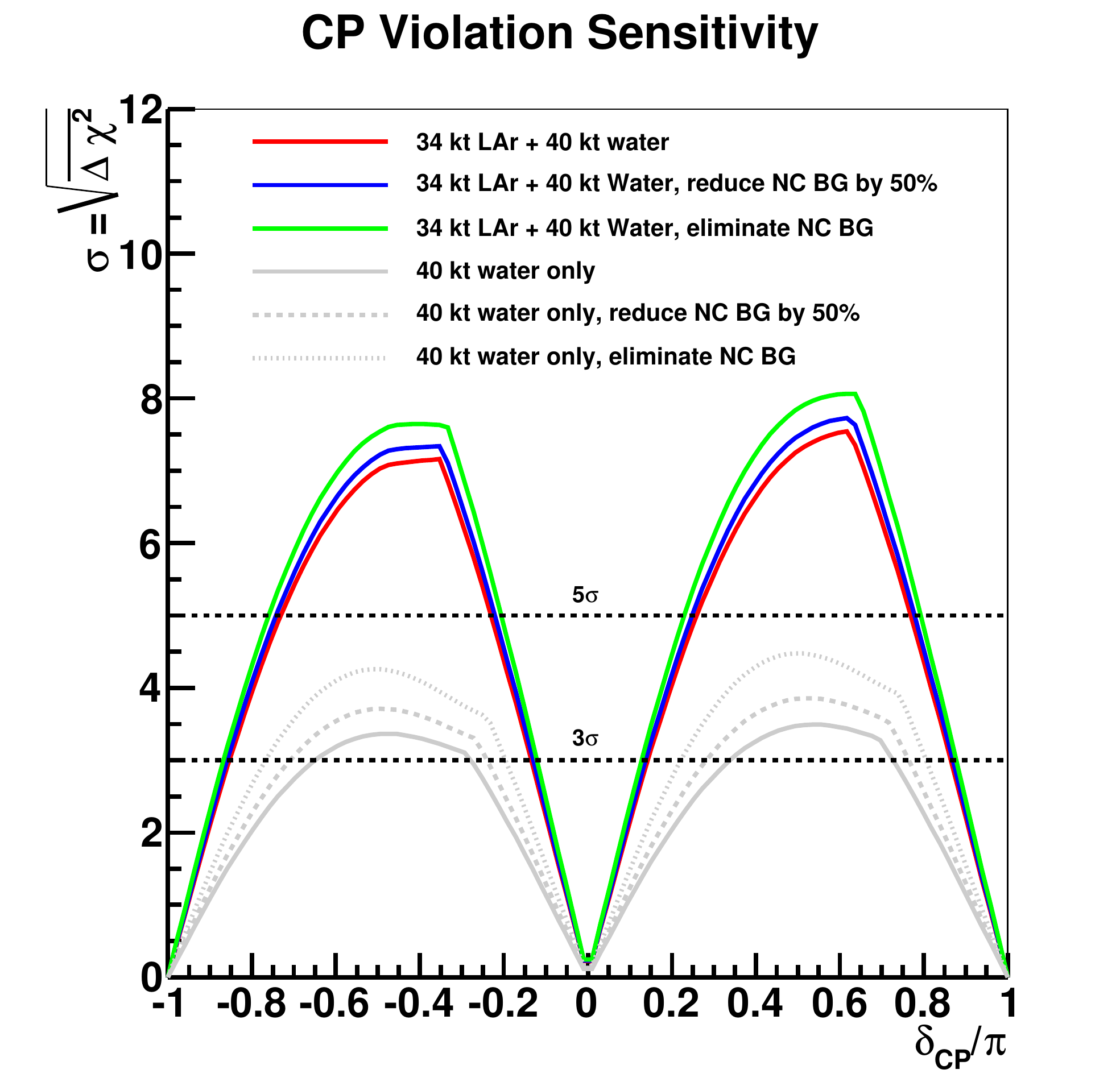}
\includegraphics[width=0.47\textwidth]{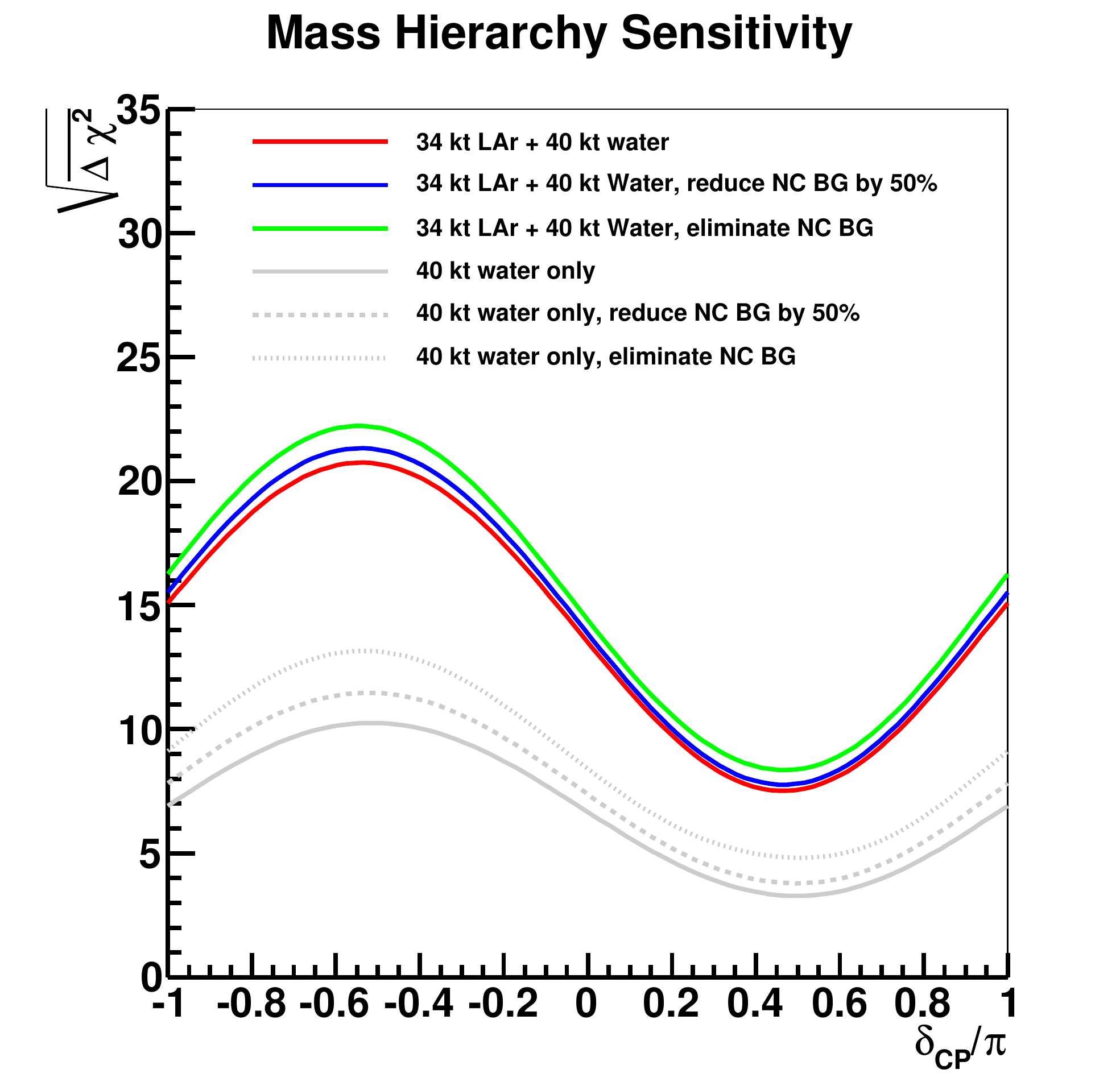}
\caption{Significance of sensitivity to CP violation (left) and neutrino
mass hierarchy (right), as a function of the true value of $\delta_{CP}$, 
for a 34-kT LAr TPC in combination with an additional 40-kT WCD when neutral
current background is reduced by 50\% (blue) or removed (green) from the 
WCD event sample. Gray curves show the sensitivity of the 40-kT WCD detector
alone. All detector masses are fiducial and all sensitivities are
calculated at a baseline of 1300 km in the LBNF nominal beam.  The normal hierarchy is assumed, and oscillation parameters and uncertainties are taken from a recent global fit~\cite{global}.   The absolute sensitivity is dependent on these parameters, in particular the choice of $\theta_{23}$, but the relative comparison is unaffected.}
\label{fig:lbne_ncbg}
\end{center}
\end{figure}

Further improvements, particularly reduction in neutral
current background, may be achieved with the excellent timing resolution
and granularity possible using LAPPDs. As shown in Fig. \ref{fig:lbne_ncbg},
removal of neutral current background from a 40-kT WCD yields an increase
in the fraction of $\delta_{CP}$ values for which CP violation can be
observed with 3$\sigma$ significance, for the WCD alone, from 37\% to 57\%.
Similarly, the minimum mass hierarchy sensitivity ($\sqrt{\Delta\chi^2 }$) for 100\% of 
$\delta_{CP}$ values, for the WCD alone, 
increases from 3.3 to 4.8. These sensitivities are
calculated for a detector at 1300 km in the nominal LBNF beam; an experiment
optimized for measurement of the second oscillation maximum could potentially 
benefit even more significantly from removal of this background.

More study is needed to understand what level of scintillator loading 
can be achieved while maintaining Cherenkov ring-imaging performance
and the extent to which timing or wavelength filtering may be used to 
separate scintillation and Cherenkov light in WbLS. 
A study of measuring directionality with scintillation detectors
concludes that selection criteria based on optical photon arrival time are 
effective at isolating Cherenkov light\cite{directionality}; it seems
reasonable to expect that this method would be effective in WbLS, which
has a lower scintillation light yield than pure liquid scintillator. 

For more quantitative predictions of the sensitivity of the ASDC to 
long-baseline neutrino oscillation, it is necessary to determine the 
following as a function of scintillator-loading fraction in WbLS:
\begin{itemize}
\item Absorption length
\item Absorption and emission spectra
\item Time distribution of direct Cherenkov, absorbed/re-emitted Cherenkov, and scintillation light
\item Efficiency of Cherenkov-scintillation light separation
\item Efficiency of electron-muon particle ID
\item Extent to which the above may be improved with adjustments to the WbLS formulation that increase the scintillation response time
\item Systematic uncertainty in LBNF with multiple far detectors
\end{itemize}
Simulation and measurements are underway to address many of these open 
questions.

\subsection{Neutrinoless double beta decay}\label{s:0nbb}

Massive neutrinos will be their own antiparticles---Majorana neutrinos---unless
lepton number conservation is more than just the accident implied by the
Standard Model Lagrangian.  Such a discovery would have far-reaching
consequences that point toward higher mass scales, and could explain the origin of
the observed matter-antimatter asymmetry.  
Should lepton number be conserved and
thus neutrinos and antineutrinos be distinct particles (`Dirac neutrinos'), we
would need a new fundamental symmetry to explain how, for example, a
right-handed antineutrino ($\bar{\nu}_R$) is different from a right-handed
neutrino ($\nu_R$).  On the other hand, if neutrinos are Majorana particles, we
have the startling consequence that the simplest mass term in the Standard Model
Lagrangian is dimension-5 and thus not renormalizable.  The Majorana mass term
would thus be akin to the old Fermi theory of $\beta$ decay, which pointed
toward our gauge theory of the weak interaction.  In addition, if neutrinos are
Majorana, and if further the additional Majorana phases were non-zero,
leptogenesis in the early Universe becomes possible.  Therefore neutrinos as
Majorana particles could explain the preponderance of  matter over
antimatter.  The most practical way to observe that neutrinos are Majorana
particles is to see the neutrinoless double beta decay process,
$\mathrm{(Z,A)\rightarrow (Z+2,A)+e^-+e^-}$, a process that explicitly violates
lepton number conservation.

The great importance of the question of the Majorana or Dirac nature of
neutrinos has led to many distinct approaches in searching for $0\nu\beta\beta$. 
Although these searches have been ongoing for many decades, only recent
measurements of the neutrino mass differences and mixing parameters allow us to
know how close we may be. If neutrino masses are degenerate, or the $\nu_3$ is
the lightest neutrino (an inverted hierarchy), a determination of whether neutrinos are Majorana or not
is within the grasp of the next generation of experiments. If the $\nu_3$ is
the heaviest and the lightest neutrino has a mass far smaller, it will take
much bigger experiments than currently envisioned to answer this question. 
While many possible classifications of such experiments can be made, for our
purposes here we group the various approaches into two categories: those that
use the $\beta\beta$ isotope as its own detector, and those in which a separate
detection medium or technology is used to observe the decays. In the first
category are experiments such as GERDA and MAJORANA, which use enriched Ge
detectors, CUORE which uses TeO$_2$ bolometers, EXO-200, nEXO, and NEXT, which
all use enriched Xenon in liquid or gaseous TPCs.  In the second category are
experiments such as NEMO and SuperNEMO, which use scintillators and tracking
detectors to observe decays of various isotopes, KamLAND-Zen, which uses enriched
Xenon dissolved in liquid scintillator, and SNO+, which will dissolve Te in a
cocktail that includes LAB scintillator, water, and a surfactant.  The recent
report from the Nuclear Sciences Advisory Committee (NSAC) Sub-Committee on
Neutrinoless Double Beta Decay Experiments is an excellent
summary of the advantages and challenges of these and other approaches.

        Experiments in the first category tend to emphasize energy resolution
as a primary (but by no means the only) background rejection technique.
Excellent energy resolution has the advantage not only of separating a
putative $0\nu\beta\beta$ peak from radioactive backgrounds, but also from
the intrinsic, and vastly larger, $2\nu\beta\beta$ continuum.  The
challenges for such experiments are primarily the cost of scaling to larger
masses, and the need for high-purity shielding to reduce ``external''
backgrounds.

Experiments in the second category tend to have noticeably poorer
energy resolution, and instead reduce most backgrounds through other
means.  For the NEMO experiments, background reduction is achieved primarily
through tracking, while for KamLAND-Zen and SNO+ background rejection is
achieved primarily through fiducialization and purification of the
scintillator-isotope mixture.  ``External'' backgrounds in the
liquid scintillator experiments are significantly reduced just by the sheer size
of the detectors.  Because many of the backgrounds in a liquid
scintillator-based experiment are intrinsic to the detector and not the isotope,
``loading'' additional isotope means that the lifetime sensitivity of the
experiment can scale like $M^{1/2}$ rather than $M^{1/4}$.

	Nevertheless, the broad energy resolution of scintillator detectors
means that the choice of isotope is critical. One solution is to use a
high-endpoint isotope, such as $^{150}$Nd, which significantly reduces the
contributing radioactive backgrounds in the region of interest around the endpoint. But as Biller and Chen have pointed out~\cite{billerchen},
an isotope like $^{130}$Te that has a small $2\nu\beta\beta$ matrix element but
a large $0\nu\beta\beta$ matrix element, and a reasonably large phase space for
the decay, is more effective than a high-endpoint isotope. On this basis alone,
$^{136}$Xe is even better, but the very high natural abundance of $^{130}$Te
makes loading large masses of isotopes more affordable.

Biller~\cite{steveb} has pointed out that the advantages of $^{130}$Te make it
in principle possible to load enough in even a $\sim$ 10~kT pure
scintillator detector that a Te-based $0\nu\beta\beta$ experiment could reach
well into the normal hierarchy regime.  The paper asserts that a loading fraction of
roughly 10\%, purification that reduces all but the $2\nu\beta\beta$ and $^8$B solar neutrino 
backgrounds to negligible levels, and a light yield of up to 2000 pe/MeV to
reduce the $2\nu\beta\beta$ background, would achieve a 90\% CL sensitivity of
$m_{\beta\beta}<3$meV.  In addition, the paper points out
that the background from $^8$B solar neutrinos should also be reduced, and that this
could in principle be done if directional Cherenkov light could be observed,
thus removing the electrons elastically scattered by such neutrinos.

\subsubsection{Neutrinoless Double Beta Decay in the ASDC}

A large-scale water-based liquid scintillator detector, like the ASDC
we envision here, could in principle realize these requirements with a much
smaller loading fraction of Te, provide good rejection of the $^8$B neutrino
events, and also allow the broad program of physics we discuss in this paper.
We assume that in the ASDC the very large size of the detector makes external
backgrounds entirely negligible, and that advances in purification techniques
reduce cosmogenic backgrounds in the tellurium to similarly low levels. Uranium and
thorium chain backgrounds are assumed to be reduced to levels similar to
Biller's assumptions ($10^{-17}$~g/g by purification), and that time-correlated cuts on $\beta-\alpha$ decays
will further reduce these backgrounds to subdominant levels.

        We are left then with the $2\nu\beta\beta$ continuum events and
elastically scattered electrons from $^8$B solar neutrinos. The approach to
reducing the former background is by maintaining a high light yield and thus as
narrow an energy resolution as possible. We have made measurements on a WbLS
cocktail that contains 4\% scintillator, and found that at low
($^{90}$Sr) energies its intrinsic light yield is roughly four times that of
water alone.  Using the SNO detector's initial light yield of 10 pe/MeV, we
would anticipate that with similar effective photocathode coverage and
photomultiplier tubes the ASDC would observe 40 pe/MeV.

        To get to higher light yields---necessary for a $0\nu\beta\beta$
search---one has two options: either increase the loading fraction, which would shorten the attenuation length and thus reduce overall light collection; or increase the effective photocathode coverage and the
capabilities of the photon detectors themselves.  The SNO detector's
effective photocathode coverage was achieved using an array of roughly 9500 8''
PMTs each surrounded by a reflector, obtaining the equivalent (for Cherenkov
light) of a detector with a geometric coverage of 55\%~\cite{snonim}. We assume
here that the ASDC will have nearly 100\% photocathode (or equivalent)
coverage, and we further assume that modern large-area high-quantum efficiency
(HQE) PMTs will be used. The Hamamatsu R11780~\cite{pmtpaper}, for example, is
a 12'' HQE PMT that could provide full coverage with roughly 100k devices in the ASDC.
Our measurements of the R11780 compared to SNO PMTs show an improvement in
detected light yield of roughly a factor of two. Thus with full coverage and
HQE PMTs, our expected light yield would be 160 pe/MeV, or an energy resolution
near the $^{130}$Te $0\nu\beta\beta$ endpoint of about 5\%, similar to existing
pure scintillator detectors. With such a resolution and an asymmetric
region-of-interest (ROI) of $-0.5\sigma$-$+1.5\sigma$ around the
$0\nu\beta\beta$ median energy like that used by SNO+, the
$2\nu\beta\beta$ background is smaller than that from the $^8$B neutrinos.

\begin{figure}[!ht]
\begin{center}
\includegraphics[width=0.57\textwidth]{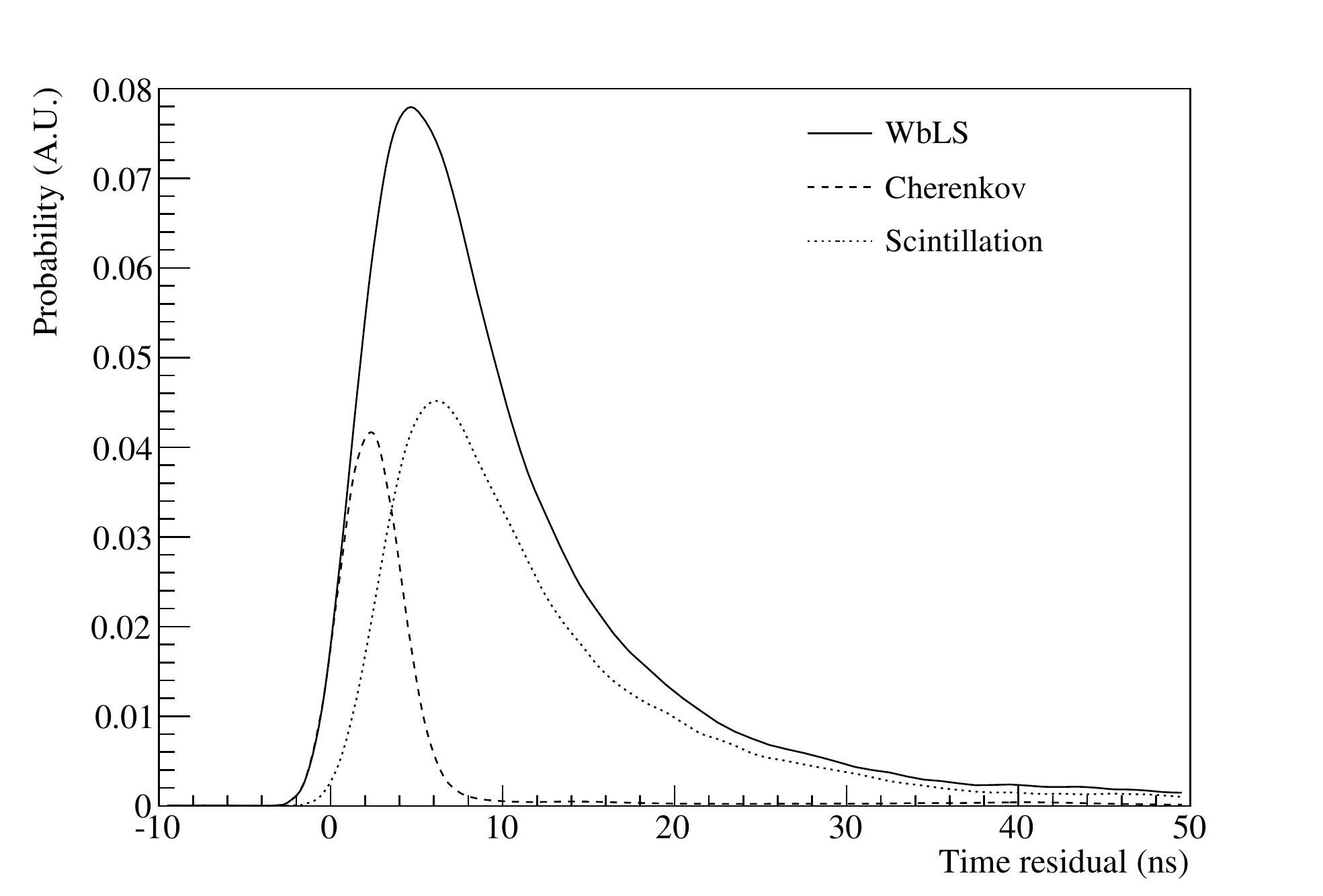}
\caption{Time profile of PMT hits from Cherenkov and scintillation light, and
their sum, modeled as a 1\% mixture of pseudocumene in water, which reproduces
the factor of four increase in intrinsic light yield we have measured for the
WbLS cocktail.  (``A.U.'' are arbitrary units for the probability scale).}
\label{fig:cherprof}
\end{center}
\end{figure}
\begin{figure}[!hb]
\begin{center}
\includegraphics[width=0.57\textwidth]{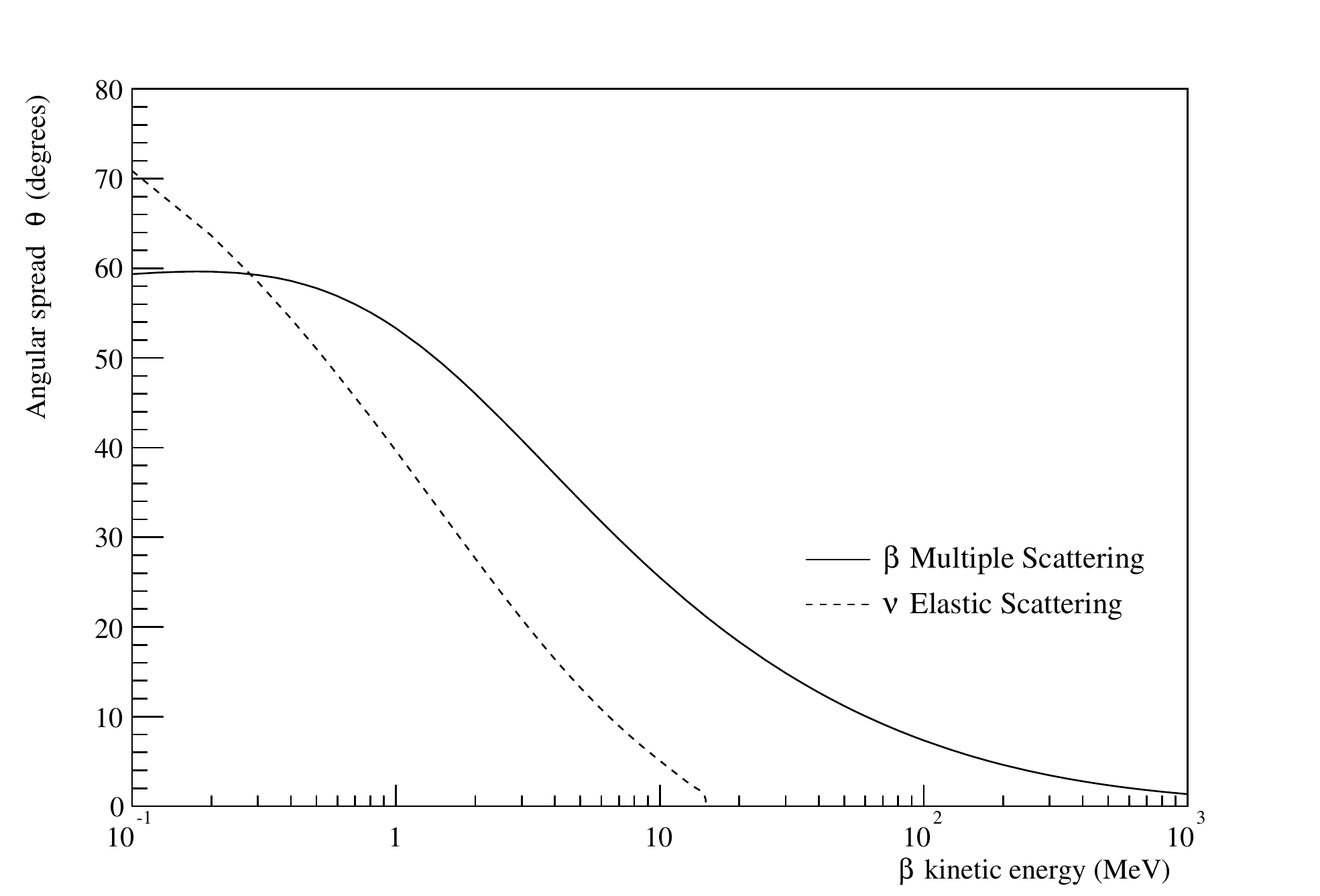}
\caption{Components of the intrinsic angular resolution of electrons due to: elastic scattering by neutrinos; and multiple scattering at 2.5~MeV initial energy.  The angular resolution is defined as the difference between the true and reconstructed directions.}
\label{fig:scattres}
\end{center}
\end{figure}
\begin{figure}[!h]
\begin{center}
\includegraphics[width=0.57\textwidth]{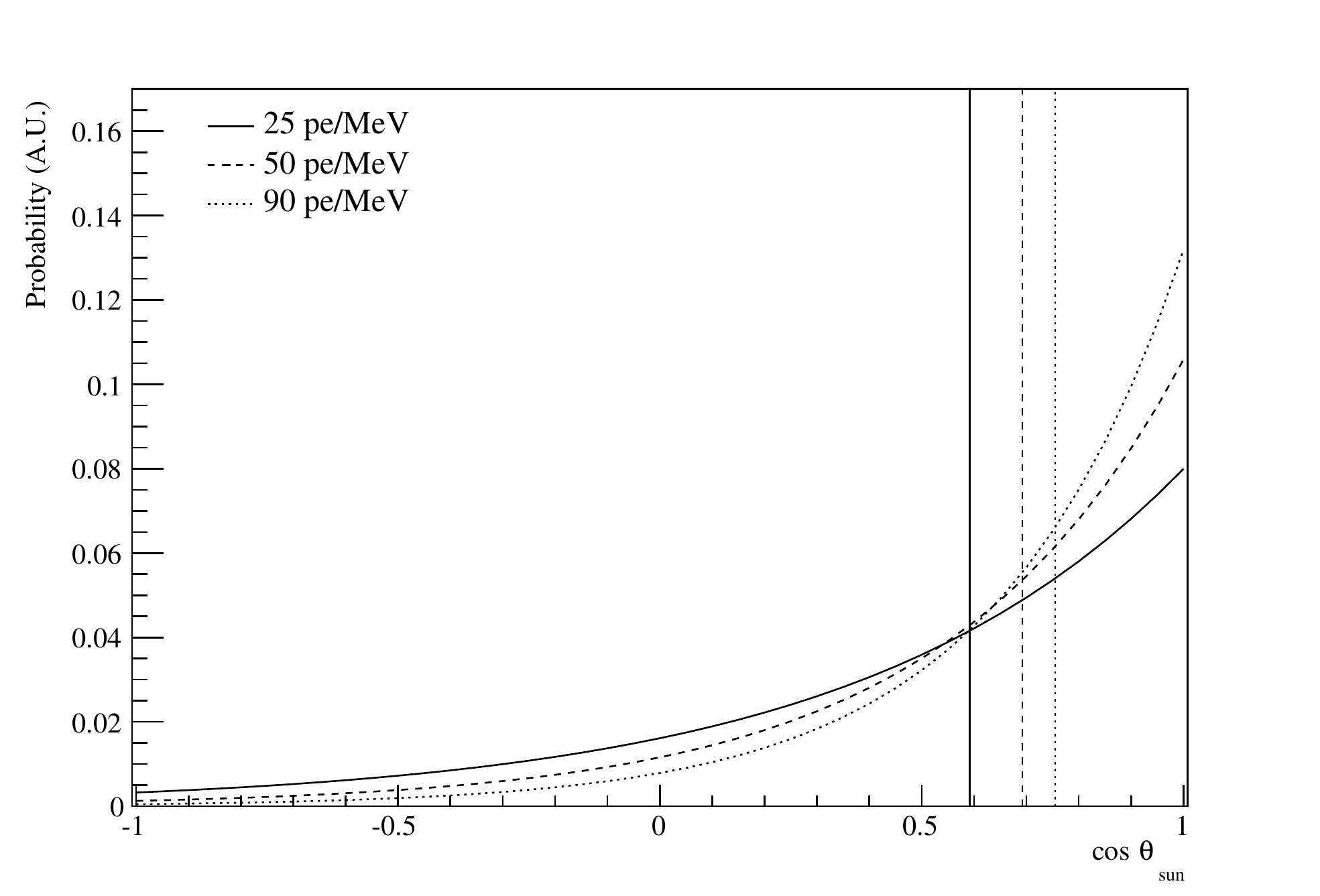}
\caption{Reconstructed directions of electrons elastically scattered by $^8$B
$\nu$s, for
various light yields. The vertical lines indicate the cuts that would have to
be placed to reduce the $^8$B background by 50\%.  The direction is obtained from a simple direction-cosine sum from the reconstructed vertex to each PMT.  The probability scale is in arbitrary units (A.U.)}
\label{fig:dirres}
\vspace{-0.3cm}
\end{center}
\end{figure}

        Reduction of the $^8$B $\nu$ events is therefore critical, making
observation of the Cherenkov light and associated reconstruction of direction
necessary. Figure~\ref{fig:cherprof} shows the time profile of observed photons
in a WbLS detector modeled as 1\% pseudocumene plus water optics, with the
smearing due to the PMT transit-time spread included. The model reproduces the
factor of four increase in total intrinsic light yield over pure water that we
have
measured for the WbLS cocktail, but includes the re-emission, scattering,
and absorption we would expect for a large WbLS detector. As can be
seen in the breakdown of the timing, about 60\% of the photons within the first
4~ns are Cherenkov photons. This fraction goes down as the fraction of
scintillator goes up, making the time window required for a ``rich'' Cherenkov
fraction smaller. Although the ASDC will therefore have a smaller number of fast
Cherenkov photons than a water detector, the resolution of neutrino direction
at these energies is already dominated by the combination of the smearing from
the elastic scattering cross section itself, and the multiple scattering of the
$\sim$2.5~MeV electrons. Figure~\ref{fig:scattres} shows the contributions to
the intrinsic angular spread from these two effects as a function of electron
energy, and in Figure~\ref{fig:dirres} we show the resultant
expected reconstructed direction distribution for various observed Cherenkov
light yields.  For full coverage and HQE PMTs, we expect 40 pe/MeV of Cherenkov
light alone (and the 160 pe/MeV total yield described above). The vertical lines
indicate the cut needed to reduce the $^8$B neutrino background by 50\%.

\begin{figure}[!h]
\begin{center}
\includegraphics[width=0.57\textwidth]{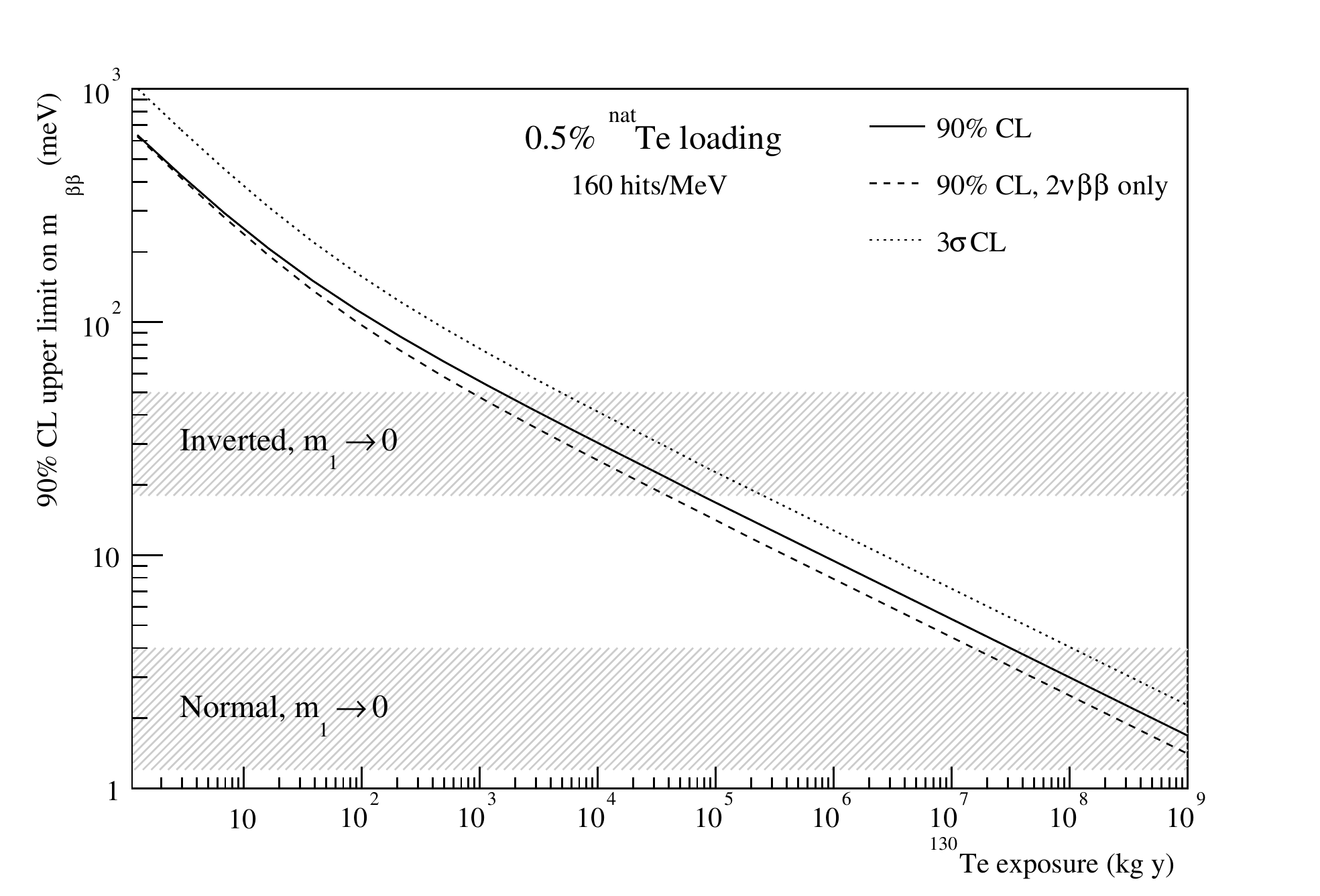}
\caption{Sensitivity as a function of $^{130}$Te exposure, for a 90\% CL limit
on on $m_{\beta\beta}$ and a 3$\sigma$ discovery of $0\nu\beta\beta$, for a 0.5\%
natural Te loading in a 50~kT WbLS-filled ASDC (50~T of $^{130}$Te), assuming a light yield of 160 pe/MeV.
Also shown is the limit achievable if all backgrounds but $2\nu$ were removed.
\label{fig:sensitivity} }
\vspace{-0.3cm}
\end{center}
\end{figure}

      Figure~\ref{fig:sensitivity} shows the sensitivity of a 50~kT
water-based liquid scintillator detector with a
0.5\% loading of natural Te, and an energy resolution of $\sim$ 5.5\% (from our
projected 160 pe/MeV light yield). We assume here that the only non-negligible
backgrounds are the $2\nu\beta\beta$ events and $^8$B neutrino events, with a
50\% reduction of the latter using the direction cuts described above. To
eliminate external backgrounds, we also use a fiducial radius restriction
requiring events to be more than 5.5~m from the PMTs, providing a total
fiducial mass of 30~kT. For the 0.5\% loading, this is therefore 150 tonnes
of natural Te, or about 50~T of $^{130}$Te.  As the figure shows the ASDC
could reach a 3$\sigma$ discovery of $0\nu\beta\beta$ in 10 years, for
$m_{\beta\beta}=15$~meV.

        One of the biggest sources of risk in our assumptions is the
bulk optical properties of the WbLS.  Absorption and scattering of photons both
reduces light yield and removes directional information from the photons;
re-emission by the scintillator or added wavelength-shifters removes
directional information.  Should any of these effects be higher than 
expectations, they could reduce the sensitivity of the WbLS
approach. There are, however, a few
mitigation strategies possible. The first is to exploit timing further by using
faster photon detectors, such as LAPPDs~\cite{lappd1,lappd2,lappd3,lappd4,lappd5}. This makes the
``Cherenkov window'' richer in Cherenkov photons, thus helping to pick out
photons that have not been scattered or re-emitted. A second approach is to use
PMTs more sensitive toward the yellow and red end of the emission
spectrum~\cite{directionality}. Such photons will not be scattered or absorbed but
come promptly from the Cherenkov production site. While there are fewer of
these than the blue end, they may nevertheless carry more useful information
for the Cherenkov measurement. Lastly, the isotope could be contained in a
smaller volume, using a balloon like that deployed by KamLAND-Zen, surrounded
either by water or a less rich WbLS mixture. 

\subsection{Solar neutrinos}\label{s:solar}	

Solar neutrinos are produced as a by-product of the nuclear fusion reactions that power the Sun.  Neutrinos are produced in two fusion cycles: the pp chain, the primary source of solar power, and the sub-dominant CNO cycle. 
The Sudbury Neutrino Observatory (SNO) experiment resolved the Solar Neutrino Problem by detecting the Sun's ``missing'' neutrinos, confirming the theory of neutrino flavor change.  
The combination of a charged-current (CC) measurement from SNO  with Super-Kamiokande's high-precision elastic scattering (ES) measurement demonstrated that the $\nu_e$ produced in the Sun were oscillating to other flavors prior to detection~\cite{snocc}, a result later confirmed (at 5$\sigma$) by SNO's measurement of the flavor-independent $^8$B flux using the neutral current (NC) interaction~\cite{snonc}.  
This opened the door to a precision regime, allowing neutrinos to be used to probe the structure of the Sun, as well as the Earth and far-distant stars.  

\begin{figure}[htbp]
\begin{center}
\includegraphics[width=0.55\textwidth]{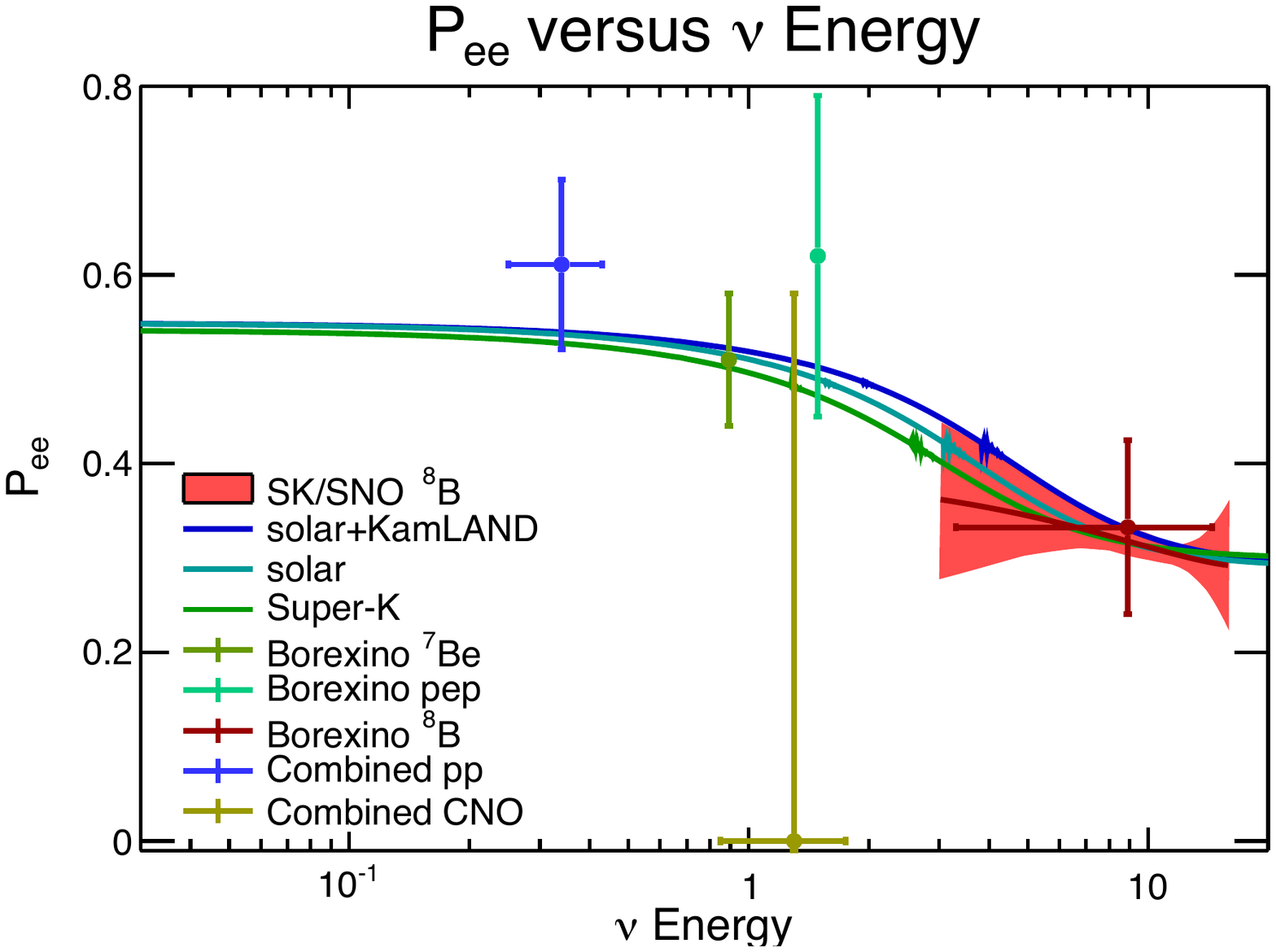} 
\caption{MSW prediction for $\nu_e$ survival probability for the
three-flavor solar+KamLAND best fit parameters, the combined
solar best fit parameters, and the SK-alone best fit parameters. Note that the pep uncertainties are
not Gaussian and the value is $\sim 2 \sigma$ from  zero. The data
point for Borexino $^8$B represents the survival probability
averaged over the measured energy range. Taken from ~\cite{lbne-sci-opp}.}
\label{f:pee}
\vspace{-0.35cm}
\end{center}
\end{figure}

The detection of solar neutrinos offers the only observed matter effect in neutrino interactions to-date.  The MSW effect causes an effective enhancement of solar neutrino oscillation at high energies ($\gtrsim 5$~MeV), with a transition region to pure vacuum oscillation between 1--5~MeV. A sensitive probe of this region not only tests our understanding of the interaction of neutrinos with matter, but is also a sensitive search for new physics effects 
such as flavor-changing neutral currents and sterile neutrinos, which can alter the shape of the survival probability in this region.  Figure~\ref{f:pee} shows a summary of solar neutrino measurements to date.  As shown in~\cite{richie}, the global data set does not yet have the precision needed for discovery.  

Another recent mystery is the claim of a surprisingly small fraction of heavy elements within the Sun. In recent years more sophisticated models of the solar atmosphere have been developed, replacing one-dimensional modeling with fully three-dimensional modeling, and including effects such as stratification and inhomogeneities.  These models produce results more consistent with neighboring stars of similar type, and yield improved agreement with absorption-line shapes~\cite{asplund}.
However, they also result in an abundance of metals (elements heavier than H or He) in the photosphere that is $\sim$30\% lower than previous values.  
One of the critical factors that engendered confidence in the Standard Solar Model~\cite{ssm} (SSM) was the excellent agreement ($\sim0.1\%$) of SSM predictions for the speed of sound with helioseismological measurements.  The speed of sound predicted by the SSM is highly dependent on solar dynamics and opacity, which are affected by the Sun's composition.  When the new metal abundances are input into the SSM it produces a discrepancy with helioseismological observations.  This new disagreement has become known as the ``Solar Metallicity Problem''~\cite{met1, met2}.
Although the impact of these changes on the pp-chain neutrinos is small relative to theoretical uncertainties, the neutrino flux from the sub-dominant CNO cycle depends linearly on the metallicity of the solar core, and the predictions for the two models differ by greater than 30\%.  The theoretical uncertainty on these predictions is roughly 14--18\%, but this can be reduced to $<$10\% using correlations in the theoretical uncertainties between the CNO and $^8$B neutrino fluxes: the two have similar dependence on environmental factors, thus a precision measurement of the $^8$B neutrino flux can be used to constrain the CNO neutrino-flux prediction.  A precision measurement of the CNO flux then has the potential to resolve the current uncertainty in heavy element abundance.

Both Borexino and SNO+ aim to measure the CNO neutrino flux, and Borexino currently holds the world's best limit~\cite{Borexpep}.  However, extraction of this flux is highly challenging due to the similarity of the spectrum of ES recoil electrons with background $^{210}$Bi decays in the target.   Borexino and SNO+ plan to extract the flux using a time-series analysis based on the short ($\sim$5-day) lifetime of $^{210}$Bi, to differentiate the (presumed) stable neutrino flux from decaying background events.  However, this method is complicated by the need to  assume either  equilibrium in the decay chain, or a lack of external sources of the daughter isotope, $^{210}$Po, which may not hold true.
A charged-current (CC) sensitive detector would have a greater chance of distinguishing these events based on energy-alone, due to the increased sensitivity to incident neutrino energy, without the need for a time-series analysis.

These open questions in solar neutrino physics require a low (sub-MeV) detection threshold, difficult to achieve in a pure water Cherenkov detector. However, the directionality of Cherenkov light provides a critical signal identification for solar neutrinos. Existing scintillator and Cherenkov experiments detect neutrinos via elastic scattering.  This interaction has a broad differential cross section, which means the outgoing electron spectrum is only weakly correlated with the underlying neutrino spectrum.  As seen by SNO, CC detection significantly improves the precision with which the neutrino spectrum can be measured.  
The LENS experiment proposes CC detection~\cite{lens} of the ultra-low-energy pp and pep neutrinos.  However, the higher-energy, lower-flux $^8$B neutrinos are only observable in a large-scale detector.  

A large-scale WbLS detector with high-precision timing, such as the ASDC, would provide unprecedented sensitivity to low-energy solar neutrinos via two channels:
\begin{enumerate}
\item Huge statistics for elastic scattering events at low energy;
\item Potential charged-current detection via isotope loading e.g. $^7$Li.
\end{enumerate}
The unique advantage of the ASDC would be the directionality information for ES events in a low-threshold detector, providing a handle for background discrimination, and in addition a precision spectral measurement using CC events.  A preliminary study has been performed (to be published).  

The sensitivity of such an experiment will depend upon the mass of target isotope loaded, its charged-current cross-section, and the detector response.  Loading of metallic ions into liquid scintillator has been demonstrated to loading fractions of several percent by mass (Section~\ref{s:wbls}).  The detector energy resolution will depend on the light yield of this isotope-loaded scintillator, whereas background rejection is highly dependent on the scintillator time profile. 

The sensitivity of a 50~kT pure scintillator detector has already been studied by the LENA collaboration~\cite{lena}.  While the threshold of a WbLS detector like the ASDC will not be as low as that of a pure scintillator detector (LENA studies assumed a 250~keV threshold), the additional information from the Cherenkov component provides a strong benefit in signal/background separation.  The further advantage of CC detection could allow spectral separation beyond that of any pure scintillator detector.

\begin{figure}[!ht]
\begin{center}
\includegraphics[width=0.47\textwidth]{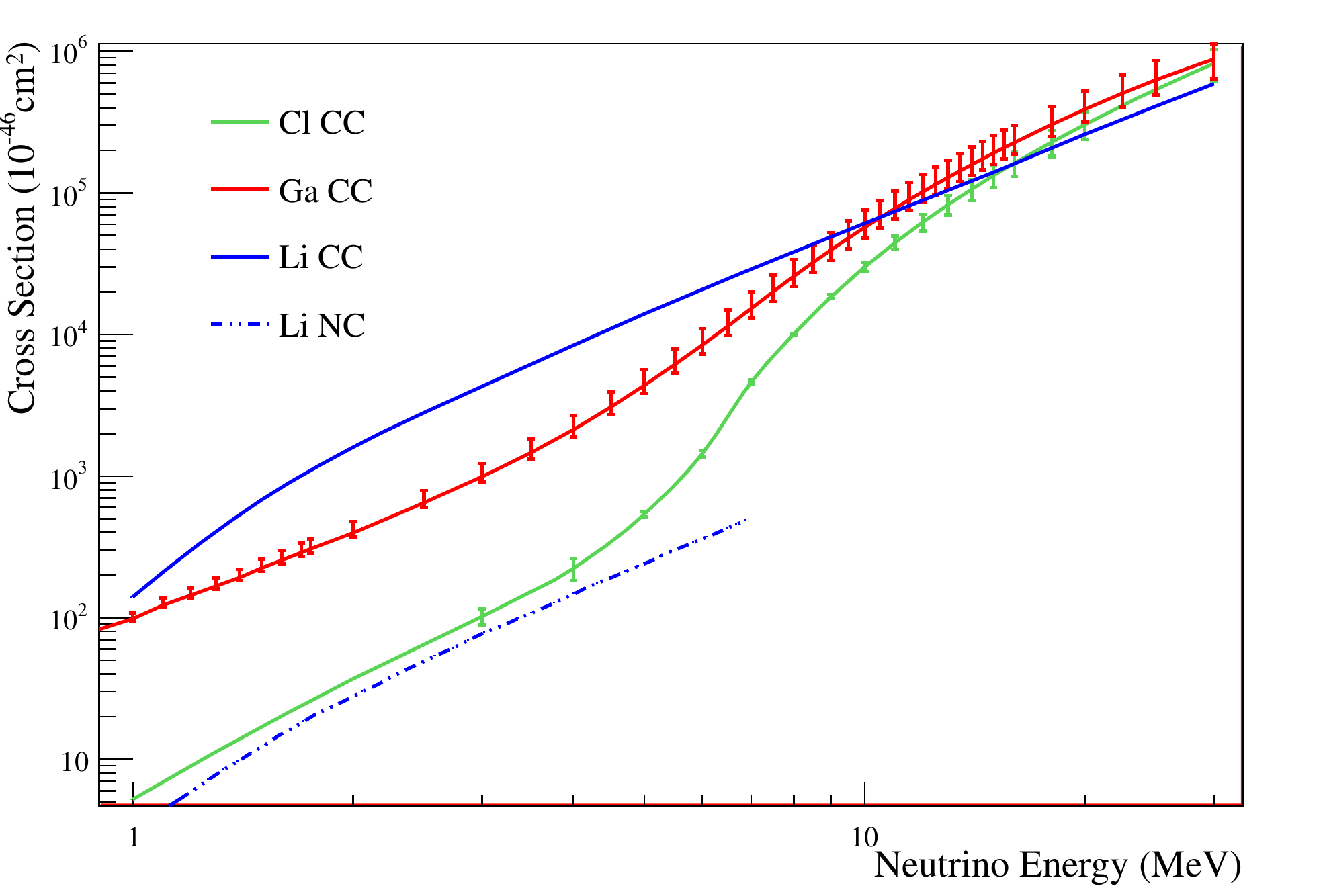}
\includegraphics[width=0.47\textwidth]{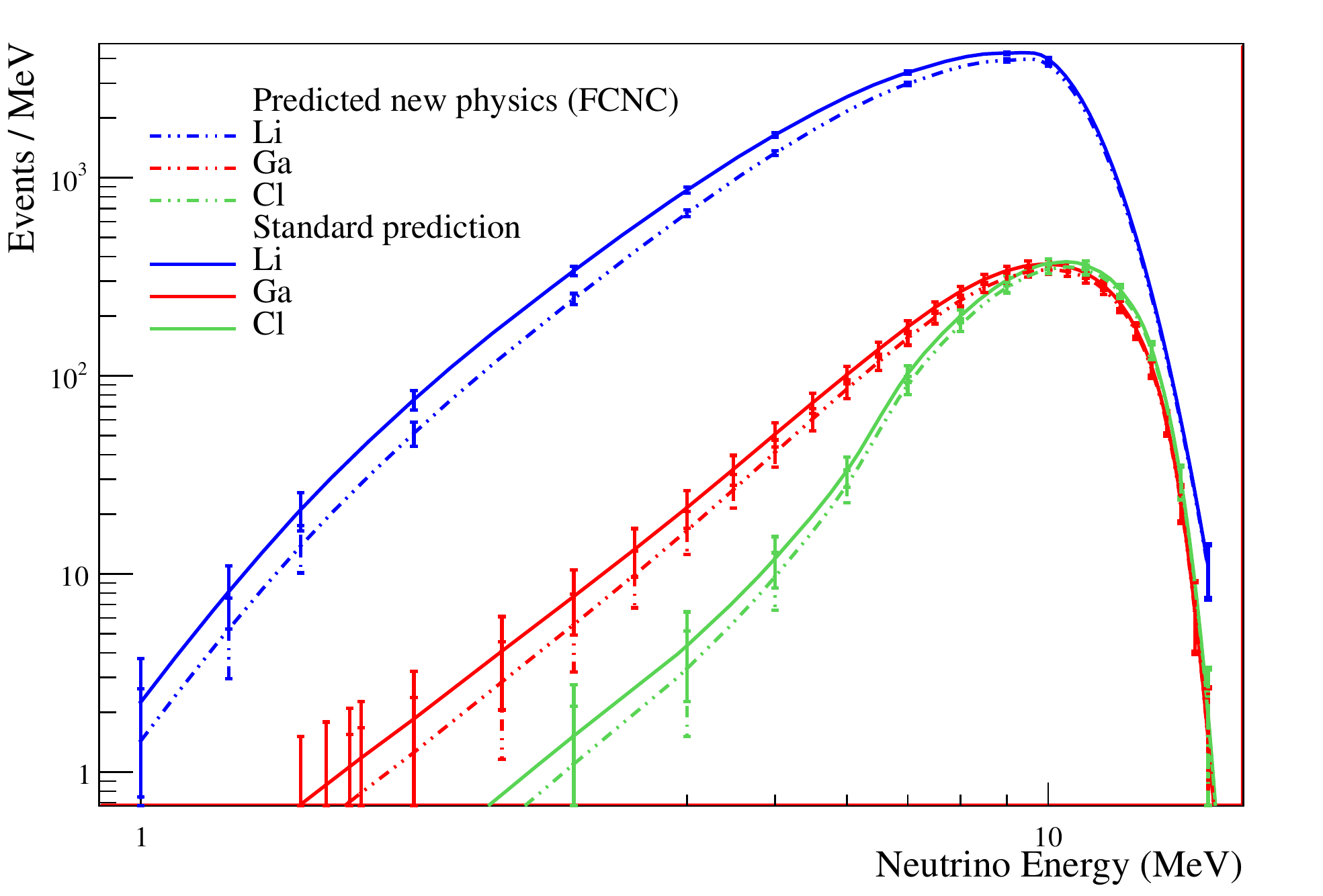}
\caption{(Left) The cross section for CC neutrino interaction on $^{37}$Cl (green), $^{71}$Ga (red), and $^{7}$Li (blue) targets and NC on $^{7}$Li (blue dashed).  Data taken from~\cite{sigcl},~\cite{sigga},~\cite{sigli}, and~\cite{signc}, respectively. Although the differential uncertainties are not shown, the uncertainty on the lithium cross section is roughly 1\%~\cite{sigd}. (Right)  Predicted solar neutrino event spectra for 5 years of data-taking, with 1\% loading by mass of candidate isotopes in a 30-kT WbLS-filled ASDC detector. Solid lines show the standard solar neutrino oscillation prediction.  Dashed lines are for a flat neutrino spectrum to low energies, indicative of new physics interactions.  $^7$Li is the most favorable choice due to a high cross section for neutrino absorption.  Five years of data taking results in over 17~$\sigma$ separation in the integral flux, and correspondingly high precision (several $\sigma$ significance) on the extracted spectrum.
\label{f:clga}}
\end{center}
\vspace{-1.\baselineskip}
\end{figure}

Several factors contribute to the choice of isotope for loading into a scintillator detector.  $^{37}$Cl and $^{71}$Ga have been used successfully by radiochemical experiments from the late 1960s to the present day.  $^{7}$Li has been considered as a favorable alternative~\cite{rcli} but such a detector was never constructed.  $^{7}$Li was also proposed as an additive to a water detector in~\cite{sigd}; such a detector would have excellent sensitivity to the high end of the $^8$B spectrum, but would be limited in threshold. 
As seen in Fig.~\ref{f:clga}, $^{71}$Ga and $^{7}$Li both have more favorable cross sections than $^{37}$Cl, particularly at low energies.    However, the cost of $^{71}$Ga would likely be prohibitive in a liquid scintillator experiment.  The relatively large differential uncertainties on the $^{71}$Ga cross section would also smear out any extracted spectrum whereas the cross sections on $^{37}$Cl and $^{7}$Li are known to extremely high precision.  The $^{37}$Cl cross section has been mapped using the $\beta$ decay of $^{37}$Ca.  
The CC interaction of $\nu_e$ on $^{7}$Li is shown in Eq.~(\ref{e:li}).
\begin{equation}\label{e:li}
^{7}Li + \nu_e \rightarrow\, ^{7}Be + e^- \quad \rm{(Q = 862~keV)}
\end{equation}
$^{7}$Li has only two significant transitions: a mixed Fermi and Gamow-Teller transition to the ground state of $^{7}$Be with a threshold of 0.862~MeV; and a super-allowed Gamow-Teller transition to the first excited state at $\sim$430~keV, which decays with a lifetime of $\tau \sim$200~fs.   The scattering is very hard, transferring almost all incident energy to the scattered electron.  If one could differentiate between the electron of the ground state and the $e^- + \gamma$ of the first excited state one would have a high-precision reconstruction of neutrino energy.  The two states also have precisely known angular distributions, which could then be used as an additional handle to differentiate signal from background.  Even without the use of particle ID to differentiate between states the contribution of the two is known precisely from theory, so the difference in threshold can be used to demonstrate that the two are being seen in the correct proportions.
There is also the potential to observe NC interactions on $^{7}$Li (Fig.~\ref{f:clga}), exciting the analog 478~keV first excited state of $^{7}$Li, which then decays with a lifetime of $\tau \sim$105~fs.  On preliminary investigation $^{7}$Li would thus appear to be the preferred isotope.  However, other factors may be important, such as the effect of isotope loading on scintillator optics.

\begin{figure}[!ht]
\begin{center}
\includegraphics[width=0.47\textwidth]{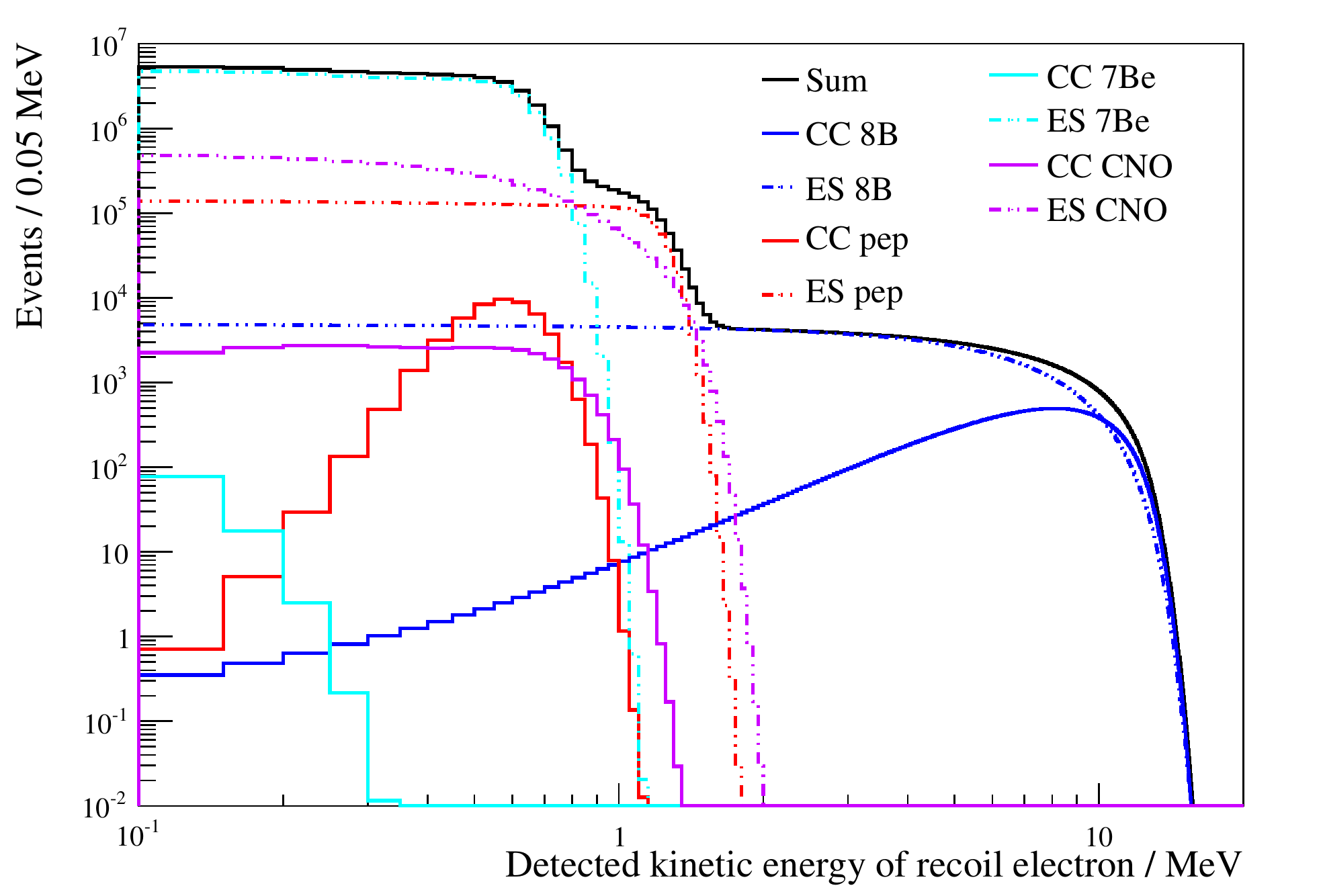}
\includegraphics[width=0.47\textwidth]{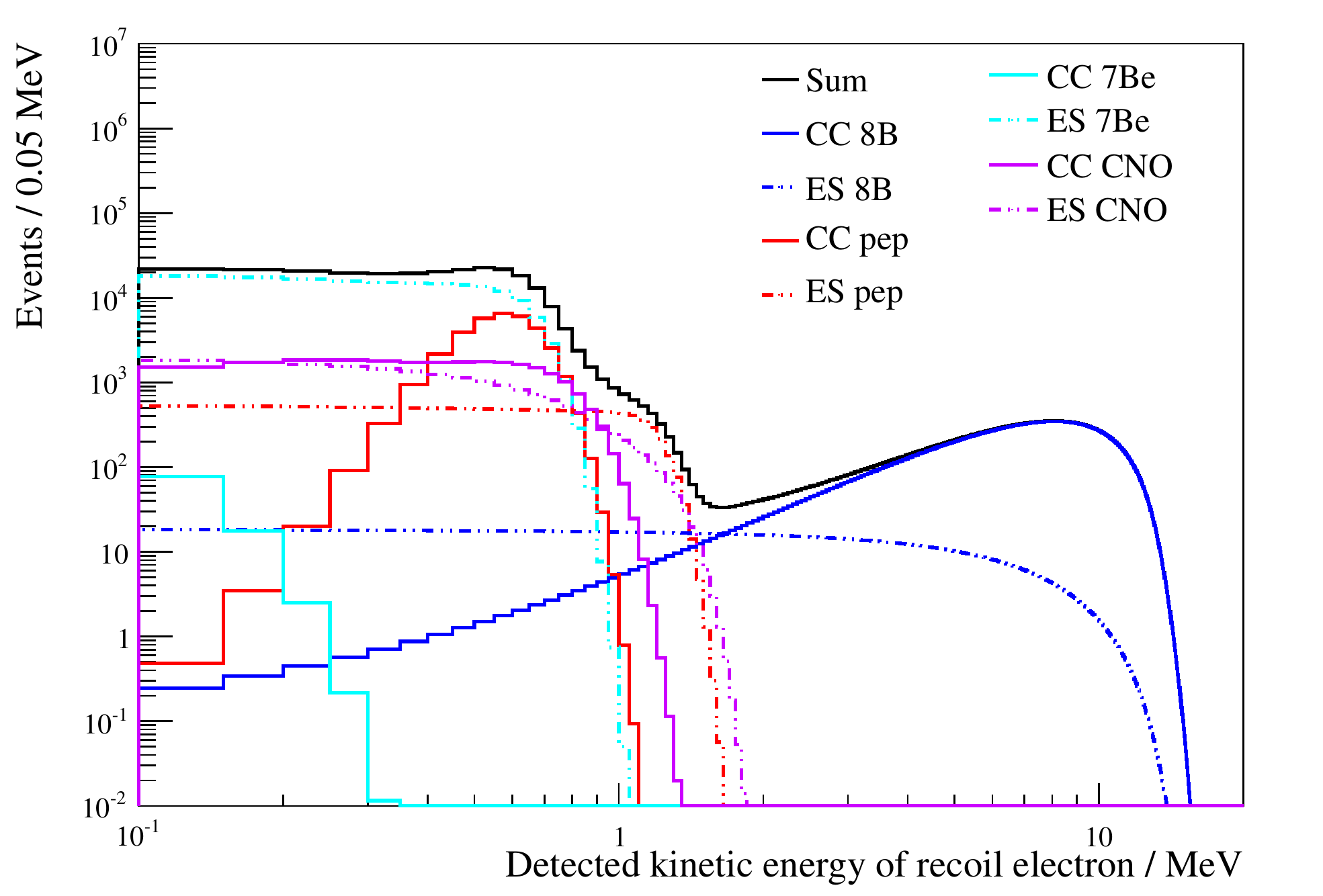}
\caption{(Left) Predicted solar neutrino spectra in a 30-kT WbLS-filled ASDC detector loaded with 1\% $^7$Li by mass.  Light yield of 100 p.e./MeV assumed.  (Right) The same spectra with a cut on $\cos\theta_{\odot}=0.4$, reducing the ES component to illustrate the power of CC detection.
\label{f:solarspec}}
\end{center}
\vspace{-1.\baselineskip}
\end{figure}

Figure~\ref{f:solarspec} shows the predicted spectrum for a WbLS-filled ASDC detector with a 30-kT fiducial volume loaded with 1\% $^7$Li by mass, and a conservative light yield of 100 photoelectrons per MeV.  Standard MSW oscillation is assumed.  Solid lines show the CC interactions and dashed lines show ES detection.  The ES statistics by far outweigh the CC (as expected at a low \%-level loading); however, the use of directionality would allow excellent separation.  The right-hand panel shows the spectrum with a cut placed on $\cos\theta_{\odot}=0.4$ (where $\theta_{\odot}$ is the angle between the event direction and a vector pointing back to the Sun), which reduces the ES signals by more than 2 orders of magnitude.  (Angular resolution equivalent to SK-III was assumed).  In practice a more sophisticated analysis would link the normalization of the ES and CC neutrino signals via their known cross sections, allowing the ES to be used to separate events from radioactive and cosmogenic backgrounds such as $^{210}$Bi and $^{11}$C, and the CC to provide the spectral sensitivity.  The power of the CC signal can be observed in particular in the $^8$B spectrum, which has a distinctive shape, and the strong peak in the $pep$ signal in comparison to the broad ES spectrum.

\begin{figure}[!ht]
\begin{center}
\includegraphics[width=0.54\textwidth]{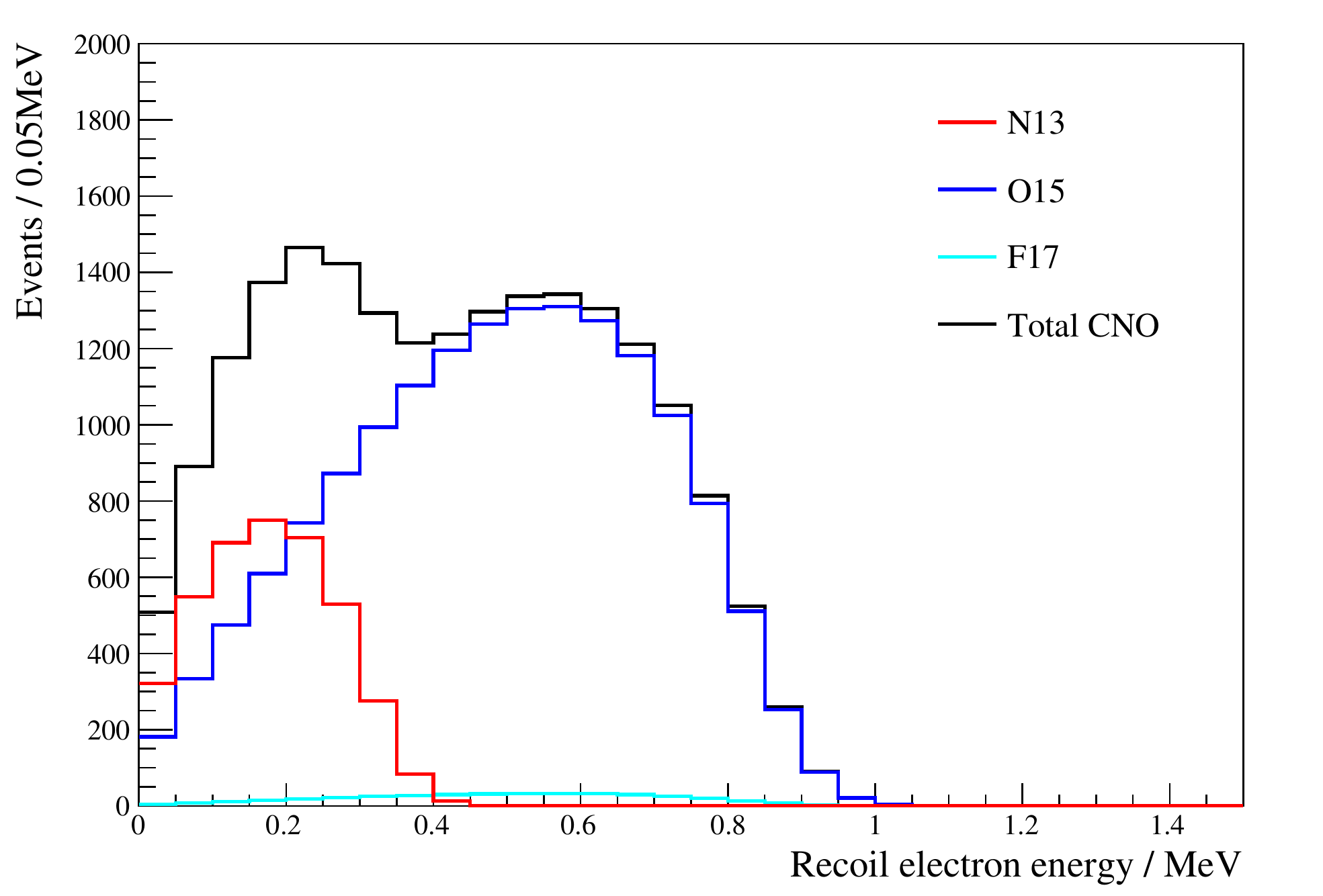}
\caption{Predicted spectrum for the individual components of the CNO neutrino flux, and the total, in a WbLS detector.
\label{f:cno}}
\end{center}
\vspace{-1.\baselineskip}
\end{figure}

Due to limited sensitivity, experiments to date have only considered detection of the sum of the three CNO lines.  The increased spectral sensitivity from isotope-loaded WbLS could allow the possibility to separate the constituent lines of the CNO neutrino flux.  The CNO cycle depends critically on temperature in the conversion of C to N, reaching equilibrium only in the most central region of the solar core, where $T > 1.33\times10^7$~K.  In this region, equal numbers of neutrinos are produced in the $\beta^+$ decay of  $^{13}$N and $^{15}$O, whereas in the cooler outer regions only $^{13}$N neutrinos are produced.  Independent measurements of the $^{13}$N and $^{15}$O neutrino fluxes would determine the separate primordial abundances of C and N~\cite{HRS}.  Figure~\ref{f:cno} shows the predicted CNO spectrum broken into its individual components.  A sufficiently sensitive detector with a low enough threshold could separate the contributions from $^{13}$N and $^{15}$O.  A separate measurement of $^{17}$F is unlikely due to the much lower flux; however, there is a strong theoretical basis for fixing the $^{17}$F component to a known fraction of the sum of $^{13}$N  and $^{15}$O.

\subsection{Geo-neutrinos}\label{s:geo}						
A large-scale WbLS detector would soon produce the largest sample of geo-neutrino events in the world. Presently the total exposure to geo-neutrinos, which has been accumulated over the last 12 years, is less than 10 kT-y. This exposure from oil-based liquid scintillator detectors in Japan and Italy, producing somewhat more than 100 events, provides an initial constraint on the radiogenic power of Earth \cite{KLgeo,BXgeo}. Deployment of a large-scale WbLS detector, such as the ASDC, would significantly improve this constraint, contributing important information on the origin and thermal evolution of our planet.

Geo-neutrinos are traditionally detected via the inverse beta decay reaction on free protons, which was successfully exploited six decades ago to discover the neutrino \cite{cow56}. The reaction produces both a positron and a neutron, providing a coincidence of signals in space and time and the efficient rejection of background. 
The combination of metallic ion loading in WbLS and fast timing detection in the ASDC will assist in directionality measurements for these low energy, primarily non-Cherenkov events. High-precision 
vertex resolution for the positron and neutron capture events can allow reconstruction of the displacement vector between the two, offering source identification for both geo-neutrinos and reactor monitoring. 
This was demonstrated by the Chooz \cite{chooz2000} and Palo Verde \cite{paloverde2000} reactor experiments and is a current avenue of active research \cite{6li}. 
The positron is emitted nearly isotropically so there is no preferential direction of the Cherenkov light relative to the neutrino direction. 
However, a distinct advantage of WbLS might be realized by reconstructing neutrino-electron elastic scattering of geo-neutrinos. The scattered electron must gain some momentum along the direction of the antineutrino. If the electron is energetic enough to emit Cherenkov light, this would enable an estimate of the antineutrino direction. There are interesting challenges associated with this unique opportunity, notwithstanding the relatively intense flux of solar neutrinos (which are of course strongly directional themselves, providing a strong handle for separation). In addition to providing directional Cherenkov light, a WbLS-based detector like the ASDC is likely to be environmentally safer and considerably less expensive than an oil-based liquid scintillator detector.

Important issues to be explored concerning the application of WbLS detectors to geo-neutrino studies are:
\begin{itemize}
\item Trigger/reconstruction efficiency for inverse beta decay reactions of antineutrinos with 1.8 - 3.3 MeV;
\item Neutron tagging efficiency and capture point reconstruction, including possible loading with Gd, $^6$Li~\cite{6li}, $^{10}$B;
\item Geo-neutrino signal sensitivity for the various deployment site options;
\item Light output and transparency of WbLS as functions of pressure and temperature;
\item Efficiency for reconstructing neutrino-electron elastic scattering reactions of antineutrinos with 1.8 - 3.3 MeV.
\end{itemize}

\newcommand{\nuebar}{\mbox{$\bar{\nu}_e$} }

\subsection{Supernova Neutrinos}\label{s:sn}

Supernova neutrinos carry unique information about one of the most dramatic 
processes in the stellar life-cycle, a process responsible for the production 
and dispersal of all the heavy elements (i.e., just about everything above helium) in 
the universe, and therefore a process absolutely essential not only to the look and feel of the 
universe as we know it, but also to life itself. As a gauge of the community's level of interest in these 
particular particles, it is worth noting that, based upon the world sample 
of twenty or so neutrinos detected from SN1987A, 
there has on average been a theoretical paper published once every 
ten days -- for the last three decades!  

The next time a Milky Way core collapse supernova goes off, an event 
expected to occur every 30 years or so~\cite{next}, it would be extremely desirable to have 
a sizable WbLS experiment in operation 
when the resulting neutrino wave sweeps across the planet.  This is primarily because the most 
copious supernova neutrino signal by far ($\sim$88\%) comes from inverse beta events. 
They are only produced by one of the six 
species of neutrinos and antineutrinos which are generated  
by a stellar collapse, and so if we could tag them 
individually by their follow-on neutron captures then we could  
extract the \nuebar time structure of the burst precisely, gaining
valuable insight into the inner dynamics of the explosion.  
What's more, we could then subtract them away from the more subtle 
non-\nuebar signals, uncovering additional information that would
otherwise be lost during this once-in-a-lifetime happening. 

Table~\ref{tab:snsig} shows the expected signals for a galactic type II supernova. 
For a stellar collapse 10~kpc distant from earth 
there will be a total of about 12,000 events recorded in a 50~kT fiducial ASDC.  
As one can see from the table, the inverse beta events 
dominate the expected signals.  However, there is valuable information to be
gained from the other signals, information which would be largely inaccessible without
neutron tagging from the scintillation light following captures on either hydrogen or a 
dissolved capture-enhancing element like gadolinium.

\begin{table}
\begin{center}
\begin{tabular}{|c|c|c|}
\hline
Neutrino & Percentage of & Type of\\  
Reaction & Total Events & Interaction \\ \hline 
$\overline{\nu}_e + p        \rightarrow  n + e^+$ & 88\% & Inverse Beta \\
$\nu_e +  e^-                \rightarrow  \nu_e + e^-$ & 1.5\% & Elastic Scattering \\
$\overline{\nu}_e +  e^-     \rightarrow  \overline{\nu}_e + e^-$ & $<$1\% & Elastic Scattering \\
$\nu_x +e^-                  \rightarrow  \nu_x + e^-$ & 1\% & Elastic Scattering\\
$\nu_e +  ^{16}O             \rightarrow  e^{-} + ^{16}F$  & 2.5\% & Charged Current \\
$\overline{\nu}_e +  ^{16}O  \rightarrow  e^{+} + ^{16}N$ & 1.5\% & Charged Current \\
$\nu_x +^{16}O               \rightarrow \nu_x +O^{*}/N^{*}+\gamma$ & 5\% & Neutral Current \\ 
\hline
\end{tabular}
\caption{Breakdown of supernova neutrino events expected in a 50-kT fiducial ASDC experiment from
a galactic supernova.  Oscillations are taken into account.
$\nu_x$ indicates the total interactions of $\nu_{\mu}$, $\nu_{\tau}$, and
their antineutrinos.}
\label{tab:snsig}
\end{center}
\end{table}

In the case of a pure water+Gd design like GADZOOKS!~\cite{gadzooks}, gadolinium's 
energetic gamma cascade following neutron capture provides enough light 
to enable efficient neutron tagging. The main advantage of also Gd loading an ASDC
experiment's water-scintillator mix -- where light levels from capture on hydrogen are not a 
problem -- in the context of a 
galactic supernova burst is that the characteristic delayed neutron 
capture signal is reduced from approximately 200 microseconds to 20 microseconds 
following the prompt positron signal.  This could prove valuable in avoiding pileup of events,
 particularly for relatively nearby explosions.  Note that during the first second alone 
 about 50\% of the total number of neutrinos will arrive, along with most of the very 
 interesting physics.

Being able to tag the \nuebar events would 
immediately double pointing accuracy back to the progenitor
star.  This is merely the result of statistics, since the elastic scatter 
events (about 3\% of the total) would no longer be sitting on a large 
background in angular phase space~\cite{raf}.  Reducing the error on this quantity by a 
factor of two, roughly from 6 degrees to 3 degrees,  
via neutron tagging would reduce the amount of sky to be searched by a factor of four.  
This could prove quite important for the wide-field
astronomical instruments which would be frantically attempting to see the 
first light from the supernova.

At the same time, this event-by-event subtraction could allow identification 
of the initial electron neutrino pulse from the neutronization of the infalling 
stellar matter, a key and as yet unobserved input in understanding supernova dynamics.

What's more, the neutral current events, which may be easily identified by 
their mono-energetic gammas between 5 and 10 MeV once the \nuebar events 
are subtracted, are very sensitive to the temperature of the burst and 
the subsequent neutrino mixing~\cite{snnc}. 

With a neutrino energy production threshold of 15.4 MeV, the weakly
backward-peaked charged current events are even more sensitive to the burst 
temperature and the subsequent mixing~\cite{sncc} than the NC events.  

If the exploding star was big and rather close (say, like Betelgeuse at 500 light years) 
we would get an early warning of its impending collapse~\cite{siburning}.  Approximately
a week before exploding, the turn-on of silicon fusion in the core would
raise the temperature of the star sufficiently that electron-positron annihilations
within its volume would begin to produce \nuebar just above inverse beta threshold.  

As early as one week before collapse we would 
see a monotonically increasing, many sigma excursion in our tagged event rate.  
The continuing increase in neutron capture event rate would clearly indicate a 
coming explosion, ensuring that we did not intentionally turn off the ASDC for 
calibration or maintenance and thereby miss the coming explosion. What's more, 
study of this pre-explosion Si-induced neutrino data would provide a unique window into the 
dying star's last days, hours, and minutes as an end-state fusion reactor.

Given a sufficiently high light yield, the ASDC could also detect elastic scattering of neutrinos on protons.  Determining the sensitivity to this channel would require a detailed simulation, along with measurements of proton quenching in the final WbLS target cocktail.  However, should it be possible, the use of a WbLS target could provide a handle for differentiating the primary background of $^{14}$C, which gives Compton electrons above Cherenkov threshold,  from the isotropic scintillation of proton scatters.

In addition, a large ASDC experiment would be sensitive to quite late 
black hole formation following a supernova explosion within our galaxy, 
since the distinctive coincident inverse beta signals from the cooling phase 
could be distinguished from the usual singles backgrounds. An abrupt cutoff 
of these coincident signals occurring up to a minute after the main burst 
would be the conclusive signature of a singularity being born~\cite{vogel}. Direct observation
of such an event would clearly be of great value, especially when correlated with
electromagnetic signals from X-ray or gamma-ray observatories as well as gravitational 
wave observations.

Finally, such an advanced detector would allow us to be absolutely 
certain that a supernova was occurring the moment the data started to arrive.
This is because the distinctive torrent of correlated \nuebar event pairs -- thousands per 
second for an explosion anywhere in our galaxy -- cannot be 
faked by electronic noise, flashing PMT's, or other physics. 
Clearly, the surer we are that a supernova has just occurred, the faster 
we should be able to get word out to the community, whether that means to another neutrino 
detector in a neighboring tunnel or to astronomers on the other side of the planet.

So, from extracting the neutronization signal, to deconvolving the main burst, to 
pointing back at the progenitor star, to observing late black hole formation, to getting the word out, 
having a large ASDC detector in operation would positively impact just about every 
physics topic connected to the detection of a galactic supernova.
Without the ASDC, much of this physics would be buried in background, degraded in precision, or 
wholly inaccessible.  In summary, the key benefits are as follows:

\begin{enumerate}

\item Neutron tagging of the inverse beta events would allow the de-convolution of a galactic 
supernova's various signals, which in turn would allow much more detailed interpretation of 
the physics of the burst.  This will result in significantly improved pointing back to the exploding star.

\item Early warning (hours before the arrival of the supernova neutrino 
wave) of large, relatively nearby supernovas would be possible via the observation of 
silicon-powered fusion in the dying stellar core.

\item Neutron tagging would allow very late time black hole formation -- minutes after the initial 
explosion -- to be observed.  This critically important signal could otherwise be hidden 
by background events.

\item The characteristic signal of prompt events rapidly followed by delayed events in about the 
same locations would make the arrival of a genuine neutrino burst instantly identifiable, vital 
for getting word out to the rest of the world in a timely fashion.

\end{enumerate}

\subsection{Diffuse Supernova Background Neutrinos }\label{s:dsnb}
A large-scale WbLS detector like the ASDC might prove to be an almost ideal observatory for diffuse supernova neutrinos as the technique can provide large detection volumes as well as excellent background rejection, allowing for a nearly optimum use of the available target mass.

As in the case of Super-Kamiokande, detection of the electron antineutrino component will be achieved by the inverse beta decay on the free protons in the liquid \cite{sk-dsnb}. Provided the ASDC features a sufficiently high scintillation light yield, the 2.2\,MeV $\gamma$-ray emitted by neutron capture on hydrogen can be exploited to obtain a coincidence signature between prompt positron and delayed capture signal. As a result the single-event background that currently constitutes the main obstacle to detection of the DSNB in Super-Kamiokande could be efficiently suppressed \cite{sk-dsnb,gadzooks}. This opens up an energy window for detection that reaches from about 10\,MeV to 30\,MeV in neutrino energy, limited by reactor antineutrino fluxes at the low end and atmospheric neutrino charged-current reactions towards higher energies.

The WbLS-based ASDC will have several advantages over either a pure water Cherenkov or pure scintillator detector: 
\begin{enumerate}
\item A pure water Cherenkov detector can be doped with gadolinium to enhance the gamma energy released in neutron capture, shifting the capture signal above detection threshold \cite{gadzooks}. Studies for this approach in the frame of the EGADS test facility and the GADZOOKS! project seem indeed very promising. However, apart from technical problems (that are largely solved) the efficiency for the detection of the delayed event in a detector the size of Super-Kamiokande will be on the order of 60\%, most likely lower for larger target volumes. A WbLS detector providing sufficiently high scintillation light yield can be expected to reach a significantly higher detection efficiency.

\item A very large liquid-scintillator detector like JUNO or LENA will easily reach detection efficiencies close to 100\% for the bulk of its active volume \cite{lena-wp}. However, the KamLAND collaboration reported~\cite{kl-ncbg} an unexpected correlated background mimicking neutrino signals in the energy region of interest. It is induced by neutral-current reactions of atmospheric neutrinos on carbon nuclei in the scintillator: The prompt signal from the emission of an energetic neutron (and other nucleons) from the nucleus is followed by its subsequent capture. The observed background rate is 1--2 orders of magnitude larger than the current expectation for the DSNB signal. A sensitivity study backed up by laboratory measurements of the scintillator response carried out in the context of the LENA project has shown that these background events can be efficiently discriminated by the differing scintillation pulse shape of the prompt events. However, the corresponding loss in DSNB signal events reduces the detection efficiency to about 40\% \cite{lena-phd}.
\end{enumerate}

The WbLS-based ASDC will suffer much less from the NC backgrounds: while positrons in this energy range will emit a sizable Cherenkov signal, nucleons or nuclear fragments of comparable energies will be well below the Cherenkov threshold. As long as Cherenkov and scintillation light emission are clearly separated (e.g.~by timing), a high efficiency for signal event detection (prompt positron with Cherenkov+scintillation signal) accompanied by a strong background suppression (scintillation only from proton recoils and similar) can be expected. Additional background rejection capability is obtained from the Cherenkov hit pattern, since the NC background produces multiple gammas, which will result in a more isotropic hit pattern.  This alone can suppress the NC background by roughly an order of magnitude~\cite{sk-dsnb}.

Based on these arguments, it seems likely that the ASDC has the capability of outperforming not only pure-water Cherenkov detectors but potentially also the much better adopted Gd-doped variant as well as purely organic liquid-scintillators. The relative fractions of Cherenkov and scintillation light would have to be carefully tuned to allow for an efficient detection of the delayed neutron due to increased scintillation while at the same time not overwhelming the Cherenkov signal vital for the NC background discrimination. The corresponding signal and background efficiencies have to be investigated.

\subsection{Proton decay}\label{s:decay}
\begin{figure}[!h]
\begin{center}
\includegraphics[width=0.47\textwidth]{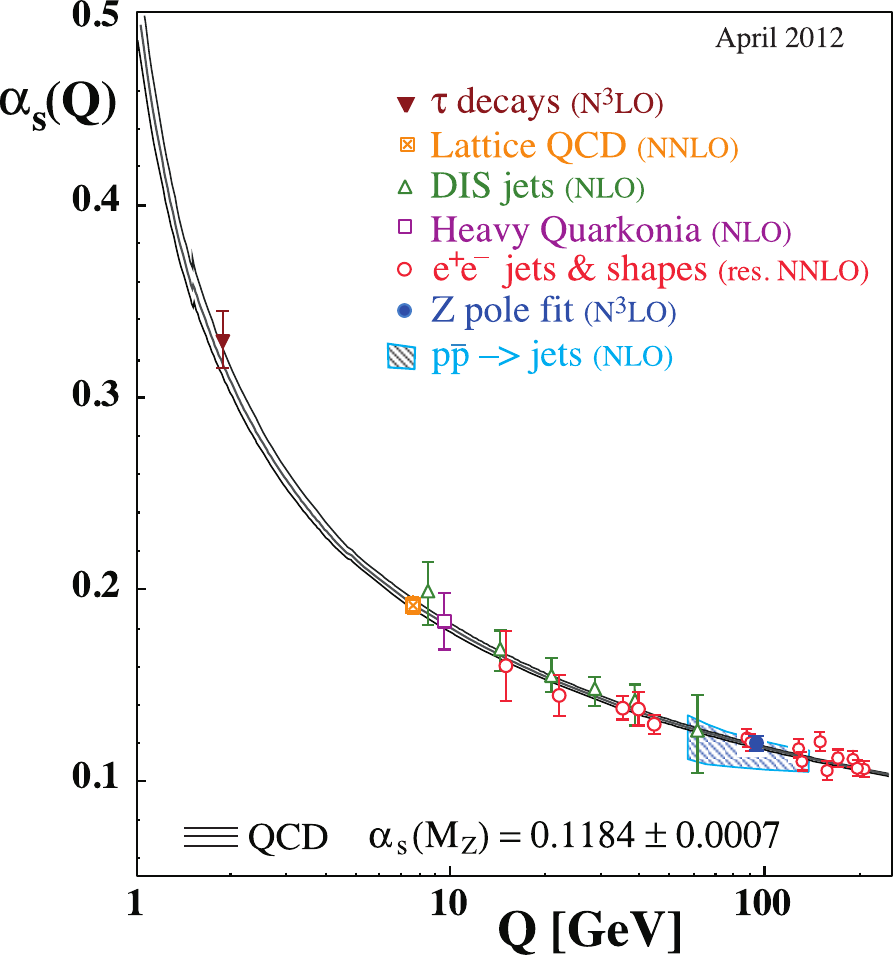}
\includegraphics[width=0.47\textwidth]{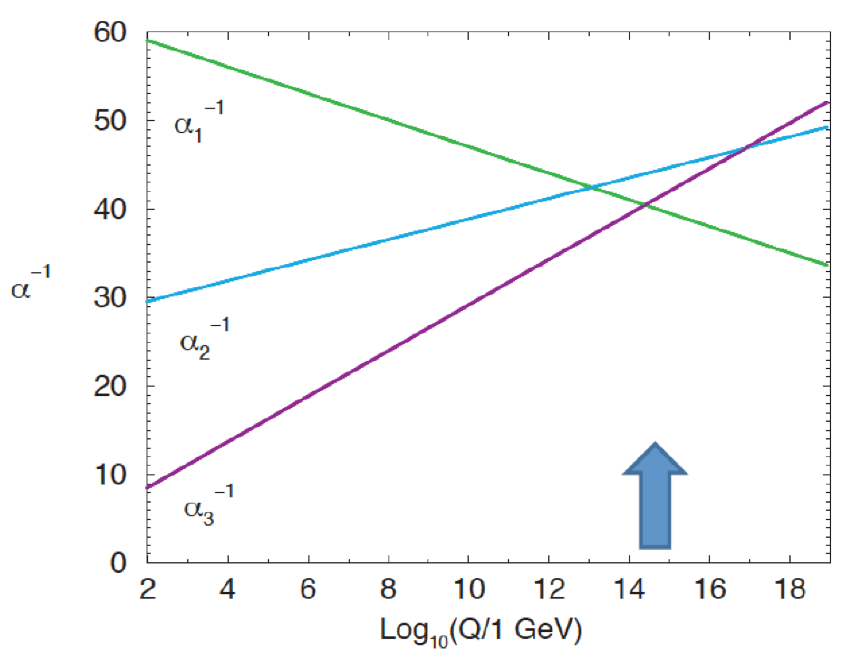}
\caption{The running of $\alpha_{S}$ with energy scale ($Q$)(left).~\cite{r:PDG2013} The running constants
for the strong ($\alpha_{3}$, weak($\alpha_{2}$), and electromagnetic ($\alpha_{1}$) forces extrapolated to the vicinity of the unification scale (indicated by blue arrow).}
\label{f:RunningCC}
\end{center}
\end{figure}

One of the Big Ideas in particle physics is the notion that at higher energies, the laws of physics become more and more symmetric and simple. The complex phenomena we see at our low energy scale are then ascribed to variations in how certain reactions freeze out at our scale. For example, the production of virtual gluon pairs via loops in the vicinity of a quark screens the true value of the color charge. As probes interacting with the quark have higher and higher momentum transfer they penetrate farther and farther into this screening field and hence interact with a smaller and smaller effective color charge. Thus the coupling constant for the strong interaction ($\alpha_{S}$) is a function of  the energy scale $Q$. This running coupling constant $\alpha_{S}(Q)$ decreases roughly exponentially between 2 and 100 GeV/c (Fig.~\ref{f:RunningCC}). Similar behavior is measured for the weak ($\alpha_{W}$) and electromagnetic ($\alpha_{EM}$) coupling constants, but the $Q$ dependence is quite different. Barring perturbations from other processes that might enter, these three running coupling constants become similar in strength at $Q$ in the range of $10^{13}$ -- $10^{16}$ GeV. This hints that these three forces may actually be a single forceâ with the differences at low energy being due to the details of the exchange particle properties and the resulting vacuum polarization. This so-called Grand Unified Theory (GUT) is a touchstone of current thinking in particle physics. Theories ranging from Supersymmetry (SUSY) to a wide class of string theories all have this basic Big Idea.  Determining if this unified view is correct is a major challenge for experimental physics. 

The only known way to address this challenge experimentally is via looking for proton decay. The seeming convergence of the coupling constants at a very high unification energy of $10^{15}$ to $10^{16}$ GeV implies that there may be a single force that could connect quarks and leptons at that scale. Such reactions would violate baryon (B) and lepton (L) number by the exchange of very heavy bosons with masses in the range of the unification energy scale. Since that energy is far beyond the reach of any conceivable accelerator, they would only manifest themselves at our low energy scale via virtual particle exchange leading to rare B- and L-violating reactions. This would mean that normal matter (e.g., protons, either free or in nuclei) would not be stable but would decay with some very long lifetime. This phenomenon, generically called {\it proton decay} although neutrons in nuclei are included, has been searched for in a series of experiments of increasing sensitivity.

\begin{figure}[!hb]
\begin{center}
\includegraphics[width=0.47\textwidth]{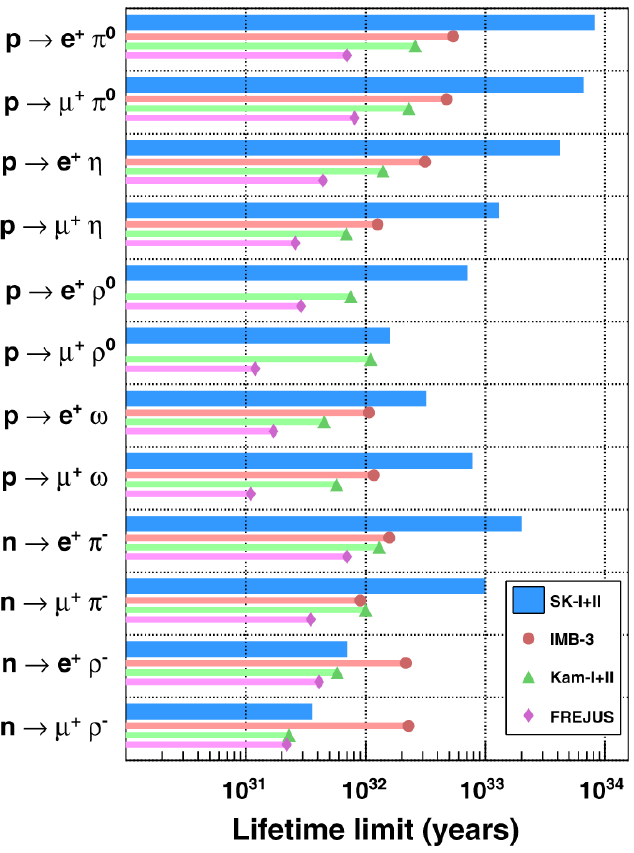}
\caption{Current limits on lepton plus meson modes of proton and neutron decay~\cite{r:nishino2012}. Five of twelve of these
modes have one or more background events. This will be an issue that can be addressed by neutron tagging.}
\label{f:leptonmeson}
\end{center}
\end{figure}

How do we know the appropriate lifetime range to look in? At the simplest level, proton decay could proceed via exchange of a heavy boson X. In this case, the emission of an X-boson converts a $u$ quark to a positron; thus it carries charge, lepton, flavor, and baryon number. Absorption by a $d$ quark converts it into an $ \bar{\mathrm{u}}$ quark, which forms a bound state with the remaining $u$ quark to make a $\pi^{0}$. This process has a lifetime given by a combination of the phase space, exchange boson, and coupling factors: $\tau \simeq m_{X}^{4}\hbar/(\alpha^{2} M_{X}^{5})$.  Using this as a guide, one can identify an approximate range of values in which the constants are close together ($\alpha\simeq 1/45$) and use the unification mass as the exchange particle mass. The very rough lifetime range is then $10^{32} -- 10^{36}$ years. This is the region currently being explored by Super-Kamiokande (SK), which with a 22.4~kT fiducial mass has the current best sensitivity for most decay modes of this type.  Figure~\ref{f:leptonmeson} shows current limits, which range from  $3.6\times 10^{31}$ to $8.2\times 10^{33}$ years.~\cite{r:nishino2012} These limits are based on a 141~kT-year exposure.~\footnote{SK presented updated limits based on 260~kT-years at TAUP2013. The limit is $1.4\times 10^{34}$ years with an expected background of 0.7 events.} 
Currently, five of the twelve modes shown in Figure~\ref{f:leptonmeson} have one or more background events from atmospheric neutrinos, and background expectations for both the $e^{+}\pi^{0}$ modes are roughly one event. Clearly, both size and backgrounds will be an important issue for future large detectors searching for modes
of this type. 

Of course, there is no guarantee that proton decay proceeds via simple tree-level particle exchange. There are many theories with more complicated processes that alter this estimate significantly by changing the unification energy and/or higher order effects. For example, in many SUSY models this tree-level proton decay is suppressed by R-parity, and proton decay proceeds by second order dimension five processes, the most well-known example being  $p\rightarrow \overline{\nu}K^{+}$. This is a difficult process for  water Cherenkov detectors to look for, as the $K^{+}$ is below the Cherenkov threshold and thus can only be detected via the decay daughters - mainly $K^{+}\rightarrow \mu^{+}\nu_{\mu}$ (64\%) and $K^{+}\rightarrow \pi^{+}\pi^{0}$ (21\%). The backgrounds from atmospheric neutrino interactions for these modes is very large. In Super-Kamiokande there are literally hundreds of possible candidates for decays into this mode. To reject these background events requires using the fact that a proton disappearing from an oxygen nucleus leaves a hole that roughly half the time gives a 6 MeV de-excitation gamma. The
average branching ratio weighted efficiency for this mode is therefore only 15.1\%~\footnote{SK4 has improved this to 19.1\%}. SK has set a limit of $5.9\times 10^{33}$ years for this mode based on a 260~kT-year exposure, with an expected background of 1 event.~\cite{r:masatontau2013}

Proton decay final states depend on the details of a given theory. While the modes $p\rightarrow e^{+}\pi^{0}$ and $p\rightarrow \overline{\nu}K^{+}$ are commonly used as benchmarks, there are many other
possible modes that conserve spin, momentum, and charge. One particularly intriguing possibility is the decay to only neutrinos, e.g. $n\rightarrow 3\nu$.~\cite{r:mohapatra2003} The basic idea motivating a search for this decay mode (and others involving 4- and 5-body decays with leptons, neutrinos, and pions) is that space has extra dimensions and a natural energy scale of ($M_{*}$) 10-100 TeV instead of the GUT scale of $10^{16}$ GeV. In this case, the decay of the proton depends on both the compactification scale ($R$) of the extra dimensions and $M_{*}$. Thus, while calculation of the lifetime is difficult, determination of the decay modes is more certain. In many cases, the models involving charged particles are difficult to search for, as there are at least two neutrino (or sterile neutrino) daughters in each mode. This ruins the invariant mass reconstruction - a key method in discriminating against atmospheric neutrino background. Thus while the $3\nu$ final state may seem difficult at first glance, it is actually made easier due to the fact that there is essentially no atmospheric neutrino background.
	An effective way to search for $n\rightarrow 3\nu$ is to look for the de-excitation of the parent nucleus following the creation of a neutron hole. To look for de-excitation products requires a low threshold, a radiologically clean detector medium, and a deep depth to suppress muon spallation backgrounds. Thus the Borexino~\cite{r:back2003},
SNO~\cite{r:ahmed2004}, and KamLAND~\cite{r:araki2006} detectors have searched for this mode, while Super-Kamiokande has not. It should also be emphasized that any model predicting neutron decay to invisible neutrinos will have the same search methodology - it doesn't depend on the details of the specific model.
	In KamLAND, the search was for neutron disappearance from the fully-occupied $1-s$ shell in $^{12}C$, which
results in energetic (up to 45 MeV) neutron emission 5.8\% of the time, followed by the neutron capture, and decay of the $^{10}C$ ground state ($t_{1/2} = 19.3\; s$). This triple coincidence has a small background, dominated by the tails internal radiological contamination distributions. They reported one candidate event in a two year exposure of a fiducial mass of 0.4 tons. Taking into account the expected background of 0.8 events (from the tails of the internal radioactive contaminant distributions) they get a lifetime limit $\tau > 5.8\times 10^{29}$ years. SNO obtained a similar limit of 
$\tau > 2\times 10^{29}$ years by looking for 
a 6.3 MeV gamma that occurs from the disappearance of a proton from $^{16}O$ about half the time (the same technique used by SK described above). The  major background in this case was from solar neutrino initiated NC breakup of deuterium, followed
by a 6.3 MeV gamma from $^{2}H(n,\gamma)^{3}H$. This large background explains the rather poor sensitivity. This confusion
of the solar NC capture gamma with proton decay would not be a problem in a light water detector.

An ASDC experiment would extend the reach for proton decay in all three of these basic search methodologies by:
\begin{enumerate}
\item Increased size, up to about 100~kT (four times larger than Super-Kamiokande).
\item Improved neutron tagging via use of WbLS, gadolinium doping, or both. This would improve atmospheric background rejection and improve efficiency for invisible modes.
\item Detection of below-Cherenkov threshold particles via use of WbLS. This would allow detecting low energy kaons, protons, and pions for signal efficiency and background reduction.
\item Increased depth and cleanliness, allowing Super-Kamiokande scale water detectors to perform invisible mode searches.
\end{enumerate}

For illustration of the effect in sensitivity, it will be assumed that depth and cleanliness are sufficient to render internal contaminants and muon spallation products irrelevant (as was the case for SNO). In addition, it is assumed that
the addition of a small amount of scintillation light from WbLS, when viewed with LAPPDs does not degrade the current
reconstruction efficiency realized by Super-Kamiokande. 

{\it $p\rightarrow e^{+}\pi^{0}$ and related modes:} In this mode, the reconstruction of free protons is important, in that in $^{16}O$ roughly 60\% of $\pi^{0}$s scatter or are absorbed before exiting the nucleus. For Super-K, Hyper-K, and an ASDC experiment this implies roughly the same overall efficiency of:

\begin{center}
$\epsilon_{average} = \frac{8}{10}\epsilon_{O}+\frac{2}{10}\epsilon_{H} = (0.8)(28\%)+(0.2)(87\%) = 40\%$
\end{center}

Note that for free protons, only 13\% are lost in reconstruction, while for oxygen 72\% are lost -- dominated by the 60\% lost in internuclear scattering.

\begin{figure}[!h]
\begin{center}
\includegraphics[width=0.45\textwidth]{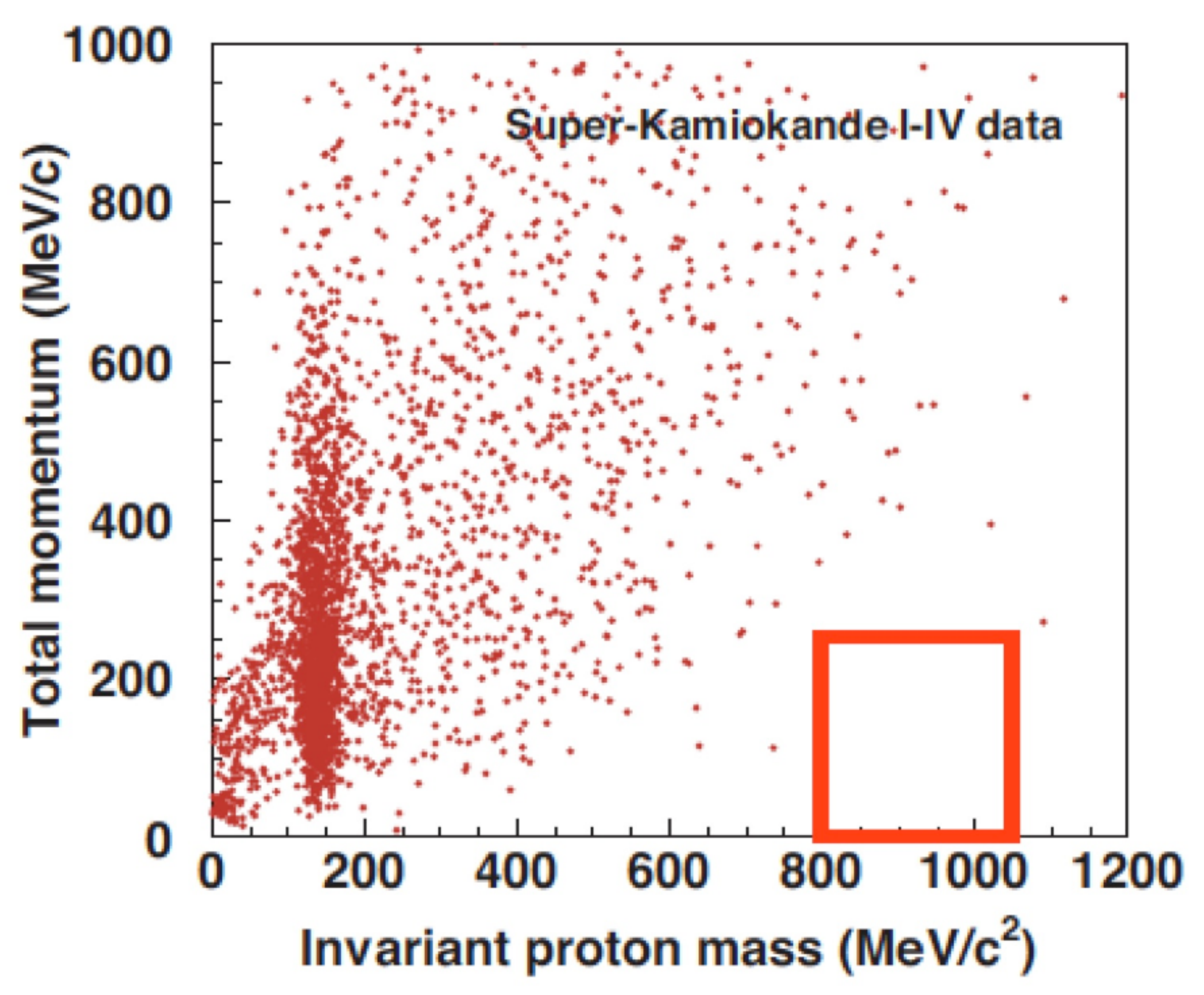}
\includegraphics[width=0.45\textwidth]{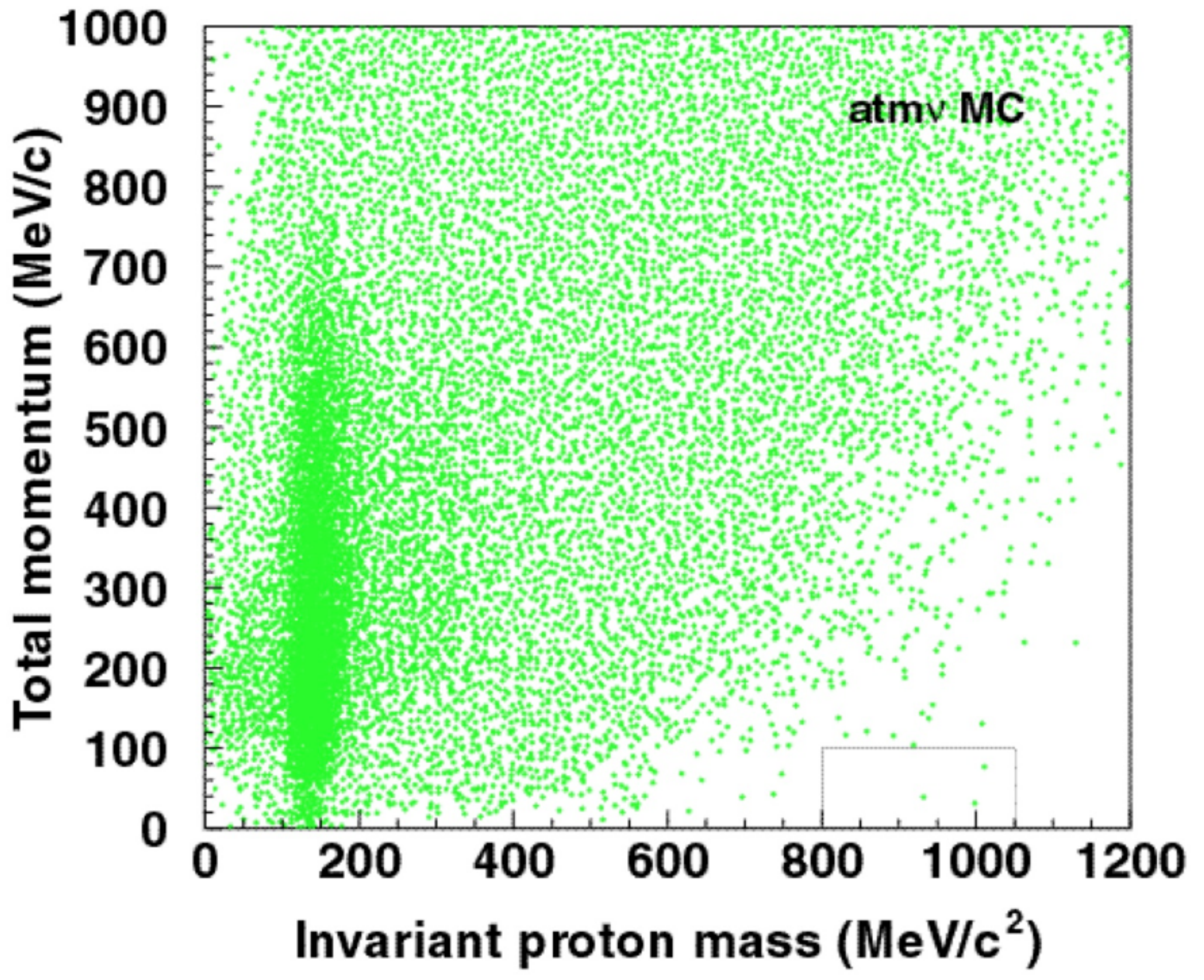}
\caption{Reconstructed invariant mass and momentum for Super-K I-IV data (left) and atmospheric neutrino simulation events (right). The effect of nuclear scattering can be clearly seen. Many of these background events could rejected
by neutron tagging.}
\label{f:SKPDK}
\end{center}
\end{figure}

	While an ASDC experiment would not be much better in signal reconstruction, the potential to reject atmospheric neutrino background is considerable. It is instructive to describe the analysis currently being used by SK. This analysis consists of (1) selection of events in the detector that have three showering tracks, (2) a requirement that at least one combination of tracks gives an invariant mass close to that of the $\pi^0$ (85--185 MeV), (3) a requirement that there was no follow-on Michel electron (indicating that there was a muon in the event), and (4) that the invariant mass be near that of the proton (800--1050 MeV) and the unbalanced momentum be less than 250 MeV/c. Figure~\ref{f:SKPDK} shows the invariant mass unbalanced momentum distributions for Super-Kamiokande for the atmospheric neutrino background MC (right) and data (left). At 260 MT-years there are no candidates.~\cite{r:masatontau2013}.

The 40\% selection efficiency or the SK analysis has an uncertainty of 18\%, dominated by pion nuclear effects.  Looking at the center plots, the incursion of background events into the signal region is clearly seen. The MC gives a background estimate of $2.1\pm 0.9$ events/Mton-year, which is consistent with the direct measurements made in the K2K 1-kT near detector~\cite{r:hayato1999} ($1.63^{+0.42}_{-0.33}$(stat)$^{+0.45}_{-0.51}$(syst) events/Mton-year). According to the SK MC, about 81\% of the background events are CC, with 47\% being events with one or more pions, and 28\% being quasi-elastic. In some cases, a $\pi^0$ is produced by an energetic proton scattering in the water. It may be that neutron tagging in an ASDC experiment could be a key method for doing this. There is a reasonable expectation that many of these background-producing events should be accompanied by one or more neutrons in the final state. This is because to look like proton decay there needs to be significant hadronic activity in the event, and there are many ways to produce secondary neutrons:

\begin{itemize}
\item direct interaction of an anti-neutrino on a proton, converting it into a neutron;
\item secondary (p,n) scattering of struck nucleons within the nucleus;
\item charge exchange reactions of energetic hadrons in the nucleus (e.g. $\pi^{-}+p\rightarrow n + \pi^{0}$);
\item de-excitation by neutron emission of the excited daughter nucleus;
\item capture of $\pi^-$ events by protons in the water, or by oxygen nuclei, followed by nuclear breakup;
\item secondary neutron production by proton scattering in water.
\end{itemize} 

The proposed ANNIE experiment~\cite{ANNIE} plans to study these effects further.

This is to be contrasted with real proton decay events, which are expected to produce very few secondary neutrons. Using very general arguments, there is an expectation that more than 80\% of all proton decays should {\emph not} have an accompanying neutron. The argument is as follows:

\begin{itemize}
\item For water, 20\% of all protons are essentially free. If these decay, there is no neutron produced as the $\pi^0$ would decay before scattering in the water, and 400 MeV electrons rarely make hadronic showers that result in free neutrons.
\item Oxygen is a doubly-magic light nucleus, and hence one can use a shell model description with some degree of confidence. Since two protons are therefore in the p$_{1/2}$ valence shell, if they decay to $^{15}$N, the resultant nucleus is bound and no neutron emission occurs except by any final state interactions  (FSI) inside the nucleus.
\item Similarly, if one of the four protons are in the p$_{3/2}$ state decays, a proton drops down from the p$_{1/2}$ state emitting a 6 MeV gamma ray, but the nucleus does not break up except by FSI.
\item Finally, if one of the two $s_{1/2}$ protons decay, there {\emph is} a chance that the nucleus will de-excite by emission of a neutron from one of the higher shells.
\end{itemize}

	There may be FSI-induced neutrons in some cases and for some modes 
(e.g., $\pi^{0}$ scattering in the nucleus could occur, but $K^{+}$ scattering would be rare), but it is also expected that not all nuclear de-excitations from $s_{1/2}$ states will give neutrons. In fact, more detailed nuclear calculations by Ejiri~\cite{r:ejiri1993} predict that only 8\% of proton decays in oxygen will result in neutron emission. This means that only 0.80 x 0.08 = 6\% of all proton decays in water should result in neutrons (ignoring FSI production by proton decay daughters). Thus neutron tagging may be an effective way to tag atmospheric neutrino backgrounds for all modes of proton decay where significant momentum is transferred to the nucleus. For ASDC we have assumed the extreme cases of 90\% and 0\% reduction to see the effect of neutron tagging. Since currently SK has only an 18\% efficiency for detecting neutrons with 40\% coverage, it is assumed that neutron tagging in HK with the planned 20\% coverage is negligible. If HK added gadolinium this would change, however.

\begin{figure}[!hb]
\begin{center}
\includegraphics[width=0.55\textwidth]{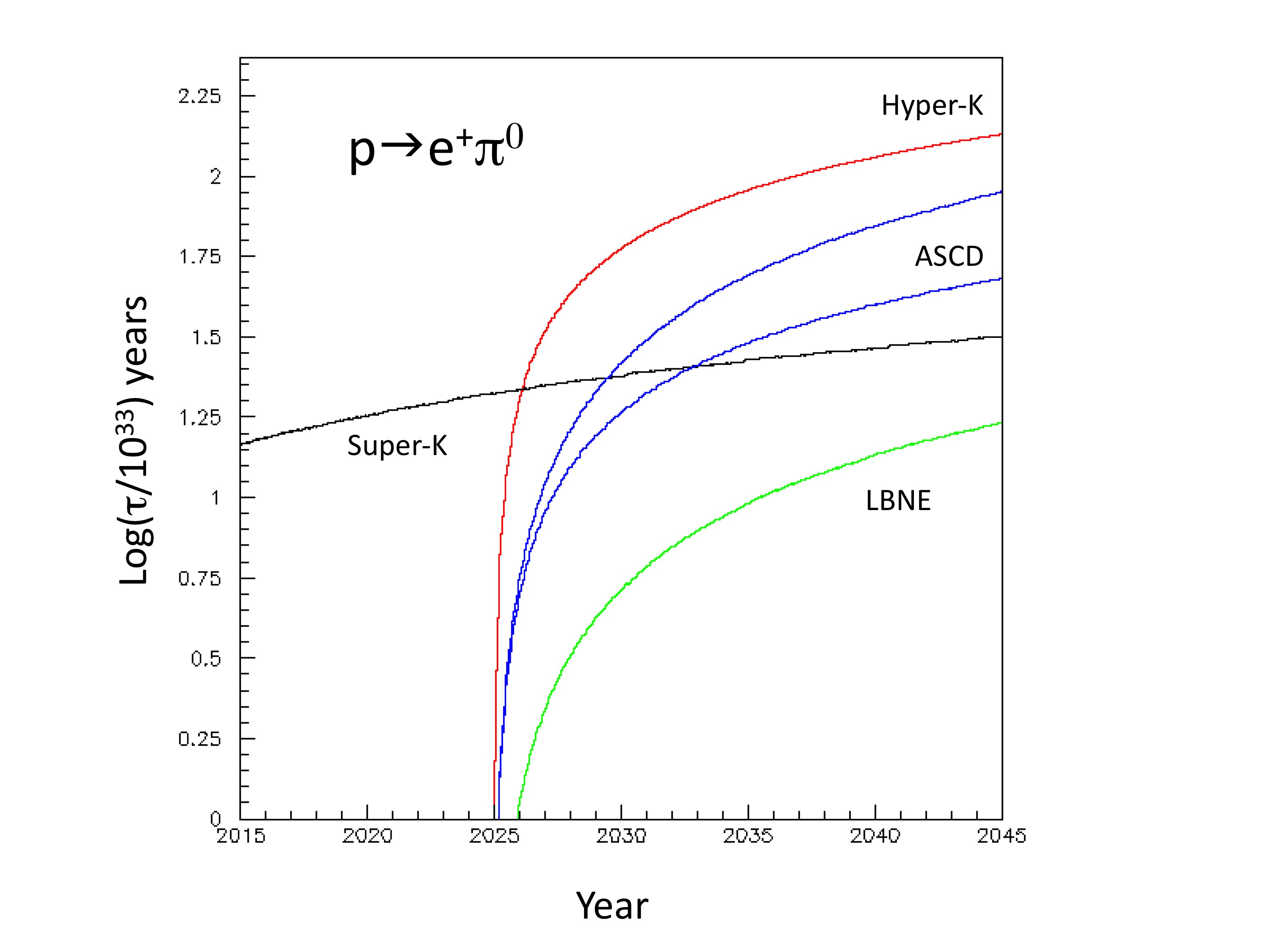}
\caption{Estimated sensitivity of an ASDC experiment compared to Super-K. The improvement is due both to larger
size and improved background reduction. If proposed long baseline detectors are built, Hyper-K would be better but LBNE worse for detecting this mode of proton decay. The upper ASDC curve assumes 90\% background reduction due to neutron tagging, whereas the lower curve assumes no neutron tagging.}
\label{f:epi0}
\end{center}
\end{figure}

	Thus we estimate that backgrounds in an ASDC with very efficient ($\simeq$100\%) neutron tagging via the
2.2 MeV gamma from will be reduced a factor of 10 compared to SK. Figure~\ref{f:epi0} shows the expected sensitivity at 90\% c.l. for detecting proton decay via this channel in SK and in an ASDC experiment with neutron tagging and with no neutron tagging.
Somewhat arbitrarily, a 2025 start date is assumed. Thus in this mode a 100~kT ASDC experiment would catch up with SK in sensitivity in a little over three years, despite the fact that SK would have been running for over thirty years at that point.
If Hyper-Kamiokande is built, it would be better in this particular mode, but an ASDC experiment could be competitive due to excellent background rejection. For comparison, a 34~kT LBNE, which has no free protons and a heavy nucleus with more pion scattering~\footnote{Note that Bueno, et al.~\cite{r:bueno2007} gives an overall efficiency for this mode of 45\%, which would mean the absorption of pions in argon would have to be {\it less} than that assumed by SK for the lighter oxygen nucleus. In order to make an apples to apples comparison we have scaled the SK absorption number to argon, which gives a 29\% survival rate (as compared to 40\% for oxygen). 100\% reconstruction efficiency is then assumed for LBNE.} is not competitive in this mode.

\begin{figure}[!h]
\begin{center}
\includegraphics[width=0.55\textwidth]{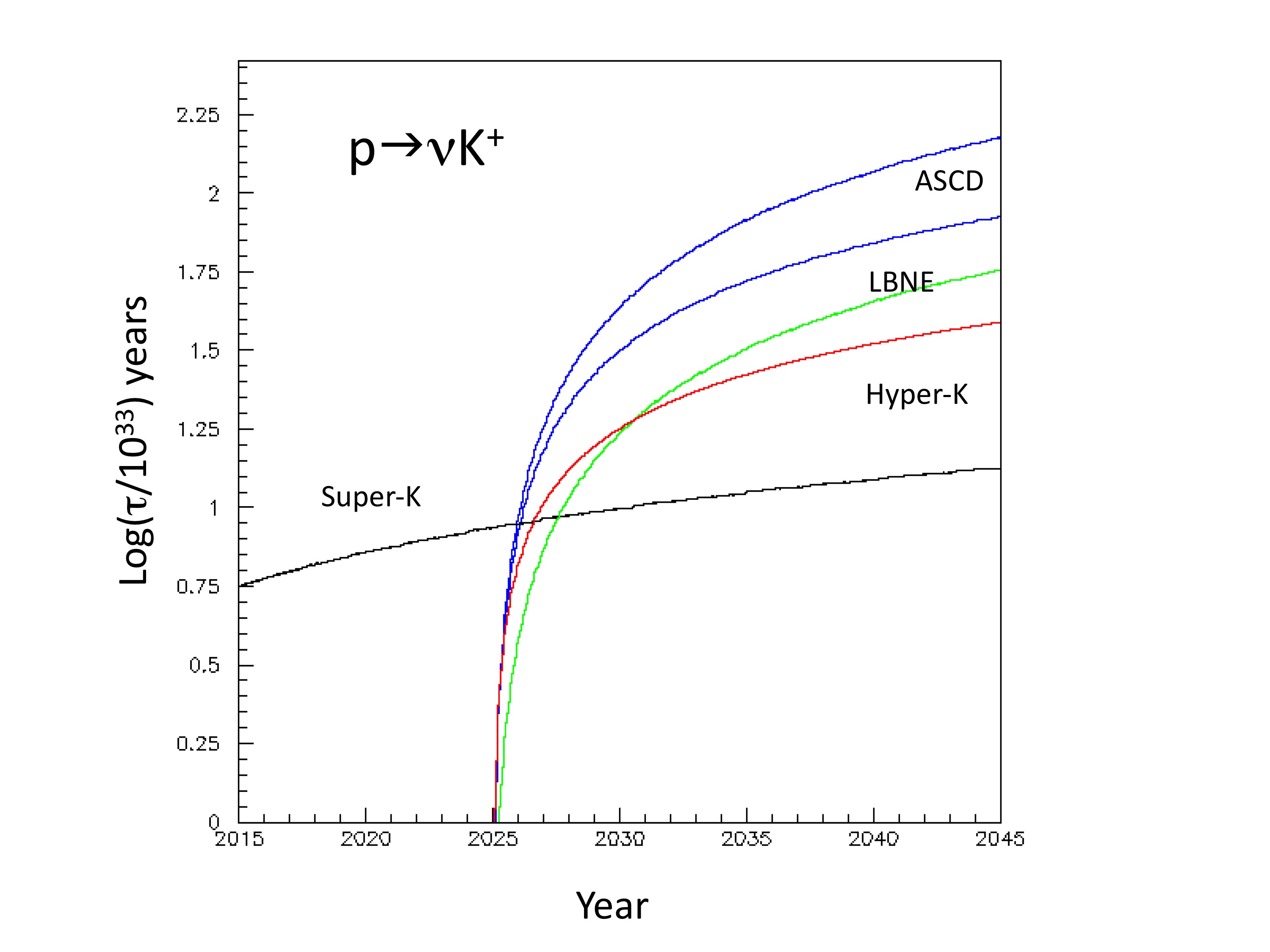}
\caption{Estimated sensitivity of an ASDC experiment compared to Super-K. The improvement is due to larger
size, much improved efficiency, and improved background reduction. The upper ASDC curve assumes 90\% background reduction due to neutron tagging, whereas the lower curve assumes no neutron tagging. LBNE efficiency and background numbers are from Bueno, et al.~\cite{r:bueno2007}}
\label{f:nuK}
\vspace{-0.3cm}
\end{center}
\end{figure}

{\it $p\rightarrow \overline{\nu}K^{+}$ and related modes:} For this mode internuclear scattering is much less important due to conservation of strangeness. Estimates are that only a few percent of $K^{+}$s are lost due to scattering. Due to the need to use the de-excitation gamma to reduce background, SK overall efficiency (averaged over four run configurations) is only 15.3\% (includes branching ratios). Recent improvements in software and electronics have increased this to 19.1\%, and so estimates for SK future sensitivity use this number. For HK, the SKII efficiency (20\% coverage) of 13.0\% was used, but an additional 3.5\% was added to estimate the effect of the SK improvements. SKII backgrounds of 6.2 events/Mton-year were used, however, as the software and electronics improvements left these relatively unchanged in SK. Figure~\ref{f:nuK} shows the expected sensitivity at 90\% c.l. for detecting proton decay via this channel in SK and in an ASDC experiment with neutron tagging and with no neutron tagging.  Of course, JUNO would also be sensitive to this mode, and has neutron tagging, but a five year run with JUNO would only be equivalent to a one year run of a 100~kT ASDC experiment.

{\it $n\rightarrow 3\nu$ and related invisible modes:} This mode is difficult in that the only detectable daughter is the excited parent nucleus. This means that the detector energy threshold must be very low as compared to to other modes, typically a few MeV. The detectors must be deep enough underground such that cosmic ray muon-induced backgrounds are not a limiting factor. In addition, other backgrounds such as solar and reactor neutrinos and internal radioactive contaminants become important backgrounds. This means that relatively shallow detectors such as JUNO, or detectors with significant radiological and spallation backgrounds such as Super-K are not as sensitive as deep, high resolution detectors, low threshold detectors such as SNO and KamLAND.

	SNO searched for this decay mode by looking for a 6.18 MeV de-excitation gamma from the neutron hole created in an oxygen nucleus. The most significant background was from single neutrons, which emit a 6 MeV gamma when captured on deuterium. Due to the large number of neutrons from NC-induced  breakup of deuterium it was only possible to set a limit of $\tau > 2\times 10^{29}$ years. This will be improved on by SNO+, which has an initial light water phase, and therefore no 6 MeV capture gammas. In this case, the main backgrounds are from solar and reactor neutrinos. In the case of SNO+, the expectations are that U/Th contamination and spallation products will be negligible, even with relatively poor energy resolution. Using directional cuts to reduce the solar neutrino contribution and an asymmetric cut of 6-9 MeV, the expected number of solar neutrino events is about 10/year. A background of about 2/year is expected from reactor neutrinos, which cannot be reduced due to the lack of neutron tagging capability. The expected sensitivity is about an order of magnitude larger than SNO, or about $2\times 10^{30}$ years.
	
\begin{figure}[!h]
\begin{center}
\includegraphics[width=0.47\textwidth]{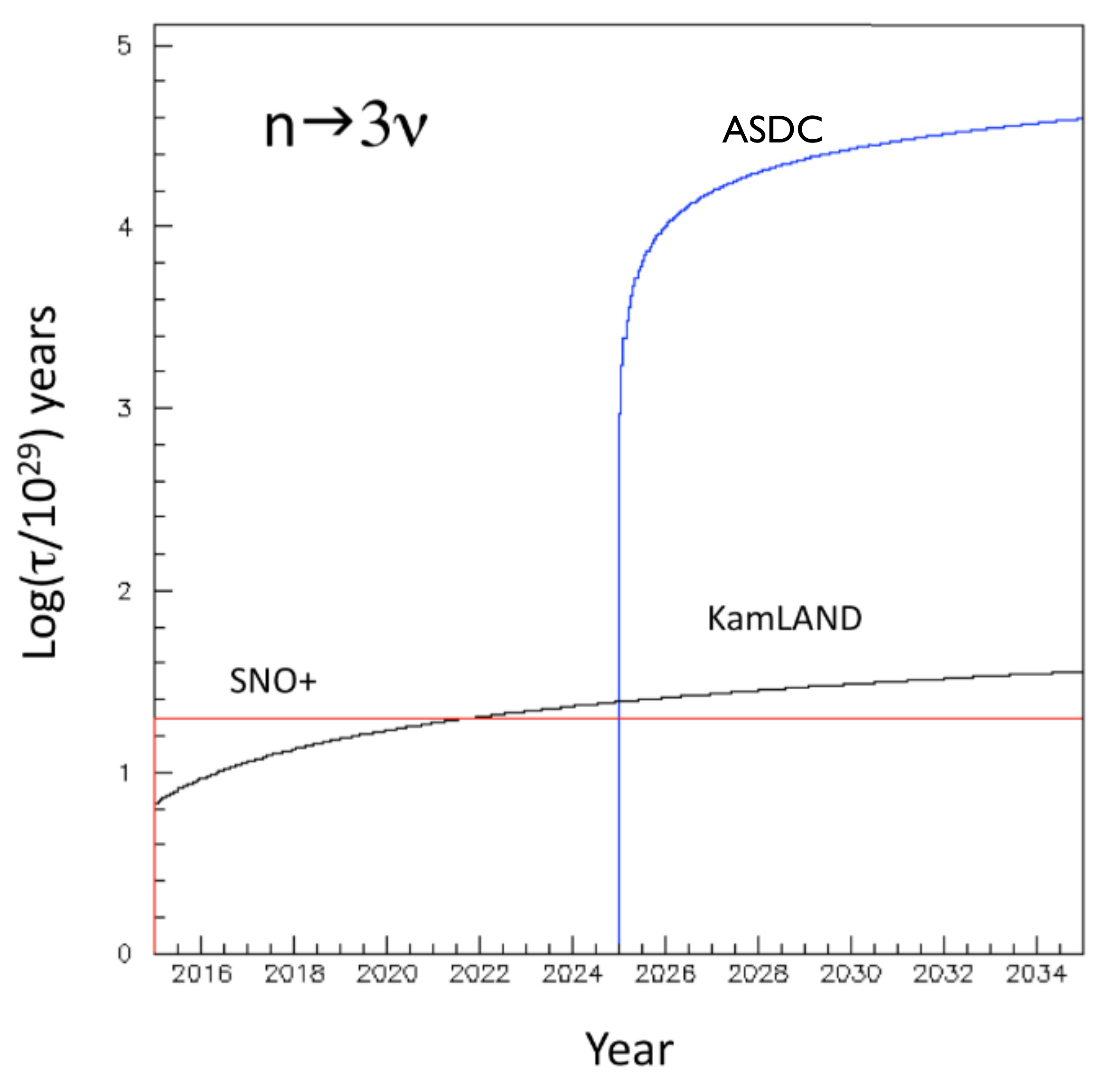}
\caption{Estimated sensitivity of an ASDC experiment compared to an extrapolated KamLAND sensitivity, plus the sensitivity expected from the water phase of SNO+. Sensitivity is clearly background limited, so background reduction by neutron tagging and directional reconstruction in an ASDC is significant.}
\label{f:3nu}
\end{center}
\end{figure}

	KamLAND's strategy of looking for nuclear de-excitation of  carbon following disappearance of a $1s$ neutron makes a clear signature via a triple coincidence: high energy neutron scattering a proton, capture of the neutron after stopping, and EC decay of the resultant nucleus - typically $^{9}C$, $^{10}C$, or $^{8}B$. The relatively long half life of these daughters (127 ms, 19.3 s, and 770 ms, respectively) and
low energy of the decay means that detectors with significant cosmic ray backgrounds (e.g.JUNO) will have difficulty in tagging them. In addition, the branching ratio into such nuclei, and the fact that only two of the 6 neutrons in $^{12}C$ are $1s$ also degrades the sensitivity. KamLAND has published a limit of $5.8\times 10^{29}$ years in 2006, with only one background event. Since the detector and has been collecting more data since then and was not background limited, the expected improvement of sensitivity with time is taken into account in Figure~\ref{f:3nu}, along with the expected sensitivity of SNO+.

	For an ASDC, it is assumed that the depth is sufficient (for example, Homestake) such that spallation events are not a problem background. In addition, the expected improved resolution over SNO+ would make U/Th contamination insignificant, and neutron tagging would remove almost all the reactor neutrino events. To estimate the backgrounds, we will adopt estimates from the SNO+ water phase, adjusted for expected detector improvements. Thus it is assumed that an asymmetric cut of 6-9 MeV on the 6.3 MeV de-excitation gamma from $^{15}O$ coupled with a factor of three better resolution can make the U/Th contribution (already small in SNO+) negligible. In addition, it is estimated that roughly 97\% of solar neutrinos in the interval 6-9 MeV visible energy can be rejected by directional cuts, with a loss of only 20\% of the 6 MeV $^{15}O$ de-excitation signal. Finally, the ability to tag neutrons via the use of WbLS will reduce the reactor neutrino background (about 0.2~/kT/year at Homestake) by 99\%, with the loss of only 1\% of the signal. Figure~\ref{f:3nu} shows a summary of the final results. An ASDC as described in this section could improve sensitivity by two to three orders of magnitude compared to existing experiments, and be the first high-sensitivity search for this important extra dimension inspired mode.
	
In conclusion, an ASDC experiment would greatly enhance sensitivity for proton decay over existing detectors such as Super-Kamiokande and KamLAND, plus meet or exceed sensitivity in some modes for planned detectors such as LBNE or Hyper-Kamiokande. It is the only proposed detector with significantly improved sensitivity in {\it all three} of the decay modes discussed above.

\subsection{Sterile neutrinos}\label{s:sterile}		
The Reactor Antineutrino Anomaly (RAA) came about as a result of detailed recalculation of the reactor neutrino fluxes in 2011~\cite{bib:Mueller, bib:Mention2011}. Contrary to the previous agreement between the number of detected and expected reactor antineutrinos, the new calculation implied a common deficit of the detected reactor antineutrinos at the level of 2.7~$\sigma$ for several previous reactor antineutrino experiments. Possible explanation of the RAA deficit may be the oscillation of electron antineutrinos into a new 4$^{th}$ neutrino species, the so-called sterile neutrino. The existence of sterile neutrinos has previously been suggested by the accelerator neutrino experiments LSND~\cite{bib:LSND} and MiniBooNE~\cite{bib:miniboone1, bib:miniboone2}. 

The nature of the RAA defines features of the proposed sterile neutrino in order to fit experimental data from reactor antineutrino detectors.  All reactor antineutrino experiments exhibit a constant deficit of detected reactor antineutrinos with respect to the new calculations, which implies that the oscillation pattern has been washed out and  is only measurable at distances of less than 10~m. Such a short oscillation length indicates a massive sterile neutrino, with the best fit value around 1~eV$^2$.

\begin{figure}[htbp]
\begin{center}
\includegraphics[width=0.48\textwidth]{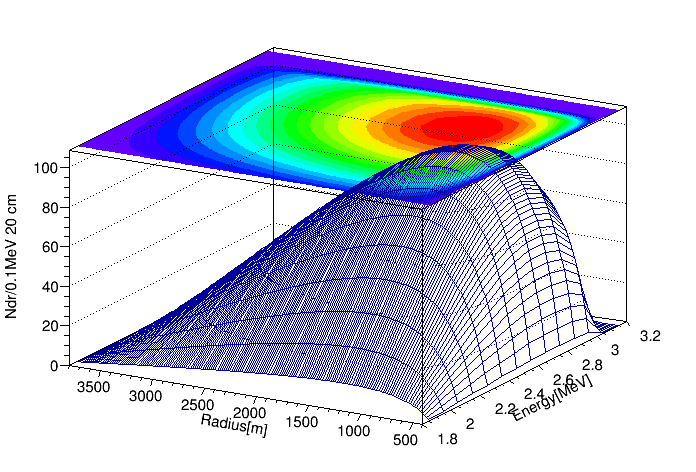}
\includegraphics[width=0.48\textwidth]{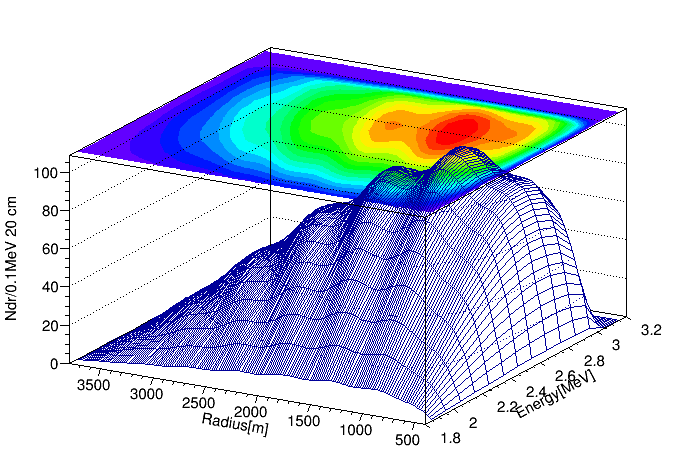}
\caption{Illustration of the oscillated and unoscillated distance- and energy-dependent flux seen in a large WbLS-filled detector such as the ASDC.}
\label{fig:surfosc}
\end{center}
\end{figure}

Three direct independent methods to test sterile neutrino hypothesis include: a very short baseline reactor experiment; an accelerator experiment; and use of neutrino and antineutrino generators in the vicinity of a large liquid scintillator or WbLS detector. A large LS or WbLS detector represents a particularly promising venue for a decisive measurement, as these detectors provide an opportunity to observe a distance-dependent neutrino flux from the source at distances on the order of the oscillation length. This can be seen in Fig.~\ref{fig:surfosc}.

The neutrino oscillation length is given by the following formula:
\begin{equation} 
L_{osc}[m] = 2.48 \frac{E_{\bar \nu_e} [MeV]}{\Delta m^2_{new} [eV^2]}.
\end{equation}

In the case of a sterile neutrino with $\Delta m^2 \sim 1-2$~eV$^2$,  the oscillation distance of interest is of the order of couple of meters.  A large LS/WbLS detector would span over 10 meters, and the source can be placed within a few meters of the target, creating a baseline comparable to the sterile neutrino oscillation length.  Observation of the oscillation pattern within the target  would represent the most convincing proof of the existence of sterile neutrinos and their oscillation with the other three flavors.

\begin{figure}[htb]
\centering
\includegraphics[width=0.6\textwidth]{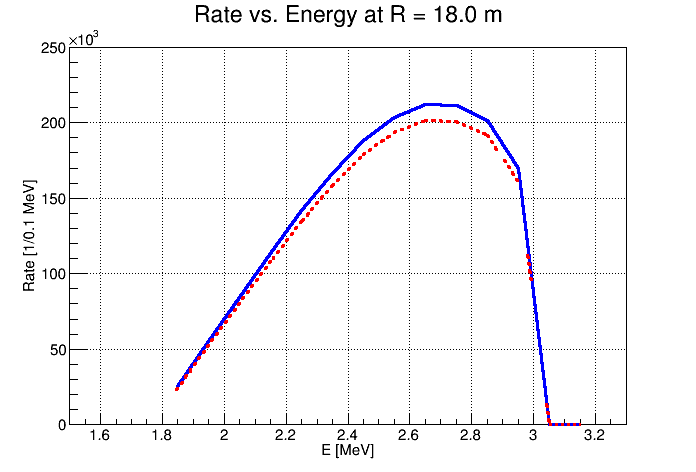}
\caption{The antineutrino rate as a function of energy for the  $^{144}$Ce - $^{144}$Pr antineutrino generator with low endpoint at 3~MeV.  The blue line is the spectrum in the absence of sterile neutrinos, and the red dashed line assumes oscillations to a sterile neutrino with $\Delta m^2 = 1$~eV$^2$ and $\sin ^2 2\theta = 0.1$.}
\label{fig:Pr_spec}
\end{figure}

The current generation of sterile neutrino searches at large LS detectors is being realized with SOX: the deployment of a 5 PBq $^{144}$Ce - $^{144}$Pr double beta decay antineutrino generator in the Borexino detector in late 2015, followed by later deployment of a $^{51}$Cr neutrino source. The $^{144}$Ce-$^{144}$Pr antineutrino source has an endpoint at 3~MeV (Fig.~\ref{fig:Pr_spec}), while $^{51}$Cr emits a mono-energetic  753~keV gamma.  SOX requires a sub-MeV energy threshold for the $^{51}Cr$ measurement and relatively good energy resolution, but suffers from relatively low statistics due to the detector size, allowing a $\sim$2~$\sigma$ C.L. test of the RAA. In 18 months of running around 12,000 antineutrino events are expected from a  5 PBq $^{144}$Ce - $^{144}$Pr source in the case of no oscillation, and a few hundred events fewer in the case of sterile neutrino oscillations ($\Delta m^2 \sim 1-2$~eV$^2$), making the statistical error the largest contributor to the overall error budget. Based on the studies done for the SOX detector, the main effect of low statistics is loss of sensitivity to small values of the mixing angle.

The current generation of sterile neutrino experiments aims to test the RAA at the 2$\sigma$ C.L. Future experiments must be designed  to measure sterile neutrino oscillations at the 5$\sigma$ level. 
The ASDC will provide an excellent opportunity to search for the 1 eV$^2$ sterile neutrinos at 5$\sigma$ C.L. With its large target mass of 20 -- 50~kT  a tremendous number of IBD interactions will be collected, allowing detailed, high statistics studies of the position-dependent neutrino flux. 

In an expanded attack on the sterile neutrino question,    the IsoDAR collaboration proposes to pair with the ASDC detector in a program to decisively address all of the short baseline anomalies.      In addition to the RAA, described above, this also includes the LSND and MiniBooNE appearance anomalies.

The IsoDAR/DAE$\delta$ALUS collaboration is developing a high-intensity $^8$Li decay-at-rest source consisting of a driver that produces neutrons that irradiate a FLiBe (a mixture of fluorine, lithium and beryllium)  sleeve containing 99.995\% isotopically pure $^7$Li \cite{daedaluswhitepaper}.   Studies have indicated that the optimal driver for the project is likely to be a cyclotron.  The first choice of beam is 60 MeV/amu H$_2^+$, although a 40 MeV/amu deuteron beam is also under consideration.     As an example of the design,  Table~\ref{parSterile} describes the H$_2^+$ running parameters, and the description of the target is the same for both beams.   
This work is progressing steadily, and the system can be produced in three years in a technically-driven schedule.

\begin{table}
{\footnotesize
\begin{center} 
      \begin{tabular}{|c|c|} \hline
        Accelerator  & 60~MeV/amu of H$_2^+$  \\  
        Beam Current  & 10~mA of protons on target  \\  
        Beam Power (CW)  & 600~kW  \\  
        Duty cycle  & 90\%  \\  
        Protons/(year of live time)  & 1.97$\times 10^{24}$ \\ 
        Run period  & 5~years  \\  
        Live time  & 5~years$\times$0.90=4.5 years  \\  
        Target   & $^9$Be with FLiBe sleeve (99.995\% pure $^7$Li)  \\  
        Sleeve diameter and length & 100~cm and 190~cm \\ 
        $\overline{\nu}$ source  & $^8$Li $\beta$ decay (6.4 MeV mean
energy flux)  \\  
        Fraction of $^8$Li produced in target &10\% \\  
        $\overline{\nu}$ flux during 4.5~years of live time  &
1.3$\times 10^{23}$~$\overline{\nu}_e$ \\  
        $\overline{\nu}$ flux uncertainty  & 5\% (shape-only is also
considered) \\  \hline
      \end{tabular}
\end{center}
\caption{\label{parSterile}Summary of IsoDAR H$_2^+$ machine
  parameters.}}
\end{table}

There are several advantages of this source over conventional sources.  
First, it produces antineutrinos with an endpoint of 13 MeV,  which is well above environmental backgrounds.   
Since the energy threshold of the ASDC is not established (and will depend on the final target WbLS cocktail), use of the DAR $^8$Li source is favorable. The $^{144}$Ce - $^{144}$Pr antineutrino generator spectrum has an end point at 3~MeV, and therefore requires an energy threshold of no more than 1.8~MeV (inverse beta decay energy threshold), while for the $^{51}$Cr the energy threshold should be below 750~keV.  
For sterile neutrino studies, which have designs driven by the $L/E$ dependence of the oscillation,  the high energy means the source can be located outside of the detector.      
   In addition, due to the smooth features of the $^8$Li spectrum the ASDC will not require high energy nor vertex resolution. Studies performed for a 1~kT detector show that energy resolution up to 15\% is fine, while vertex resolution better than 50~cm is desired. The studies also brought up two very important conclusions: knowledge of the absolute flux is necessary to search for the higher $\Delta m^2$; and the distance between the source to the detector should be kept as small as possible in order to provide sensitivity to small mixing angles.  
IsoDAR can run for many years,  as opposed to radioactive neutrino/antineutrino sources where the activity declines with the decay constant.      The installation does not require transport of radioactive material below ground.    And the source can be turned on and off allowing for background measurements and complementary physics programs.

The IsoDAR source has the capability of addressing {\emph both} the RAA and the appearance anomalies, assuming  $CPT$ invariance, making this a very powerful experiment.     To explain this latter point,  consider that 
$CPT$ invariance requires identical probabilities for $\bar \nu_\mu \rightarrow \bar \nu_e$ and $\nu_e \rightarrow \nu_\mu$ oscillations.   Also, 
probability for $\nu_e$ disappearance must exceed the probability  for $\nu_e \rightarrow \nu_\mu$ oscillations, since appearance is only one channel and disappearance sums over all channels.   Keep in mind, also,  that  
$CPT$ invariance also requires that the probability of $\bar \nu_e$ disappearance equal
the probability for $\nu_e$ disappearance.   Based on this,  one can conclude that if the sensitivity for $\bar\nu_e$ disappearance 
entirely covers the $\bar \nu_\mu \rightarrow
\bar \nu_e$ signal region,  then either a signal must be observed with mixing
angle such that
$\sin^2 2\theta_{\bar{ee}} > \sin^2 2\theta_{\bar{\mu e}}$, or else all models based that assume $CPT$ invariance must be ruled out.    

\begin{figure}[!h]
\begin{center}
\includegraphics[width=0.57\textwidth]{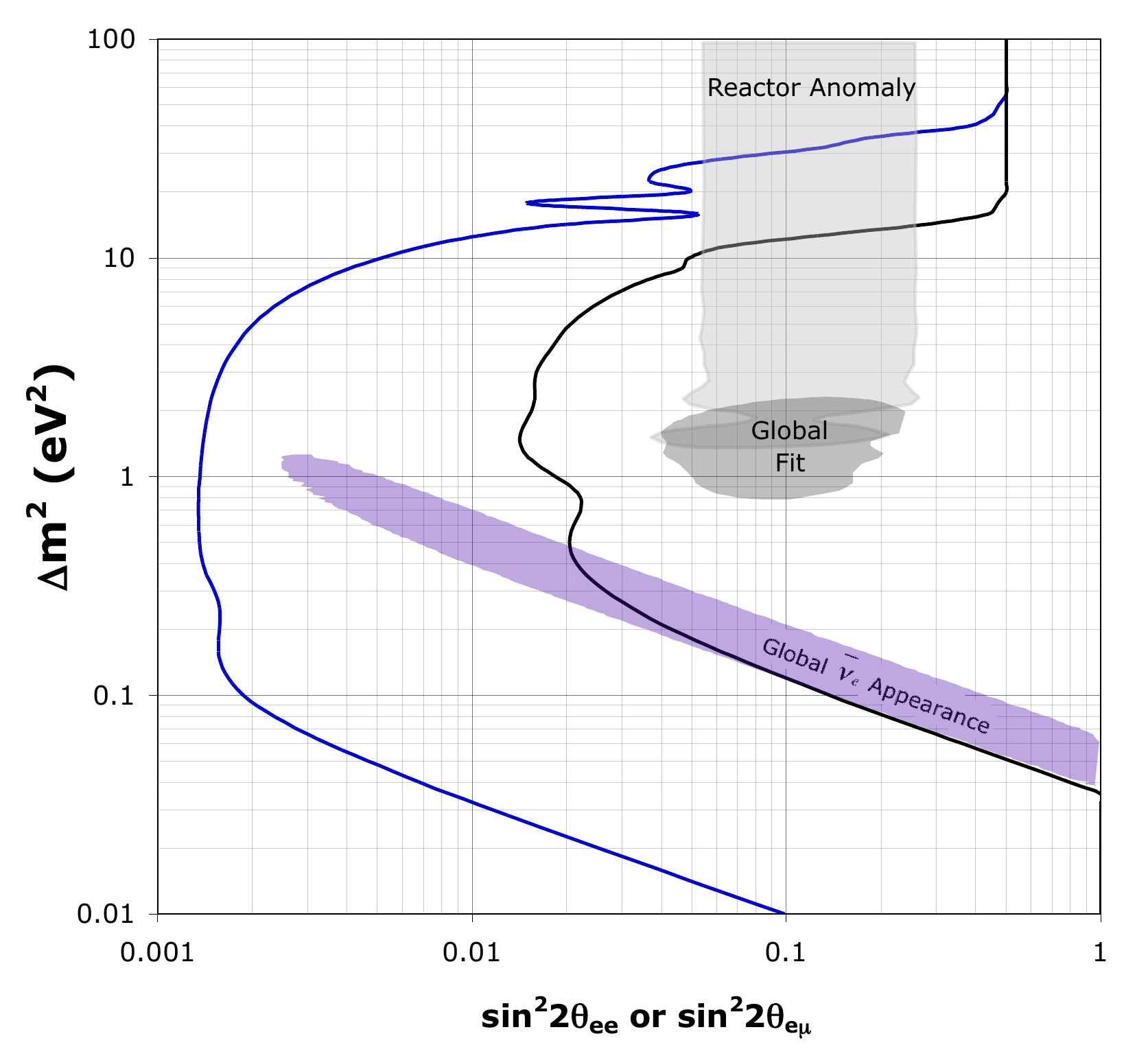}
\caption{The disappearance sensitivity with respect to the short baseline anomalies of a 20 kT (blue curve) and 1~kT (black curve) highly-doped ASDC detector paired with the IsoDAR antineutrino source .   Assuming $CPT$ is a good symmetry,  this program will decisively address all of present indications of sterile neutrinos.  The shaded areas indicate the allowed regions for the RAA (gray), the global oscillation fit (dark gray), and the global appearance fit (purple).}
\label{oscCPT}
\vspace{-0.5cm}
\end{center}
\end{figure}

For the 5$\sigma$ sensitivity shown in Fig.~\ref{oscCPT},  we have assumed 5 years of running with a 20~kT  ASDC detector, with maximal scintillant doping.    The allowed region for appearance is indicated by the purple region.    The RAA is indicated by the gray region.     One can see that this design will  will provide irrefutable evidence for the existence of sterile neutrinos as an explanation for the anomalies, or decisively refute that these data sets are due to sterile neutrinos.   Moreover, this system can distinguish whether there are more than one type of sterile neutrinos as the oscillation waves can be reconstructed with very high precision.      

At the same time, this system can be used for other physics studies.   The source will also give an unprecedented data set of $\bar \nu_e e$ scattering for electroweak studies.   The well-understood flux spectrum and cross section for inverse beta decay also makes this an excellent calibration source for the detector up to 13 MeV.

Lastly, one could begin on this program early in the development of the ASDC.   Fig.~\ref{oscCPT} also shows the sensitivity for running with a 1~kT maximally-scintillator-doped detector.      This could be a run with an existing detector, like KamLAND, in the near future,  or with a 1~kT prototype detector for the larger ASDC, for example WATCHMAN, if this is constructed.    This will decisively address the RAA signal and will give the experiment experience in running the IsoDAR source.

In conclusion, the ASDC will provide an exceptional playground for a detailed study of sterile neutrino oscillations, placing only modest requirements on detector performance:  energy and vertex resolutions equivalent to those already achieved by pure water Cherenkov detectors, and an energy threshold as high as several MeV.  Additional important ingredients include knowledge of the absolute incoming antineutrino flux and a few-meter distance between the source and the detector.

\begin{figure}[!h]
\begin{center}
\includegraphics[width=0.6\textwidth]{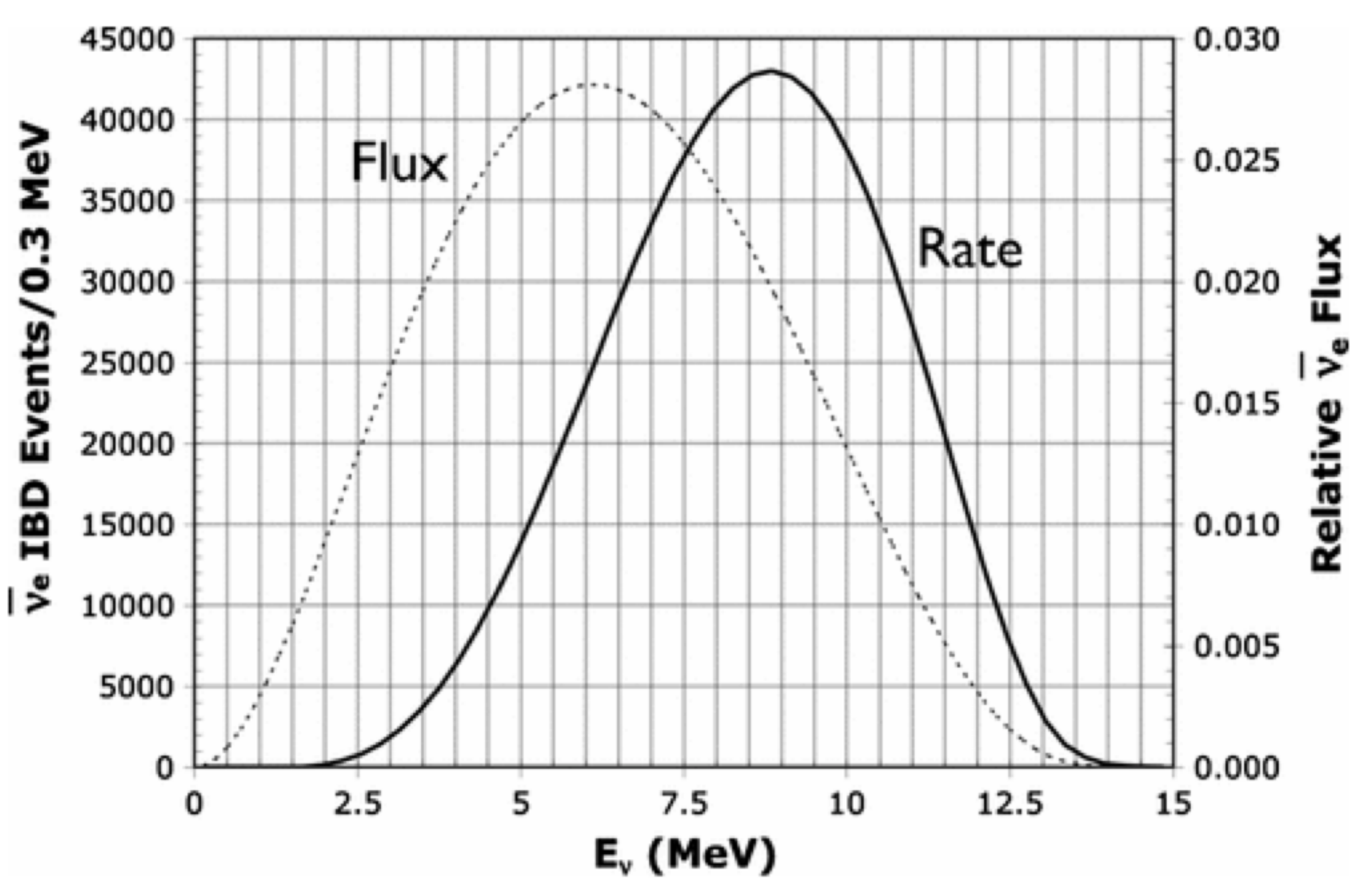}
\caption{The $^8$Li antineutrino spectrum for a $^8$Li IsoDAR source for 5 years of data.}
\label{fig:8Li}
\vspace{-0.5cm}
\end{center}
\end{figure}

\section{Detector Configuration}\label{s:detector}

Although the optimal detector configuration is under study and will depend on physics priorities, some requirements are already clear:
\begin{itemize}
\item Excellent separation of Cherenkov and scintillation light -- enables ring imaging for the long-baseline program, and directionality information for background rejection in the low-energy program.
\item A fiducial volume of at least 20--100~kT.
\item A light yield of at least 150 photoelectrons per MeV is required for the minimal resolution needed for a sensitive $0\nu\beta\beta$ search.  (This places a requirement on the intrinsic light yield and attenuation of the cocktail).
\item The effect of isotope loading ($e.g.$ $^{130}$Te for $0\nu\beta\beta$ or $^7$Li for solar) must not detrimentally affect the optics in any significant way.
\end{itemize}

In order to further optimize the detector design for particular physics goals, we will need to understand the impact of each property on the final physics sensitivity in detail.  
For this purpose a full Monte Carlo simulation of the proposed detector is under development.  
Results from R\&D at BNL and other institutions will be incorporated in order  to define the properties of the WbLS target and the behavior and response of different photon detection methods.  The simulation will be used to determine the impact of each detector design choice on the sensitivity to a particular physics goal.  This will allow us to optimize the detector design and WbLS target configuration in order to maximize physics output from the ASDC.

The proposed 1~T prototype at BNL will allow full characterization and optimization of the WbLS target on a small scale.  A proposed deployment of WbLS in a second phase of the $\sim$~kT-scale WATCHMAN detector would allow complete characterization on the scales required for a convincing demonstration of event reconstruction and signal separation, including a possible test of advanced photon detection methods such as LAPPDs.

\subsection{Target Cocktail}
Extensive development of the WbLS chemical cocktail is ongoing at BNL.  
The most critical issue is a complete understanding of the 
optical properties and stability of WbLS, the effect of various isotope loadings, and optimization for precision particle detection and maximal physics impact.    
To fully characterize this material we need to understand several features, including: 
\begin{itemize}
\item Light production level; 
\item Loss of light (attenuation) due to absorption and scattering; 
\item The fraction of Cherenkov to scintillation light produced; 
\item The time profile of the emitted scintillation light (for Cherenkov/scintillation separation, and particle ID);
\item The effect of isotope loading on each of these (as a function of loading fraction);
\item The stability of each of these properties over time.  
\end{itemize}

Detector energy resolution depends on the number of  photons detected, which is determined by the initial light production and the level of attenuation in the target. 
The fraction of Cherenkov light determines the accuracy of direction reconstruction, which impacts signal-background separation.  The scintillation timing also provides a handle for background discrimination due to known differences in the timing response for different particle types. 
Each of these properties is critical for high-precision measurement.  Each can be affected by variables in the target, including: the ratio of water to scintillator; the addition of wavelength shifter to move away from a region with high absorption; and the level of isotope loading.

\subsubsection{Isotope Loading}
There are many options for isotopic loading, depending on the relative prioritization of the physics goals.  These include $^7$Li, $^{37}$Cl or $^{71}$Ga for charged-current detection of solar neutrinos, $^{130}$Te and other candidates for a $0\nu\beta\beta$ search, Gd for enhanced neutron detection, as well as many other options.  Our working group will evaluate these options on the basis of the scientific goals and technical requirements in order to select the one or two isotopic loadings that will best advance the overall physics program.

\subsection{Detector Geometry}
While it is most natural to first consider a monolithic detector, similar to the Super-Kamiokande~\cite{sk}, SNO~\cite{sno}, KamLAND~\cite{kl}, and Borexino~\cite{Bor} detectors so well known, we should investigate other possibilities which have emerged since the design of those instruments, and which may open new avenues of sensitivity.  In particular a scaled up version of the geometry of the LENS~\cite{lens} solar neutrino experiment might provide an opportunity for increasing sensitivity and false event rejection for a variety of tasks in this WBLS application.  The LENS scheme depends on the Raghavan Optical Lattice, whereby the volume becomes subdivided into a cubical lattice by having totally internally reflecting boundaries, which channel light from one cell to the six detectors along the three cardinal directions from that cell.  The net effect is to cleanly segment the detector volume into a large number of well-resolved, independent cells without the limitations of using timing, or loss of light.

In practice, for example, this geometry could be realized through the use of long thin parallel acrylic planes, sealed to provide a 5 mil air gap, as have been built (in small scale) for LENS.  The number of optical detectors need not be much different from an open detector design, so the costs would be similar except for the optical lattice.  Clearly liquid circulation would not be possible, but some argue that it is neither necessary nor wanted (as shown by experience in KamLAND and Borexino). 

A modest 1~m-scale detector NuLat is now under construction for purposes of neutrino study near nuclear reactors, and a Whitepaper for that project is available~\cite{NuLat}.

The final decision about such a radical geometry change will have to await studies, though some collaborators claim that it may have significant positive effect on the accelerator and atmospheric neutrino experiments, as well as possibly solar neutrino studies.  Detailed work is needed to study this question further.

\subsection{Photon Detection Methods}
For the low-energy program we envision, high-efficiency light
collection will be necessary, and for much of the program the separation of
Cherenkov and scintillation light is critical.  To collect enough photons to
achieve thresholds near or below 1~MeV, off-the-shelf large-area high-quantum
efficiency photomultiplier tubes could be used, which have peak efficiencies
around 34\%.  The 12'' version of these, the R11780, also have transit timing
jitter distributions whose half-width at half-maximum is around 1.2~ns,
possibly fast enough to do reasonable scintillation/Cherenkov separation. To
reduce costs, either reflectors or wavelength-shifting plates could be used,
increasing the effective collection area (thus reducing the number of PMTs)
with only small effects on the timing.

\subsubsection{LAPPDs}

Large Area Picosecond Photodetectors (LAPPDs) are imaging, vacuum photosensors based on microchannel plate (MCP) technology~\cite{wiza}. The LAPPD collaboration~\cite{LAPPD} has developed new fabrication techniques for making large-area (8" x 8") MCP photodetectors using low-cost materials, well established industrial batch techniques, and advances in material science~\cite{lappd1}.

\begin{figure}[!h]
	\begin{center}
	    \includegraphics[width=0.6\linewidth]{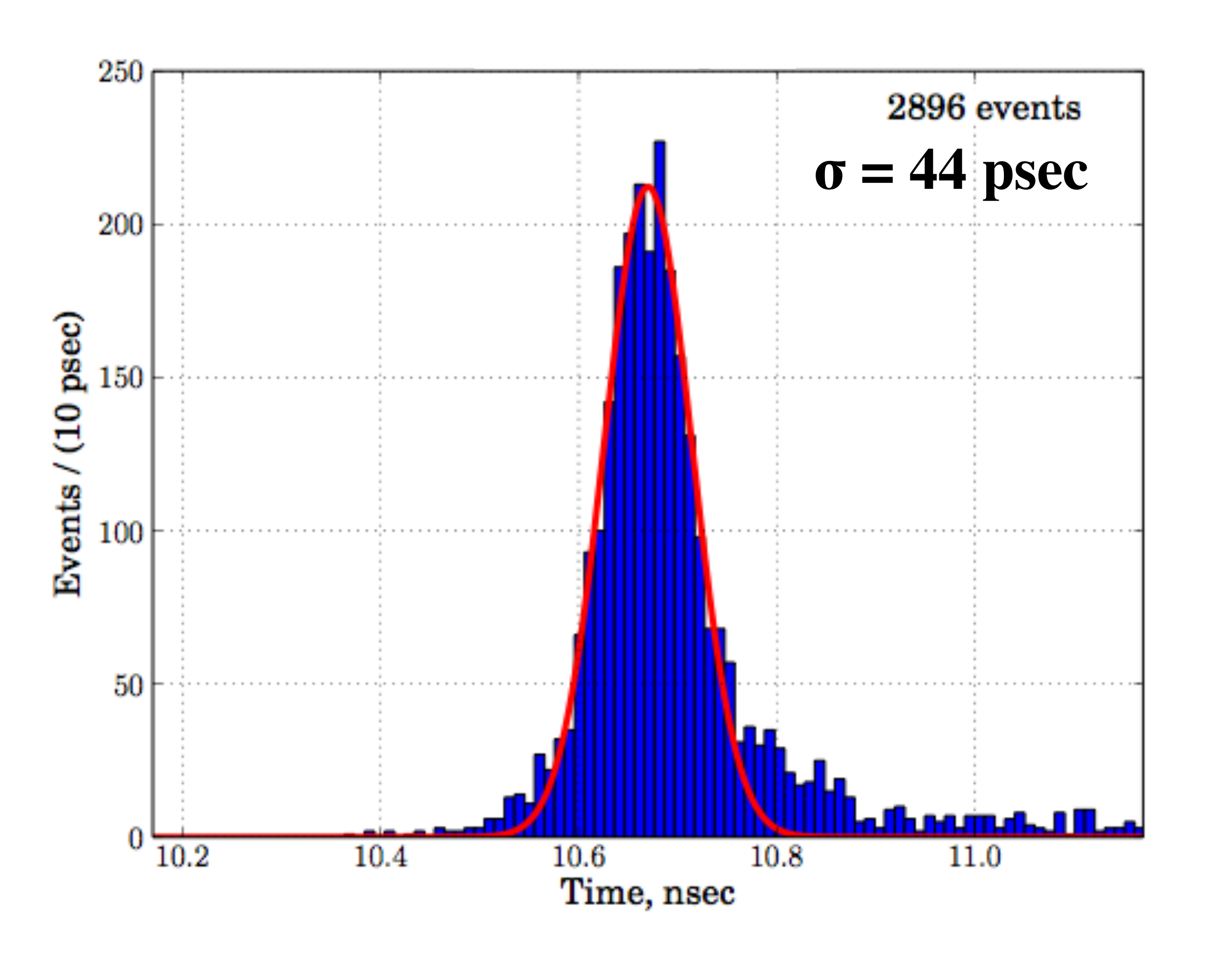}	
	\caption{(Preliminary) The transit-time spread (TTS) of a pair of 8''x 8'' Microchannel plates, measured from an LAPPD designed 8'' anode. (b) Pulse height distribution, showing the spread of single photoelectron gains from an 8''x8'' MCP pair.}
	\label{TTS}
		\end{center}
		\vspace{-0.5cm}
\end{figure}

LAPPDs offer excellent time resolution. Prototype systems consistently demonstrate time resolutions at or below 100 picoseconds. Figure~\ref{TTS} shows the 44 psec spread of measured arrival times of single photoelectrons in one configuration of an LAPPD system. These resolutions represent a significant improvement on the $>$ 1 nsec resolutions of conventional phototubes. The segmented microstripline anode of the LAPPD design allows for sub-centimeter spatial resolution. With gains of order $10^7$ for single photoelectrons and large area photocathodes shown to provide $>20\%$ quantum efficiency, LAPPDs are also effective single-photon counting devices. A major milestone of the project was the production of low power, fast pulse-digitization electronics to fully exploit the capabilities of these photosensors~\cite{{lappd3},{lappd4}}. A complete end-to-end detector system at the Argonne LAPPD test facility has been tested, using a so-called ``demountable" LAPPD and the full PSEC-based front-end~\cite{APS}. 

\begin{figure}[!h]
	\begin{center}
		\includegraphics[width=0.6\linewidth]{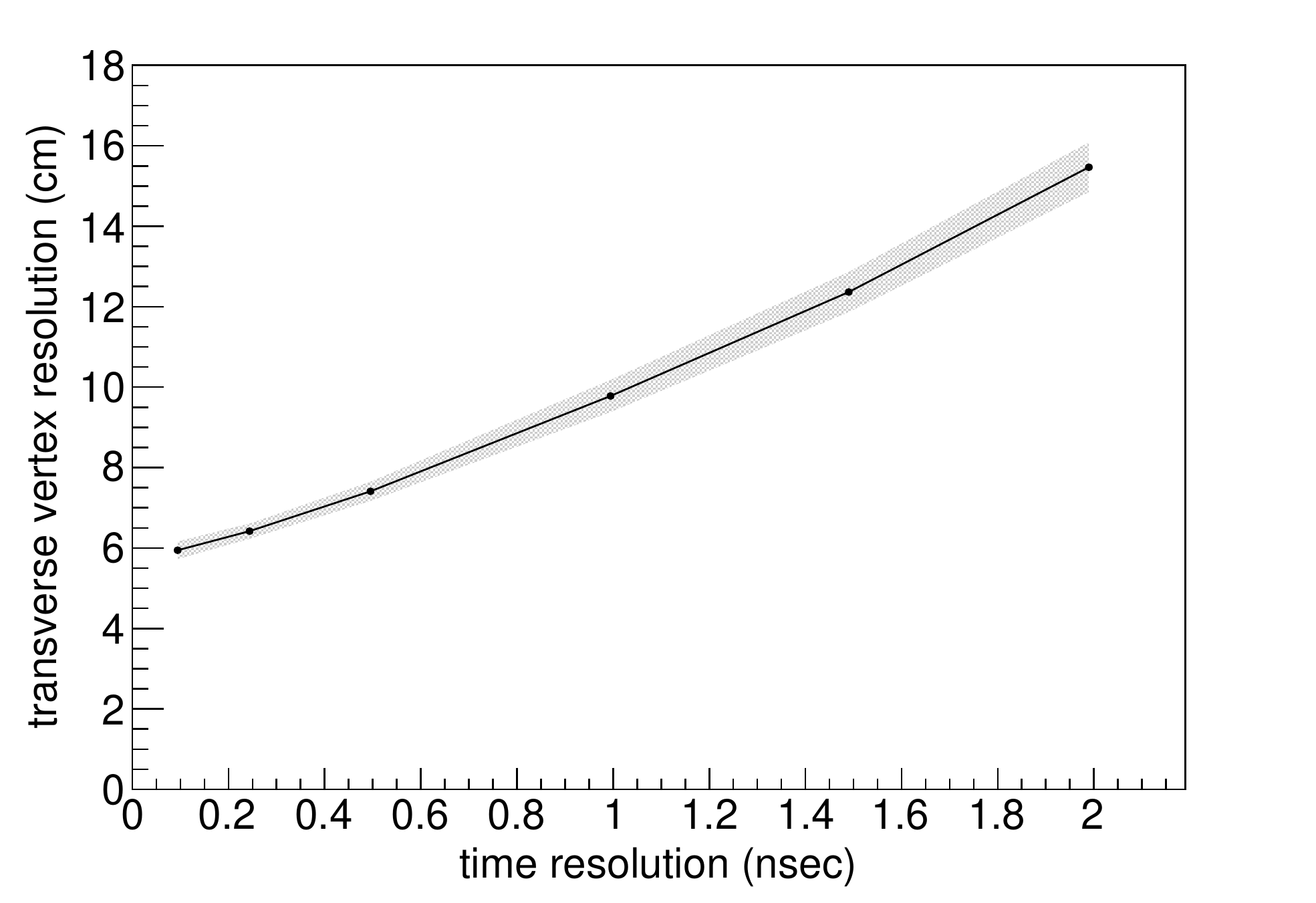}
	\caption{Resolution of timing-based muon vertex fits in the direction transverse to the track direction, as a function of photosensor resolution. There is nearly a factor of 3 improvement as the photosensor resolution goes from 2 nsec to 100 psec.}
	\label{transverseres}
	\end{center}
		\vspace{-0.75cm}
\end{figure}

Fast-timing capabilities such as those demonstrated by LAPPDs will be essential in separating between the Cherenkov and scintillation light in the ASDC~\cite{directionality}. In rejecting backgrounds, such as high energy neutral pions decaying to two very forward gammas, the ability to reconstruct particle tracks with much finer granularity is also critical. Preliminary Monte Carlo studies indicate that measurements of Cherenkov photon arrival space-time points with resolutions of 1 cm and 100 psec will allow the detector to reconstruct track features approaching size scales of just a few centimeters, even in large ($>$200 kton) detectors with coverages at or below 20$\%$ (Fig~\ref{transverseres})~\cite{fastneutrino}. Moreover, While conventional photomultipliers are single-pixel detectors, LAPPDs are imaging tubes, able to resolve the position and time of single incident photons within an individual sensor. This would enable reconstruction of events very close to the wall of the detector, where the light can only spread over a small area, thereby making better use of the target volume.

Critical to the success of LAPPDs in future neutrino experiments will be continued simulation work and, above all, successful demonstration in small-scale experiments. The $\sim$30 ton ANNIE experiment~\cite{ANNIE} and 1 kton WATCHMAN detector~\cite{wm} will provide a pathway for these technical demonstrations.

\newpage
\section{Conclusions}

This paper describes the Advanced Scintillation Detector Concept (ASDC) - a powerful scientific instrument for the next-generation of neutrino physics and rare-event searches.  The ASDC combines the newly-developed water-based liquid scintillator (WbLS) medium with ultra-fast and highly efficient photon detection methods,  and the potential for isotopic doping to enhance the physics program. 
Deployment of such a detector at a deep underground laboratory with careful controls of low-energy radioactive backgrounds could have major impacts in a broad range of physics topics, with the potential to revolutionize next-generation frontier experiments.  

WbLS is a new, cost-effective target for massive detectors with the unique capability of exploring physics below the Cherenkov threshold. 
A WbLS target allows for the high light yield
and correspondingly high energy resolution and low threshold of a scintillator
experiment in a directionally sensitive detector.  Directionality adds a powerful handle for signal separation to a low-threshold, high-precision detector.  The admixture of water also increases the attenuation length, thereby allowing the possibility of a very
large detector. The development of new, high-precision timing devices in combination with control over the timing of different scintillation components in the target via various additives could enable a high degree of separation between the prompt Cherenkov
light and the slower scintillation light.  This offers many advantages, including background rejection for directional signals such as solar neutrinos even at low energies, and a powerful long-baseline program.  The ability to load hydrophilic ions further
increases the physics program, including the potential for a $0\nu\beta\beta$
search with an isotopic mass on the scale of tens of tons.  

The Long Baseline Neutrino Facility (LBNF), a \$1B flagship high-energy physics program in the US, is currently considering choices for an additional detector module to complement the planned underground liquid argon detector.  The addition of a 30--50~kT WbLS module would add sensitivity equivalent to or greater than an additional 10~kT of LArTPC, with the advantage of an independent, complementary measurement and a broad program of additional physics topics.  

In order to realize this goal, a vigorous and forward-looking program of R\&D is called for targeting the properties of the WbLS, along with further development of fast-timing solutions and the associated readout electronics.  Such a program is already underway, with the goal of developing the WbLS target and associated detector technology for broad physics goals, from long-baseline to low-energy physics.  
A full technical design and cost estimate is  under consideration.  Many of the risks and challenges associated with the ASDC have been addressed in the development of the Technical Design Report for the water Cherenkov detector proposed by the LBNE collaboration~\cite{lbne-wcd-cdr}.   Additional risks are associated with the use of new technology: the WbLS target, and new ultra-fast timing photosensors.  WbLS has been well studied on a bench-top scale; the primary risk remaining is in the attenuation length and stability of optical properties over long time periods.  These will be addressed with high priority.  While the use of new technology such as LAPPDs could significantly enhance performance, the baseline detector design assumes conventional high-QE PMTs, which are well characterized and understood.  If LAPPDs or an alternative technology are available and tested prior to construction, then the possible replacement could be of benefit to the physics program.  Planned tests of new technologies at Brookhaven National Laboratory, and in the ANNIE and WATCHMAN detectors will be critical steps in the developing R\&D program.

The ASDC offers a unique opportunity to combine conventional neutrino physics with rare-event searches in a single, large-scale detector.  
Use of the novel and inexpensive WbLS target  could signify a breakthrough in background-rejection capability and signal detection efficiency, allowing detectors to be scaled by an order of magnitude or more, which would revolutionize the field.  
In addition,  the ASDC would have flexibility to adapt to new directions in the scientific program as the field evolves, making it a unique instrument of discovery.


\end{document}